\documentclass[longauth]{aa}

\usepackage{graphicx}
\usepackage{booktabs}
\usepackage{txfonts}
\usepackage[dvipsnames]{xcolor}

\newcommand{\trades}[0]{\textsc{TRADES}}
\newcommand{\pyde}[0]{\textsc{PyDE}}
\newcommand{\plb}[0]{\mathrm{b}}
\newcommand{\plc}[0]{\mathrm{c}}
\newcommand{\unif}[2]{\ensuremath{\mathcal{U}(#1,#2)}}

\usepackage{hyperref}
\hypersetup{
    colorlinks=true,
    linkcolor=blue,
    citecolor=blue,
    filecolor=blue,      
    urlcolor=blue,
    breaklinks=true,
}

\renewcommand{\arraystretch}{1.3}

\begin{document} 

   \title{
   A joint effort to discover and characterize two resonant mini Neptunes around TOI-1803 with TESS, HARPS-N and CHEOPS
   \thanks{This study uses CHEOPS data observed as part of the Guaranteed Time Observation (GTO) program CH\_PR100031 and the  observations made with the Italian \textit{Telescopio Nazionale Galileo} (TNG) operated by the \textit{Fundaci\`{o}n Galileo Galilei} (FGG) of the \textit{Istituto Nazionale di Astrofisica} (INAF) at the \textit{Observatorio del Roque de los Muchachos} (La Palma, Canary Islands, Spain).}
   } 
   

\author{
T.~Zingales\inst{\ref{inst:1},\ref{inst:2}}\,$^{\href{https://orcid.org/0000-0001-6880-5356}{\protect\includegraphics[height=0.19cm]{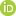}}}$, 
L.~Malavolta\inst{\ref{inst:1},\ref{inst:2}}\,$^{\href{https://orcid.org/0000-0002-6492-2085}{\protect\includegraphics[height=0.19cm]{orcid.jpg}}}$, 
L.~Borsato\inst{\ref{inst:2}}\,$^{\href{https://orcid.org/0000-0003-0066-9268}{\protect\includegraphics[height=0.19cm]{orcid.jpg}}}$, 
D.~Turrini\inst{\ref{inst:17}}\,$^{\href{https://orcid.org/0000-0002-1923-7740}{\protect\includegraphics[height=0.19cm]{orcid.jpg}}}$, 
A.~Bonfanti\inst{\ref{inst:3}}\,$^{\href{https://orcid.org/0000-0002-1916-5935}{\protect\includegraphics[height=0.19cm]{orcid.jpg}}}$, 
D.~Polychroni\inst{\ref{inst:18}}\,$^{\href{https://orcid.org/0000-0002-7657-7418}{\protect\includegraphics[height=0.19cm]{orcid.jpg}}}$, 
G.~Mantovan\inst{\ref{inst:1},\ref{inst:2}}\,$^{\href{https://orcid.org/0000-0002-6871-6131}{\protect\includegraphics[height=0.19cm]{orcid.jpg}}}$
D.~Nardiello\inst{\ref{inst:1},\ref{inst:2}}\,$^{\href{https://orcid.org/0000-0003-1149-3659}{\protect\includegraphics[height=0.19cm]{orcid.jpg}}}$, 
V.~Nascimbeni\inst{\ref{inst:2}}\,$^{\href{https://orcid.org/0000-0001-9770-1214}{\protect\includegraphics[height=0.19cm]{orcid.jpg}}}$, 
A.~F.~Lanza\inst{\ref{inst:19}}\,$^{\href{https://orcid.org/0000-0001-5928-7251}{\protect\includegraphics[height=0.19cm]{orcid.jpg}}}$, 
A.~Bekkelien\inst{\ref{inst:5}}, 
A.~Sozzetti\inst{\ref{inst:17}}\,$^{\href{https://orcid.org/0000-0002-7504-365X}{\protect\includegraphics[height=0.19cm]{orcid.jpg}}}$, 
C.~Broeg\inst{\ref{inst:6},\ref{inst:7}}\,$^{\href{https://orcid.org/0000-0001-5132-2614}{\protect\includegraphics[height=0.19cm]{orcid.jpg}}}$, 
L.~Naponiello\inst{\ref{inst:17}}\,$^{\href{https://orcid.org/0000-0001-9390-0988}{\protect\includegraphics[height=0.19cm]{orcid.jpg}}}$, 
M.~Lendl\inst{\ref{inst:5}}\,$^{\href{https://orcid.org/0000-0001-9699-1459}{\protect\includegraphics[height=0.19cm]{orcid.jpg}}}$, 
A.~S.~Bonomo\inst{\ref{inst:17}}\,$^{\href{https://orcid.org/0000-0002-6177-198X}{\protect\includegraphics[height=0.19cm]{orcid.jpg}}}$, 
A.~E.~Simon\inst{\ref{inst:6},\ref{inst:7}}\,$^{\href{https://orcid.org/0000-0001-9773-2600}{\protect\includegraphics[height=0.19cm]{orcid.jpg}}}$, 
S.~Desidera\inst{\ref{inst:2}}\,$^{\href{https://orcid.org/0000-0001-8613-2589}{\protect\includegraphics[height=0.19cm]{orcid.jpg}}}$, 
G.~Piotto\inst{\ref{inst:1},\ref{inst:2}}\,$^{\href{https://orcid.org/0000-0002-9937-6387}{\protect\includegraphics[height=0.19cm]{orcid.jpg}}}$, 
L.~Mancini\inst{\ref{inst:21},\ref{inst:24},\ref{inst:25}}\,$^{\href{https://orcid.org/0000-0002-9428-8732}{\protect\includegraphics[height=0.19cm]{orcid.jpg}}}$, 
M.~J.~Hooton\inst{\ref{inst:8}}\,$^{\href{https://orcid.org/0000-0003-0030-332X}{\protect\includegraphics[height=0.19cm]{orcid.jpg}}}$, 
A.~Bignamini\inst{\ref{inst:26}}\,$^{\href{https://orcid.org/0000-0002-5606-6354}{\protect\includegraphics[height=0.19cm]{orcid.jpg}}}$, 
J.~A.~Egger\inst{\ref{inst:6}}\,$^{\href{https://orcid.org/0000-0003-1628-4231}{\protect\includegraphics[height=0.19cm]{orcid.jpg}}}$, 
A.~Maggio\inst{\ref{inst:27}}\,$^{\href{https://orcid.org/0000-0001-5154-6108}{\protect\includegraphics[height=0.19cm]{orcid.jpg}}}$, 
Y.~Alibert\inst{\ref{inst:7},\ref{inst:6}}\,$^{\href{https://orcid.org/0000-0002-4644-8818}{\protect\includegraphics[height=0.19cm]{orcid.jpg}}}$, 
D.~Locci\inst{\ref{inst:27}}\,$^{\href{https://orcid.org/0000-0002-9824-2336}{\protect\includegraphics[height=0.19cm]{orcid.jpg}}}$, 
L.~Delrez\inst{\ref{inst:9},\ref{inst:10},\ref{inst:11}}\,$^{\href{https://orcid.org/0000-0001-6108-4808}{\protect\includegraphics[height=0.19cm]{orcid.jpg}}}$, 
F.~Biassoni\inst{\ref{inst:28}}\,$^{\href{https://orcid.org/0000-0001-9113-3906}{\protect\includegraphics[height=0.19cm]{orcid.jpg}}}$, 
L.~Fossati\inst{\ref{inst:3}}\,$^{\href{https://orcid.org/0000-0003-4426-9530}{\protect\includegraphics[height=0.19cm]{orcid.jpg}}}$, 
L.~Cabona\inst{\ref{inst:29}}\,$^{\href{https://orcid.org/0000-0002-5130-4827}{\protect\includegraphics[height=0.19cm]{orcid.jpg}}}$, 
G.~Lacedelli\inst{\ref{inst:12}}\,$^{\href{https://orcid.org/0000-0002-4197-7374}{\protect\includegraphics[height=0.19cm]{orcid.jpg}}}$, 
I.~Carleo\inst{\ref{inst:30}}\,$^{\href{https://orcid.org/0000-0002-0810-3747}{\protect\includegraphics[height=0.19cm]{orcid.jpg}}}$, 
P.~Leonardi\inst{\ref{inst:14},\ref{inst:1},\ref{inst:2}}\,$^{\href{https://orcid.org/0000-0001-6026-9202}{\protect\includegraphics[scale=0.5]{orcid.jpg}}}$, 
G.~Andreuzzi\inst{\ref{inst:31},\ref{inst:32}}\,$^{\href{https://orcid.org/0000-0001-5125-6397}{\protect\includegraphics[height=0.19cm]{orcid.jpg}}}$, 
A.~Brandeker\inst{\ref{inst:15}}\,$^{\href{https://orcid.org/0000-0002-7201-7536}{\protect\includegraphics[height=0.19cm]{orcid.jpg}}}$, 
R.~Cosentino\inst{\ref{inst:19},\ref{inst:32}}\,$^{\href{https://orcid.org/0000-0003-1784-1431}{\protect\includegraphics[height=0.19cm]{orcid.jpg}}}$, 
A.~C.~M.~Correia\inst{\ref{inst:16}}\,$^{\href{https://orcid.org/0000-0002-8946-8579}{\protect\includegraphics[height=0.19cm]{orcid.jpg}}}$, 
R.~Claudi\inst{\ref{inst:2}}\,$^{\href{https://orcid.org/0000-0001-7707-5105}{\protect\includegraphics[height=0.19cm]{orcid.jpg}}}$, 
R.~Alonso\inst{\ref{inst:12},\ref{inst:33}}\,$^{\href{https://orcid.org/0000-0001-8462-8126}{\protect\includegraphics[height=0.19cm]{orcid.jpg}}}$, 
M.~Damasso\inst{\ref{inst:17}}\,$^{\href{https://orcid.org/0000-0001-9984-4278}{\protect\includegraphics[height=0.19cm]{orcid.jpg}}}$, 
T.~G.~Wilson\inst{\ref{inst:61}}\,$^{\href{https://orcid.org/0000-0001-8749-1962}{\protect\includegraphics[height=0.19cm]{orcid.jpg}}}$,
T.~B\`{a}rczy\inst{\ref{inst:34}}\,$^{\href{https://orcid.org/0000-0002-7822-4413}{\protect\includegraphics[height=0.19cm]{orcid.jpg}}}$, 
M.~Pinamonti\inst{\ref{inst:17}}\,$^{\href{https://orcid.org/0000-0002-4445-1845}{\protect\includegraphics[height=0.19cm]{orcid.jpg}}}$, 
D.~Baker\inst{\ref{inst:85}}\,$^{\href{https://orcid.org/0000-0002-4445-1845}{\protect\includegraphics[height=0.19cm]{orcid.jpg}}}$, 
K.~Barkaoui\inst{\ref{inst:9},\ref{inst:78},\ref{inst:12}}\,$^{\href{https://orcid.org/0000-0003-1464-9276}{\protect\includegraphics[height=0.19cm]{orcid.jpg}}}$, 
D.~Barrado~Navascues\inst{\ref{inst:35}}\,$^{\href{https://orcid.org/0000-0002-5971-9242}{\protect\includegraphics[height=0.19cm]{orcid.jpg}}}$, 
S.~C.~C.~Barros\inst{\ref{inst:36},\ref{inst:37}}\,$^{\href{https://orcid.org/0000-0003-2434-3625}{\protect\includegraphics[height=0.19cm]{orcid.jpg}}}$, 
W.~Baumjohann\inst{\ref{inst:3}}\,$^{\href{https://orcid.org/0000-0001-6271-0110}{\protect\includegraphics[height=0.19cm]{orcid.jpg}}}$, 
T.~Beck\inst{\ref{inst:6}}, 
C.~Beichman\inst{\ref{inst:68}}\,$^{\href{https://orcid.org/0000-0002-5741-3047}{\protect\includegraphics[height=0.19cm]{orcid.jpg}}}$,
W.~Benz\inst{\ref{inst:6},\ref{inst:7}}\,$^{\href{https://orcid.org/0000-0001-7896-6479}{\protect\includegraphics[height=0.19cm]{orcid.jpg}}}$, 
A.~Bieryla\inst{\ref{inst:79}}\,$^{\href{https://orcid.org/0000-0001-6637-5401}{\protect\includegraphics[height=0.19cm]{orcid.jpg}}}$, 
N.~Billot\inst{\ref{inst:5}}\,$^{\href{https://orcid.org/0000-0003-3429-3836}{\protect\includegraphics[height=0.19cm]{orcid.jpg}}}$, 
P.~Bosch-Cabot\inst{\ref{inst:84}},
L.~G.~Bouma\inst{\ref{inst:76}}\,$^{\href{https://orcid.org/0000-0002-0514-5538}{\protect\includegraphics[height=0.19cm]{orcid.jpg}}}$, 
D.~R.~Ciardi\inst{\ref{inst:68}}\,$^{\href{https://orcid.org/0000-0002-5741-3047}{\protect\includegraphics[height=0.19cm]{orcid.jpg}}}$, 
A.~Collier~Cameron\inst{\ref{inst:38}}\,$^{\href{https://orcid.org/0000-0002-8863-7828}{\protect\includegraphics[height=0.19cm]{orcid.jpg}}}$, 
K.~A.~Collins\inst{\ref{inst:79}}\,$^{\href{https://orcid.org/0000-0001-6588-9574}{\protect\includegraphics[height=0.19cm]{orcid.jpg}}}$, 
Ian~J.~M.~Crossfield\inst{\ref{inst:72}},
Sz.~Csizmadia\inst{\ref{inst:39}}\,$^{\href{https://orcid.org/0000-0001-6803-9698}{\protect\includegraphics[height=0.19cm]{orcid.jpg}}}$, 
P.~E.~Cubillos\inst{\ref{inst:3},\ref{inst:40}}, 
M.~B.~Davies\inst{\ref{inst:41}}\,$^{\href{https://orcid.org/0000-0001-6080-1190}{\protect\includegraphics[height=0.19cm]{orcid.jpg}}}$, 
M.~Deleuil\inst{\ref{inst:42}}\,$^{\href{https://orcid.org/0000-0001-6036-0225}{\protect\includegraphics[height=0.19cm]{orcid.jpg}}}$, 
A.~Deline\inst{\ref{inst:5}}, 
O.~D.~S.~Demangeon\inst{\ref{inst:36},\ref{inst:37}}\,$^{\href{https://orcid.org/0000-0001-7918-0355}{\protect\includegraphics[height=0.19cm]{orcid.jpg}}}$, 
B.-O.~Demory\inst{\ref{inst:7},\ref{inst:6}}\,$^{\href{https://orcid.org/0000-0002-9355-5165}{\protect\includegraphics[height=0.19cm]{orcid.jpg}}}$, 
A.~Derekas\inst{\ref{inst:43}}, 
D.~Dragomir\inst{\ref{inst:80}}\,$^{\href{https://orcid.org/0000-0003-2313-467X}{\protect\includegraphics[height=0.19cm]{orcid.jpg}}}$,
B.~Edwards\inst{\ref{inst:44}}, 
D.~Ehrenreich\inst{\ref{inst:5},\ref{inst:45}}\,$^{\href{https://orcid.org/0000-0001-9704-5405}{\protect\includegraphics[height=0.19cm]{orcid.jpg}}}$, 
A.~Erikson\inst{\ref{inst:39}}, 
B.~Falk\inst{\ref{inst:75}},
A.~Fortier\inst{\ref{inst:6},\ref{inst:7}}\,$^{\href{https://orcid.org/0000-0001-8450-3374}{\protect\includegraphics[height=0.19cm]{orcid.jpg}}}$, 
M.~Fridlund\inst{\ref{inst:46},\ref{inst:47}}\,$^{\href{https://orcid.org/0000-0002-0855-8426}{\protect\includegraphics[height=0.19cm]{orcid.jpg}}}$, 
A.~Fukui\inst{\ref{inst:86}, \ref{inst:12}}\,$^{\href{https://orcid.org/0000-0002-4909-5763}{\protect\includegraphics[height=0.19cm]{orcid.jpg}}}$, 
D.~Gandolfi\inst{\ref{inst:48}}\,$^{\href{https://orcid.org/0000-0001-8627-9628}{\protect\includegraphics[height=0.19cm]{orcid.jpg}}}$, 
K.~Gazeas\inst{\ref{inst:49}}, 
M.~Gillon\inst{\ref{inst:9}}\,$^{\href{https://orcid.org/0000-0003-1462-7739}{\protect\includegraphics[height=0.19cm]{orcid.jpg}}}$, 
E.~Gonzales\inst{\ref{inst:70}}\,$^{\href{https://orcid.org/0000-0002-9329-2190}{\protect\includegraphics[height=0.19cm]{orcid.jpg}}}$,
M.~Güdel\inst{\ref{inst:50}}, 
P.~Guerra\inst{\ref{inst:84}}\,$^{\href{https://orcid.org/0000-0002-0619-7639}{\protect\includegraphics[height=0.19cm]{orcid.jpg}}}$, 
M.~N.~Günther\inst{\ref{inst:51}}\,$^{\href{https://orcid.org/0000-0002-3164-9086}{\protect\includegraphics[height=0.19cm]{orcid.jpg}}}$, 
A.~Heitzmann\inst{\ref{inst:5}}\,$^{\href{https://orcid.org/0000-0002-8091-7526}{\protect\includegraphics[height=0.19cm]{orcid.jpg}}}$, 
Ch.~Helling\inst{\ref{inst:3},\ref{inst:52}}, 
S.~B.~Howell\inst{\ref{inst:69}}\,$^{\href{https://orcid.org/0000-0002-2532-2853}{\protect\includegraphics[height=0.19cm]{orcid.jpg}}}$, 
K.~G.~Isaak\inst{\ref{inst:51}}\,$^{\href{https://orcid.org/0000-0001-8585-1717}{\protect\includegraphics[height=0.19cm]{orcid.jpg}}}$, 
J.~Jenkins\inst{\ref{inst:69}}\,$^{\href{https://orcid.org/0000-0002-4715-9460}{\protect\includegraphics[height=0.19cm]{orcid.jpg}}}$, 
L.~L.~Kiss\inst{\ref{inst:53},\ref{inst:54}}, 
J.~Korth\inst{\ref{inst:55}}\,$^{\href{https://orcid.org/0000-0002-0076-6239}{\protect\includegraphics[height=0.19cm]{orcid.jpg}}}$,
K.~W.~F.~Lam\inst{\ref{inst:39}}\,$^{\href{https://orcid.org/0000-0002-9910-6088}{\protect\includegraphics[height=0.19cm]{orcid.jpg}}}$, 
J.~Laskar\inst{\ref{inst:56}}\,$^{\href{https://orcid.org/0000-0003-2634-789X}{\protect\includegraphics[height=0.19cm]{orcid.jpg}}}$, 
A.~Lecavelier~des~Etangs\inst{\ref{inst:57}}\,$^{\href{https://orcid.org/0000-0002-5637-5253}{\protect\includegraphics[height=0.19cm]{orcid.jpg}}}$, 
D.~Magrin\inst{\ref{inst:2}}\,$^{\href{https://orcid.org/0000-0003-0312-313X}{\protect\includegraphics[height=0.19cm]{orcid.jpg}}}$, 
R.~Matson\inst{\ref{inst:69}}\,$^{\href{https://orcid.org/0000-0002-2532-2853}{\protect\includegraphics[height=0.19cm]{orcid.jpg}}}$, 
E.~C.~Matthews\inst{\ref{inst:25}}\,$^{\href{https://orcid.org/0000-0003-0593-1560}{\protect\includegraphics[height=0.19cm]{orcid.jpg}}}$, 
P.~F.~L.~Maxted\inst{\ref{inst:58}}\,$^{\href{https://orcid.org/0000-0003-3794-1317}{\protect\includegraphics[height=0.19cm]{orcid.jpg}}}$, 
S.~McDermott \inst{\ref{inst:77}},
M.~Munari\inst{\ref{inst:19}}\,$^{\href{https://orcid.org/0000-0003-0990-050X}{\protect\includegraphics[height=0.19cm]{orcid.jpg}}}$, 
C.~Mordasini\inst{\ref{inst:6},\ref{inst:7}}, 
N.~Narita\inst{\ref{inst:86}, \ref{inst:87}, \ref{inst:12}}\,$^{\href{https://orcid.org/0000-0001-8511-2981}{\protect\includegraphics[height=0.19cm]{orcid.jpg}}}$, 
G.~Olofsson\inst{\ref{inst:15}}\,$^{\href{https://orcid.org/0000-0003-3747-7120}{\protect\includegraphics[height=0.19cm]{orcid.jpg}}}$, 
R.~Ottensamer\inst{\ref{inst:50}}, 
I.~Pagano\inst{\ref{inst:19}}\,$^{\href{https://orcid.org/0000-0001-9573-4928}{\protect\includegraphics[height=0.19cm]{orcid.jpg}}}$, 
E.~Pall\`{e}\inst{\ref{inst:12},\ref{inst:33}}\,$^{\href{https://orcid.org/0000-0003-0987-1593}{\protect\includegraphics[height=0.19cm]{orcid.jpg}}}$, 
G.~Peter\inst{\ref{inst:60}}\,$^{\href{https://orcid.org/0000-0001-6101-2513}{\protect\includegraphics[height=0.19cm]{orcid.jpg}}}$, 
D.~Pollacco\inst{\ref{inst:61}}, 
D.~Queloz\inst{\ref{inst:62},\ref{inst:8}}\,$^{\href{https://orcid.org/0000-0002-3012-0316}{\protect\includegraphics[height=0.19cm]{orcid.jpg}}}$, 
R.~Ragazzoni\inst{\ref{inst:2},\ref{inst:1}}\,$^{\href{https://orcid.org/0000-0002-7697-5555}{\protect\includegraphics[height=0.19cm]{orcid.jpg}}}$, 
N.~Rando\inst{\ref{inst:51}}, 
F.~Ratti{\inst{\ref{inst:51}}},
H.~Rauer\inst{\ref{inst:39},\ref{inst:63}}\,$^{\href{https://orcid.org/0000-0002-6510-1828}{\protect\includegraphics[height=0.19cm]{orcid.jpg}}}$, 
I.~Ribas\inst{\ref{inst:64},\ref{inst:65}}\,$^{\href{https://orcid.org/0000-0002-6689-0312}{\protect\includegraphics[height=0.19cm]{orcid.jpg}}}$, 
S.~Salmon\inst{\ref{inst:5}}\,$^{\href{https://orcid.org/0000-0002-1714-3513}{\protect\includegraphics[height=0.19cm]{orcid.jpg}}}$, 
N.~C.~Santos\inst{\ref{inst:36},\ref{inst:37}}\,$^{\href{https://orcid.org/0000-0003-4422-2919}{\protect\includegraphics[height=0.19cm]{orcid.jpg}}}$, 
G.~Scandariato\inst{\ref{inst:19}}\,$^{\href{https://orcid.org/0000-0003-2029-0626}{\protect\includegraphics[height=0.19cm]{orcid.jpg}}}$, 
S.~Seager\inst{\ref{inst:81},\ref{inst:82},\ref{inst:83}}\,$^{\href{https://orcid.org/0000-0002-6892-6948}{\protect\includegraphics[height=0.19cm]{orcid.jpg}}}$, 
D.~S\`{e}gransan\inst{\ref{inst:5}}\,$^{\href{https://orcid.org/0000-0003-2355-8034}{\protect\includegraphics[height=0.19cm]{orcid.jpg}}}$, 
A.~M.~S.~Smith\inst{\ref{inst:39}}\,$^{\href{https://orcid.org/0000-0002-2386-4341}{\protect\includegraphics[height=0.19cm]{orcid.jpg}}}$, 
J.~Schlieder\inst{\ref{inst:71}},
R.~P.~Schwarz\inst{\ref{inst:79}}\,$^{\href{https://orcid.org/0000-0001-8227-1020}{\protect\includegraphics[height=0.19cm]{orcid.jpg}}}$, 
A.~Shporer\inst{\ref{inst:78}}\,$^{\href{https://orcid.org/0000-0002-1836-3120}{\protect\includegraphics[height=0.19cm]{orcid.jpg}}}$,
S.~G.~Sousa\inst{\ref{inst:36}}\,$^{\href{https://orcid.org/0000-0001-9047-2965}{\protect\includegraphics[height=0.19cm]{orcid.jpg}}}$, 
M.~Stalport\inst{\ref{inst:10},\ref{inst:9}}, 
M.~Steinberger\inst{\ref{inst:3}},
S.~Sulis\inst{\ref{inst:42}}\,$^{\href{https://orcid.org/0000-0001-8783-526X}{\protect\includegraphics[height=0.19cm]{orcid.jpg}}}$, 
Gy.~M.~Szab\`{o}\inst{\ref{inst:43},\ref{inst:66}}\,$^{\href{https://orcid.org/0000-0002-0606-7930}{\protect\includegraphics[height=0.19cm]{orcid.jpg}}}$, 
J.~D.~Twicken\inst{\ref{inst:73}}\,$^{\href{https://orcid.org/0000-0002-6778-7552}{\protect\includegraphics[height=0.19cm]{orcid.jpg}}}$, 
S.~Udry\inst{\ref{inst:5}}\,$^{\href{https://orcid.org/0000-0001-7576-6236}{\protect\includegraphics[height=0.19cm]{orcid.jpg}}}$, 
V.~Van~Grootel\inst{\ref{inst:10}}\,$^{\href{https://orcid.org/0000-0003-2144-4316}{\protect\includegraphics[height=0.19cm]{orcid.jpg}}}$, 
J.~Venturini\inst{\ref{inst:5}}\,$^{\href{https://orcid.org/0000-0001-9527-2903}{\protect\includegraphics[height=0.19cm]{orcid.jpg}}}$, 
E.~Villaver\inst{\ref{inst:12},\ref{inst:33}}, 
N.~A.~Walton\inst{\ref{inst:67}}\,$^{\href{https://orcid.org/0000-0003-3983-8778}{\protect\includegraphics[height=0.19cm]{orcid.jpg}}}$ and
J.~N.~Winn\inst{\ref{inst:74}}\,$^{\href{https://orcid.org/0000-0002-4265-047X}{\protect\includegraphics[height=0.19cm]{orcid.jpg}}}$
}

\institute{
\label{inst:1} Dipartimento di Fisica e Astronomia ``Galileo Galilei'', Universit\'{a} degli Studi di Padova, Vicolo dell'Osservatorio 3, 35122 Padova, Italy \and
\label{inst:2} INAF, Osservatorio Astronomico di Padova, Vicolo dell'Osservatorio 5, 35122 Padova, Italy \and
\label{inst:3} Space Research Institute, Austrian Academy of Sciences, Schmiedlstrasse 6, A-8042 Graz, Austria \and
\label{inst:4} Dipartimento di Fisica e Astronomia, Universit\'{a} degli Studi di Padova, Vicolo dell’Osservatorio 3, 35122 Padova, Italy \and
\label{inst:5} Observatoire astronomique de l'Universit\`{e} de Gen\'{e}ve, Chemin Pegasi 51, 1290 Versoix, Switzerland \and
\label{inst:6} Weltraumforschung und Planetologie, Physikalisches Institut, University of Bern, Gesellschaftsstrasse 6, 3012 Bern, Switzerland \and
\label{inst:7} Center for Space and Habitability, University of Bern, Gesellschaftsstrasse 6, 3012 Bern, Switzerland \and
\label{inst:8} Cavendish Laboratory, JJ Thomson Avenue, Cambridge CB3 0HE, UK \and
\label{inst:9} Astrobiology Research Unit, Universit\`{e} de Li\'{e}ge, All\`{e}e du 6 Aout 19C, B-4000 Li\'{e}ge, Belgium \and
\label{inst:10} Space sciences, Technologies and Astrophysics Research (STAR) Institute, Universit\`{e} de Li\'{e}ge, All\`{e}e du 6 Aout 19C, 4000 Li\'{e}ge, Belgium \and
\label{inst:11} Institute of Astronomy, KU Leuven, Celestijnenlaan 200D, 3001 Leuven, Belgium \and
\label{inst:12} Instituto de Astrof\`{i}sica de Canarias, V\`{i}a L\`{a}ctea s/n, 38200 La Laguna, Tenerife, Spain \and
\label{inst:14} Dipartimento di Fisica, Universit\'{a} di Trento, Via Sommarive 14, 38123 Povo \and
\label{inst:15} Department of Astronomy, Stockholm University, AlbaNova University Center, 10691 Stockholm, Sweden \and
\label{inst:16} CFisUC, Departamento de Fisica, Universidade de Coimbra, 3004-516 Coimbra, Portugal \and
\label{inst:17} INAF - Osservatorio Astrofisico di Torino,via Osservatorio 20,10025 Pino Torinese,Italy \and
\label{inst:18} INAF - Osservatorio Astronomico di Trieste,Via Giambattista Tiepolo,11,34131 Trieste (TS),Italy \and
\label{inst:19} INAF, Osservatorio Astrofisico di Catania, Via S. Sofia 78, 95123 Catania, Italy \and
\label{inst:21} Department of Physics,University of Rome ``Tor Vergat'',Via della Ricerca Scientifica 1,00133 Rome,Italy \and
\label{inst:22} Department of Physics and Astronomy,University of Florence,Largo Enrico Fermi 5,50125 Firenze,Italy \and
\label{inst:24} INAF - Turin Astrophysical Observatory,Pino Torinese,Italy \and
\label{inst:25} Max Planck Institute for Astronomy,Heidelberg,Germany \and
\label{inst:26} INAF - Osservatorio Astronomico di Trieste,via Tiepolo 11,34143 Trieste \and
\label{inst:27} INAF - Osservatorio Astronomico di Palermo,Piazza del Parlamento 1,90134 Palermo,Italy \and
\label{inst:28} Universit\'{a} degli Studi dell'Insubria, Via Ravasi 2, 21100 Varese \and
\label{inst:29} INAF - Osservatorio Astronomico di Brera, via Bianchi 46, 23807 Merate, Italy. \and
\label{inst:30} Instituto de Astrof\`{i}sica de Canarias, Calle de la v\`{i}a L\`{a}ctea s/n, 38205 San Crist\`{o}bal de La Laguna, Santa Cruz de Tenerife, España \and
\label{inst:31} INAF, Astronomical Observatory of Rome, Via Frascati 33, 00178 Monte Porzio Catone (RM), Italy \and
\label{inst:32} Fundaci\`{o}n Galileo Galilei - INAF, Rambla Jos\`{e} Ana Fern\`{a}ndez P\`{e}rez 7, 38712 Breña Baja, Tenerife, Spain \and
\label{inst:33} Departamento de Astrof\`{i}sica, Universidad de La Laguna, Astrof\`{i}sico Francisco Sanchez s/n, 38206 La Laguna, Tenerife, Spain \and
\label{inst:34} Admatis, 5. Kand\`{o} K\`{a}lm\`{a}n Street, 3534 Miskolc, Hungary \and
\label{inst:35} Depto. de Astrof\`{i}sica, Centro de Astrobiolog\`{i}a (CSIC-INTA), ESAC campus, 28692 Villanueva de la Cañada (Madrid), Spain \and
\label{inst:36} Instituto de Astrofisica e Ciencias do Espaco, Universidade do Porto, CAUP, Rua das Estrelas, 4150-762 Porto, Portugal \and
\label{inst:37} Departamento de Fisica e Astronomia, Faculdade de Ciencias, Universidade do Porto, Rua do Campo Alegre, 4169-007 Porto, Portugal \and
\label{inst:38} Centre for Exoplanet Science, SUPA School of Physics and Astronomy, University of St Andrews, North Haugh, St Andrews KY16 9SS, UK \and
\label{inst:39} Institute of Planetary Research, German Aerospace Center (DLR), Rutherfordstrasse 2, 12489 Berlin, Germany \and
\label{inst:40} INAF, Osservatorio Astrofisico di Torino, Via Osservatorio, 20, I-10025 Pino Torinese To, Italy \and
\label{inst:41} Centre for Mathematical Sciences, Lund University, Box 118, 221 00 Lund, Sweden \and
\label{inst:42} Aix Marseille Univ, CNRS, CNES, LAM, 38 rue Fr\`{e}d\`{e}ric Joliot-Curie, 13388 Marseille, France \and
\label{inst:43} ELTE Gothard Astrophysical Observatory, 9700 Szombathely, Szent Imre h. u. 112, Hungary \and
\label{inst:44} SRON Netherlands Institute for Space Research, Niels Bohrweg 4, 2333 CA Leiden, Netherlands \and
\label{inst:45} Centre Vie dans l’Univers, Facult\`{e} des sciences, Universit\`{e} de Gen\'{e}ve, Quai Ernest-Ansermet 30, 1211 Gen\'{e}ve 4, Switzerland \and
\label{inst:46} Leiden Observatory, University of Leiden, PO Box 9513, 2300 RA Leiden, The Netherlands \and
\label{inst:47} Department of Space, Earth and Environment, Chalmers University of Technology, Onsala Space Observatory, 439 92 Onsala, Sweden \and
\label{inst:48} Dipartimento di Fisica, Universit\'{a} degli Studi di Torino, via Pietro Giuria 1, I-10125, Torino, Italy \and
\label{inst:49} National and Kapodistrian University of Athens, Department of Physics, University Campus, Zografos GR-157 84, Athens, Greece \and
\label{inst:50} Department of Astrophysics, University of Vienna, Türkenschanzstrasse 17, 1180 Vienna, Austria \and
\label{inst:51} European Space Agency (ESA), European Space Research and Technology Centre (ESTEC), Keplerlaan 1, 2201 AZ Noordwijk, The Netherlands \and
\label{inst:52} Institute for Theoretical Physics and Computational Physics, Graz University of Technology, Petersgasse 16, 8010 Graz, Austria \and
\label{inst:53} Konkoly Observatory, Research Centre for Astronomy and Earth Sciences, 1121 Budapest, Konkoly Thege Mikl\`{o}s \`{u}t 15-17, Hungary \and
\label{inst:54} ELTE E\''otv\''os Lor\'and University, Institute of Physics, P\'azm\'any P\'eter s\'et\'any 1/A, 1117 Budapest, Hungary \and
\label{inst:55} Lund Observatory, Division of Astrophysics, Department of Physics, Lund University, Box 43, 22100 Lund, Sweden \and
\label{inst:56} IMCCE, UMR8028 CNRS, Observatoire de Paris, PSL Univ., Sorbonne Univ., 77 av. Denfert-Rochereau, 75014 Paris, France \and
\label{inst:57} Institut d'astrophysique de Paris, UMR7095 CNRS, Universit\`{e} Pierre \& Marie Curie, 98bis blvd. Arago, 75014 Paris, France \and
\label{inst:58} Astrophysics Group, Lennard Jones Building, Keele University, Staffordshire, ST5 5BG, United Kingdom \and
\label{inst:60} Institute of Optical Sensor Systems, German Aerospace Center (DLR), Rutherfordstrasse 2, 12489 Berlin, Germany \and
\label{inst:61} Department of Physics, University of Warwick, Gibbet Hill Road, Coventry CV4 7AL, United Kingdom \and
\label{inst:62} ETH Zurich, Department of Physics, Wolfgang-Pauli-Strasse 2, CH-8093 Zurich, Switzerland \and
\label{inst:63} Institut fuer Geologische Wissenschaften, Freie Universitaet Berlin, Maltheserstrasse 74-100,12249 Berlin, Germany \and
\label{inst:64} Institut de Ciencies de l'Espai (ICE, CSIC), Campus UAB, Can Magrans s/n, 08193 Bellaterra, Spain \and
\label{inst:65} Institut d'Estudis Espacials de Catalunya (IEEC), 08860 Castelldefels (Barcelona), Spain \and
\label{inst:66} HUN-REN-ELTE Exoplanet Research Group, Szent Imre h. u. 112., Szombathely, H-9700, Hungary \and
\label{inst:67} Institute of Astronomy, University of Cambridge, Madingley Road, Cambridge, CB3 0HA, United Kingdom \and
\label{inst:68} NASA Exoplanet Science Institute-Caltech/IPAC, Pasadena, CA 91125, USA \and
\label{inst:69} NASA Ames Research Center, Moffett Field, CA 94035, USA \and
\label{inst:70} Department of Astronomy and Astrophysics, University of California, Santa Cruz, CA 95064, USA \and
\label{inst:71} NASA Goddard Space Flight Center, 8800 Greenbelt Road, Greenbelt, MD 22071, USA \and
\label{inst:72} Department of Physics and Astronomy, University of Kansas, Lawrence, KS 66045, USA \and
\label{inst:73} SETI Institute, Mountain View, CA 94043 USA/NASA Ames Research Center, Moffett Field, CA 94035 USA \and
\label{inst:74} Department of Astrophysical Sciences, Princeton University, Princeton, NJ 08544, USA \and
\label{inst:75} Space Telescope Science Institute, 3700 San Martin Drive, Baltimore, MD 21218, USA \and
\label{inst:76} Cahill Center for Astrophysics, California Institute of Technology, Pasadena, CA 91125, USA \and
\label{inst:77} Proto-Logic Consulting LLC
, Washington, DC, USA \and
\label{inst:78} Department of Physics and Kavli Institute for Astrophysics and Space Science, Massachusetts Institute of Technology, 77 Massachusetts Avenue, Cambridge, MA 02139, USA \and
\label{inst:79} Center for Astrophysics Harvard \& Smithsonian, 60 Garden Street, Cambridge, MA 02138, USA \and
\label{inst:80} Department of Physics and Astronomy, University of New Mexico, 210 Yale Boulevard NE, Albuquerque, NM 87106, USA \and
\label{inst:81} Department of Physics and Kavli Institute for Astrophysics and Space Research, Massachusetts Institute of Technology, Cambridge, MA 02139, USA \and 
\label{inst:82}Department of Earth, Atmospheric and Planetary Sciences, Massachusetts Institute of Technology, Cambridge, MA 02139, USA \and
\label{inst:83} Department of Aeronautics and Astronautics, MIT, 77 Massachusetts Avenue, Cambridge, MA 02139, USA \and
\label{inst:84} Observatori Astron\'{o}mic Albany\'{a}, Cam\`{i} de Bassegoda S/N, Albany\'{a} 17733, Girona, Spain \and
\label{inst:85} Physics Department, Austin College, Sherman, TX 75090, USA \and
\label{inst:86} Komaba Institute for Science, The University of Tokyo, 3-8-1 Komaba, Meguro, Tokyo 153-8902, Japan \and
\label{inst:87} Astrobiology Center, 2-21-1 Osawa, Mitaka, Tokyo 181-8588, Japan
}


\newcommand{\msun}{$M_{\odot}$\xspace}
\newcommand{\rsun}{$R_{\odot}$\xspace}
\newcommand{\mstar}{\ensuremath{M_{\star}}\xspace}
\newcommand{\rstar}{\ensuremath{R_{\star}}\xspace}
\newcommand{\feh}{\ensuremath{[\mbox{Fe}/\mbox{H}]}\xspace}
\newcommand{\teff}{\ensuremath{T_{\mathrm{eff}}}\xspace}
\newcommand{\logg}{\ensuremath{\log g}\xspace}
\newcommand{\vsini}{\ensuremath{v \sin i_\star}\xspace}
\newcommand{\vmic}{$V_{\rm mic}$}
\newcommand{\vmac}{$V_{\rm mac}$}

\date{Received \today; Accepted -}

 
  \abstract
  {The discovery and characterization of mini Neptunes has a crucial impact on planetary formation and evolution theories. A precise characterization of their orbital parameters and atmospheric properties would give us valuable hints to improve our formation and atmospheric models.}
  {We present the discovery of two mini Neptunes near a 2:1 orbital resonance configuration orbiting the K0 star TOI-1803. We describe in detail their orbital architecture and suggest some possible formation and evolution scenarios.}
  {Using CHEOPS, TESS, and HARPS-N datasets we can estimate the radius and the mass of both planets. We used a multidimensional Gaussian Process with a quasi-periodic kernel to disentangle the planetary components from the stellar activity in the HARPS-N dataset. We performed dynamical modeling to explain the orbital configuration and performed planetary formation and evolution simulations. For the least dense planet, we finally assume different atmospheric compositions and define possible atmospheric characterization scenarios with simulated JWST observations.}
  {TOI-1803\,b and TOI-1803\,c have orbital periods of $\sim$6.3 and $\sim$12.9 days, respectively, residing in close proximity to a 2:1 orbital resonance. Ground-based photometric follow-up observations revealed significant transit timing variations (TTV) with an amplitude of $\sim 10$ min and $\sim 40$ min, respectively, for planet -b and -c. With the masses computed from the radial velocities data set, we obtained a density of $(0.39 \pm 0.10)\,\rho_{\oplus}$ and $(0.076 \pm 0.038)\,\rho_{\oplus}$  for planet -b and -c, respectively. TOI-1803\,c is among the least dense mini Neptunes currently known, and due to its inflated atmosphere, it is a suitable target for transmission spectroscopy with JWST. With NIRSpec observations, we could understand whether the planet kept its primary atmosphere or not, constraining our formation models.}
   {We report the discovery of two mini Neptunes close to a 2:1 orbital resonance. The detection of significant TTVs from ground-based photometry opens scenarios for a more precise mass determination. TOI-1803\,c is one of the least dense mini Neptune known so far, and it is of great interest among the scientific community since it could constrain our formation scenarios. JWST observations could give us valuable insights to characterise this interesting system.}

   \keywords{stars: individual: TOI-1803 (TIC 144401492, Gaia DR3 4031575166693272832) – techniques: photometric – techniques: radial velocities – planets and satellites: fundamental parameters}

\titlerunning{Discovery of two mini Neptunes around TOI-1803}
\authorrunning{Zingales et al}

\maketitle

%
\section{Introduction}

The term mini Neptunes usually refers to planets similar in size to Neptune (which is about 4 times the size and about 17 times the mass of Earth) but with a lower mass than Neptune. They usually have a size between 2 to 4 Earth radii. Mini Neptunes are often found orbiting close to their host stars, where they are subjected to intense radiation and heat \citep{miguel2015,venturini2016,carleo2020,lacedelli2021,leleu2021,lacedelli2022}. Thanks to Kepler/K2 data we know that mini Neptunes may be among the most common planetary types in our Galaxy \citep{fulton2017,sheng2021,fressin2013,beleznau2022}. Despite their abundance, we do not know for certain how they form, nor their atmospheric compositions or interior structures. 
Super-Earths are smaller than mini Neptunes and have a size between 1.2 to 2 Earth radii. The prevailing idea is that mini Neptunes and super Earths have a common origin; probably they both had a gas envelope consisting of a few percent of their mass \citep{rogers2011,lopez2014,wolfgang2015}. Depending on the stellar irradiation, some of these planets could lose their atmospheres because of photoevaporation \citep{lopez2012,owen2013,ehrenreich2015,modirroustagalian2023} or because of core-powered mass loss \citep{ginzburg2018,gupta2019}. The observation of the atmospheric types could shed light on the planetary formation and evolution history.
Primary atmospheres are acquired during the first planetary formation stage when the planet accumulates the gas from the protoplanetary disk. The main atmospheric components during this phase are primarily light elements such as hydrogen and helium. The elemental composition of the protoplanetary disk strongly influences the composition of the primary atmosphere. In contrast, secondary atmospheres are developed or modified after the formation via mechanisms like outgassing from the planetary core, impacts by asteroids and comets, or processes of atmospheric loss and accretion \citep{bean2021,kite2019,kite2020}.

The discrimination between primary and secondary atmosphere is crucial for understanding the evolutionary path of exoplanets. A combination of observational and theoretical approaches is fundamental to distinguishing between these two atmospheric types. With JWST \citep{greene2016} it is possible to distinguish between the two atmospheric types and its observation could constrain the formation and evolution scenarios. Current observations of WASP-39b \citep{jwst2023} already confirmed the presence of a complex atmosphere around this hot Jupiter and \citet{kite2021} demonstrated that it is possible to distinguish also the presence of complex molecules around smaller-sized planets.

The expectation for multi-planet systems hosting sub-neptune planets is for their planets to have experienced disk-driven migration during their growth \citep{nelson2017}. As these planets experience different migration rates based on their masses, they can undergo convergent migration and become locked into resonant architectures \citep{malhotra1993,inamidar2015,owen2018,morrison2020,izidoro2022}. These resonant architectures, however, can be broken by the interaction with non-resonant planets \citep{cimerman2018, turrini2023}. The existence of resonances in mature planetary systems, therefore, provides a strong indication that their architecture is primordial and the direct result of their formation process.

In this context, the detection by TESS of the two Neptune-sized candidate planets in a near 2:1 resonance around the $\lesssim 1$~Gyr old star TYC 2526-1545-1 (TOI-1803) represents a unique prospect to investigate how these systems could evolve to their current configuration, what is their main composition and how do they form. The orbital resonance could be a hint of planet-planet interaction, leading the system to evolve to an equilibrium configuration with an orbital resonance. We started the radial velocity follow-up with high-accuracy measurements, using the HARPS-N spectrograph mounted at Telescopio Nationale Galileo, within the framework of the Global Architecture of Planetary System consortium (GAPS, \citealt{naponiello2022,naponiello2023,covino2013}). At the same time, the target was selected by the CHEOPS Science Team as a suitable candidate for Transit Time Variation (TTV) studies. In this paper, we present the confirmation and characterization of this planetary system using photometry, with CHEOPS and TESS, and radial velocity using HARPS-N. The simultaneous analysis of TESS and CHEOPS allowed for a determination of the radius with a precision better than 3\%. The presence of stellar activity allowed only a marginal detection of the masses of the planets, despite the use of state-of-the-art methods for stellar activity modeling.

Finally, the discrimination between primary and secondary atmospheres could give us strong constraints on our planetary evolution and formation models. Mini Neptunes with a primary atmosphere may suggest an embryonic stage in planetary formation, whereas those with a secondary atmosphere could describe subsequent volatile-rich accretion or outgassing events. The possibility of having such different atmospheric scenarios provides a key point to deciphering complex evolutionary paths of mini Neptunes, shedding light on the potential outcomes of planetary formation and subsequent atmospheric evolution \citep{scheucher2020,kasper2020}.

In this work, we fully characterize the planetary system around TOI-1803. In Sect. \ref{sec:observations} we described all the instruments used in the analysis. 
The adopted stellar model is detailed in Sect. \ref{sec:star_pars}.
The photometric and radial velocity and TTV analysis are described in Sect. \ref{sec:phot_rv}. We describe the planetary formation and evolution in Sect. \ref{sec:planetary_formation}. Sect. \ref{sec:internal_structure} describes the internal structure models for both planets in the TOI 1803 system. Finally in Sect. \ref{sec:JWST} we demonstrate how it is possible to distinguish between the two atmospheric types on the least dense planet using JWST/NIRSpec instrumentation.

\section{High Resolution Imaging}
    As part of our standard process for validating transiting exoplanets to assess the possible contamination of bound or unbound companions on the derived planetary radii \citep{ciardi2015}, we observed TOI~1803 with near-infrared adaptive optics imaging on Keck and optical speckle imaging at Gemini-North.  The near-infrared and optical imaging complement each other with differing resolutions and sensitivities.
	
   	Keck Observations of TOI-1803 were made on 2020-05-28UT with the NIRC2 instrument on Keck-II (10m) behind the natural guide star AO system \citep{wizinowich2000} in the standard 3-point dither pattern that is used with NIRC2 to avoid the left lower quadrant of the detector which is typically noisier than the other three quadrants. The dither pattern step size was $3\arcsec$ and was repeated twice, with each dither offset from the previous dither by $0.5\arcsec$.  NIRC2 was used in the narrow-angle mode with a full field of view of $\sim10\arcsec$ and a pixel scale of approximately $0.0099442\arcsec$ per pixel.  The Keck observations were made in the narrow-band Br-$\gamma$ filter $(\lambda_o = 2.1686; \Delta\lambda = 0.0326~\mu$m).  Flat fields were taken on-sky, dark-subtracted, and median averaged, and sky frames were generated from the median average of the dithered science frames. Each science image was then sky-subtracted and flat-fielded.  The reduced science frames were combined into a single mosaiced image, with final combined resolution 0.058\arcsec.  
	
\begin{figure}
    \centering
    \includegraphics[width=0.5\textwidth]{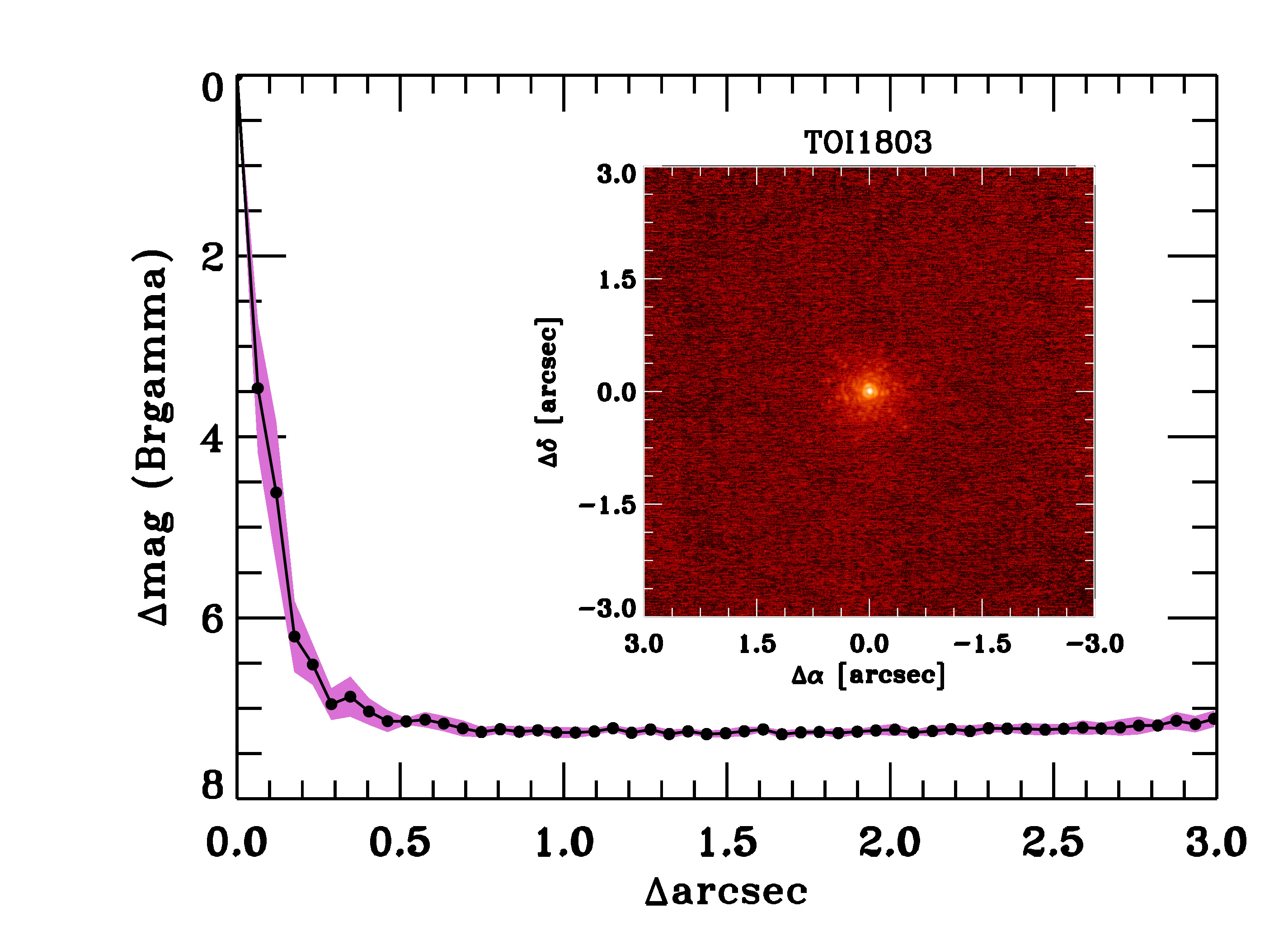}
    \caption{NIR AO imaging and sensitivity curve. {\it Insets:} Images of the central portion of the images.
    }
    \label{fig:ao_imaging}
\end{figure}

	The sensitivity of the final combined AO image was determined by injecting simulated sources azimuthally around the primary target every $20^\circ $ at separations of integer multiples of the central source's FWHM \citep{furlan2017}. The brightness of each injected source was scaled until standard aperture photometry detected it with $5\sigma $ significance.  The final $5\sigma $ limit at each separation was determined from the average of all of the determined limits at that separation and the uncertainty on the limit was set by the rms dispersion of the azimuthal slices at a given radial distance; sensitivities are shown in (Figure~\ref{fig:ao_imaging}).  
	
\begin{figure}
    \centering
    \includegraphics[width=0.95\linewidth]{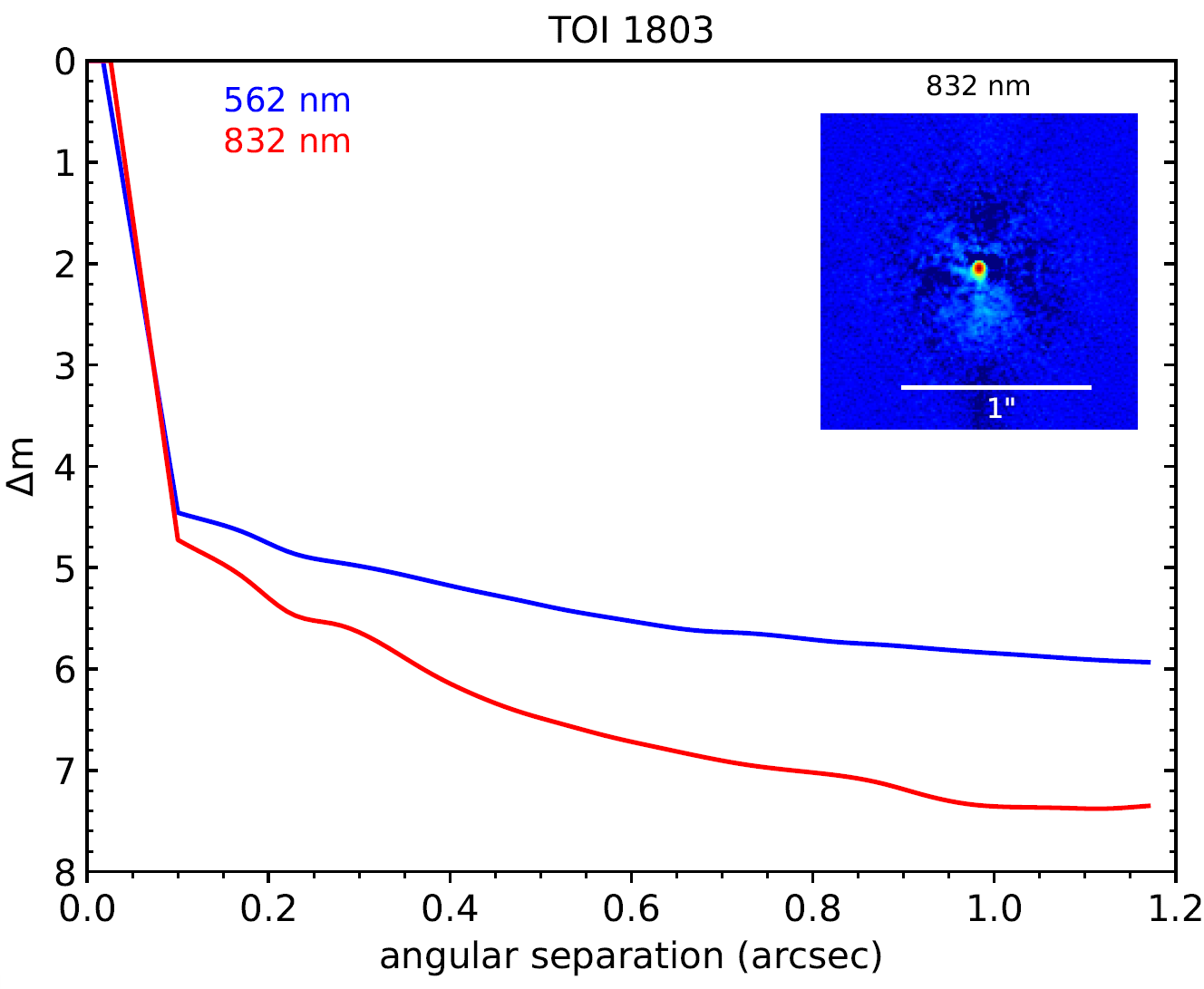}
    \caption{Companion sensitivity (5$\sigma$ limits) for the Gemini-North speckle imaging.  The inset image is of the primary target at 832nm showing no additional companions.}
    \label{fig:speckle_image}
\end{figure}

	TOI-1803 was also observed with the optical speckle imager ‘Alopeke on Gemini-North \citep{scott2021}.  Simultaneous observations in the narrow-band filters centered at 562nm ($\Delta\lambda = 54$nm) and 832nm ($\Delta\lambda=40$nm) were obtained on 2020-Jun-08 UT with a standard of 1000 frames each taken with an exposure time of 60ms.  The data were reduced with the standard speckle data reduction pipeline \citep{howell2011} that produces sensitivity curves and a final image (Fig~\ref{fig:speckle_image}) constructed from the interferometric specklegram.    The final speckle image has a field of view of $\sim2\arcsec$ with a resolution of 0.01\arcsec; the imaging was sensitive to $\Delta mag = 5.4$mag at 0.5\arcsec (562nm filter) and $\Delta mag = 6.5$mag at 0.5\arcsec\ (832filter).  Neither infrared adaptive imaging nor the speckle optical imaging detects a stellar companion.

\section{Validation analysis}
\label{sec:validation_analysis}
TOI-1803 (TIC 144401492) was observed by the Transiting Exoplanet Survey Satellite \citep[TESS,][]{ricker2015} in Sectors 22 and 49. The 2-minute cadence data were processed in the TESS Science Processing Operations Center \citep[SPOC,][]{jenkins2016} at NASA Ames Research Center. The SPOC conducted a transit search of the Sector 22 light curve on March 29, 2020, with an adaptive, noise-compensating matched filter \citep{jenkins2002,jenkins2010,jenkins2020}, producing Threshold Crossing Events with orbital periods of 12.9 d and 6.3 d. A limb-darkened transit model was fitted \citep{li2019} and a suite of diagnostic tests were conducted to help assess the planetary nature of the signals \citep{twicken2018}. The transit signatures passed all diagnostic tests presented in the SPOC Data Validation reports. The TESS Science Office reviewed the vetting information and issued alerts for TOI 1803.01 (period ~ 12.8912 +- 0.0027 d and transit model fit S/N ~ 12.9$\sigma$) and TOI 1803.02 (period ~ 6.2944 +- 0.0014 d and S/N ~ 11.1$\sigma$) on April 15, 2020 \citep{guerrero2021}. The SPOC difference image centroid offsets localized the source of the TOI 1803.01 transit signal within 1.4 +- 2.7 arcsec of the target star and the source of the TOI 1803.02 transit signal within 6.1 +- 3.0 arcsec; this excludes all TESS Input Catalog (TIC) objects other than TOI 1803 as potential transit sources. 
This system was also analysed in \citet{giacalone2021} and classified as a ``likely planet'' with \texttt{TRICERATOPS}.
As a first check of the quality of both candidate exoplanets TOI-1803.01 and TOI-1803.02, we performed a probabilistic validation study to rule out any false positive (FP) scenario that could mimic the transit signals identified by the TESS official pipelines. Some of the objects initially identified as sub-stellar candidates might be FPs due to the low spatial resolution of TESS cameras ($\approx$ 21 arcsec/pixels). The analysis we followed uses photometric data provided by \textit{Gaia} and is fully described in \cite{mantovan2022}. First, we conducted a stellar neighborhood analysis to find any potential contaminating stars capable of being the origin of blended eclipsing binaries (BEBs). This crucial study allowed us to rule out each resolved \textit{Gaia} neighborhood star as the source of the transit signals. To corroborate that the two candidates are not FPs, we used the VESPA software \citep{morton2012,morton2016}. In particular, we followed the procedure adopted by \citet{mantovan2022}, which proactively addresses the major concerns reported by \citet{morton2023} and ensures reliable results when using VESPA. We found a false positive probability (FPP) of $3.54\times 10^{-3}$ and $2.04\times 10^{-6}$ for TOI-1803.01 and TOI-1803.02, respectively -- enough to claim a statistical vetting for both candidates \citep{morton2012}. It is important to note that candidates associated with a star having more than one transit candidate are more likely to be genuine planets than similar candidates associated with stars having no other transit candidates \citep{latham2011,lissauer2012,valizadegan2023}. This statistical validation triggered the follow-up observations described in Section \ref{sec:observations}, which ultimately confirmed the planetary nature of the two candidates.

\section{Observations and data reduction}
\label{sec:observations}

\subsection{HARPS-N}
The HARPS-N spectrograph \citep{cosentino2012,cosentino2014} is a high-precision radial velocity instrument, with a wavelength range between 383-693 nm and a resolving power of R $\sim 115\,000$, mounted at Telescopio Nazionale Galileo (TNG), a 3.58m telescope located on the island of La Palma in the Canary Islands.
We collected a total of 127 observations of TOI-1803 with HARPS-N, during three observational seasons: the first one from May 2020 to July 2020, the second one from December 2020 to July 2021, and the third one from December 2021 to June 2022. The exposure time was set to 1800s to reach an average SNR of 34, corresponding to an average error in the RV data set of $4$ m/s when using the K5 mask of the HARPS-N DRS pipeline. 

\begin{figure*}[!htbp]
    \centering
    \includegraphics[width=\textwidth]{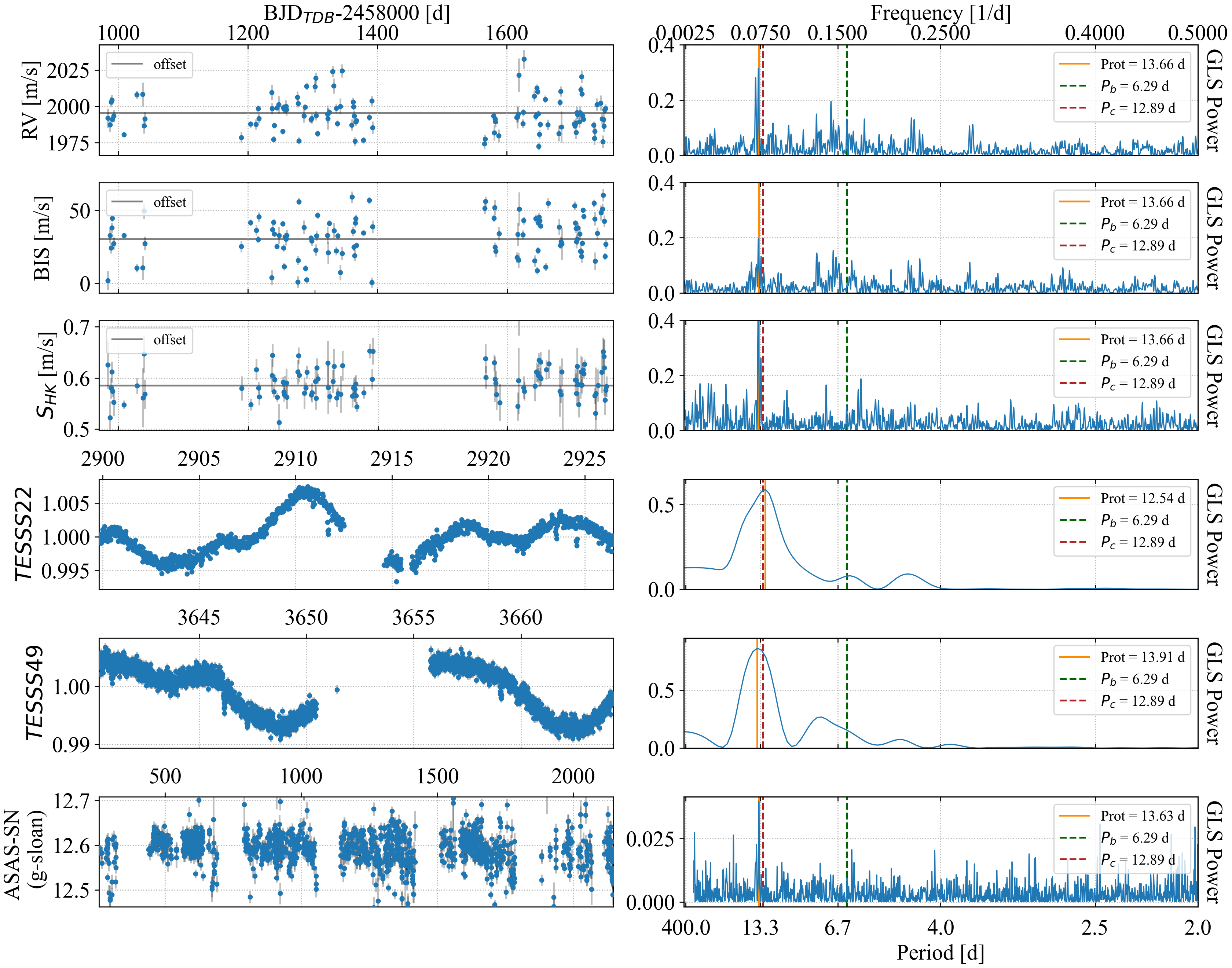}
    \caption{GLS Periodograms, in order from the top to the bottom panel: HARPS-N spectrograph RVs, BIS and $S_{HK}$  series, TESS photometric time series (Sectors 22 and 49), and ASAS-SN. The vertical orange lines represent the main peak of the periodogram, at $\sim13.4$ days, corresponding to the stellar rotation period $P_{\rm{rot}}$. The green and the red lines represent the orbital period of, respectively, planets -b and -c.}
    \label{fig:periodogram}
\end{figure*}

We extracted radial velocities from HARPS-N spectra using the TERRA pipeline \citep{anglada2012} and calculated the time series over the $S_{\rm HK}$ and the bisector inverse slope (BIS) indices to investigate stellar activity \citep{lovis2011}. Additionally, we computed the generalized Lomb-Scargle (GLS) periodogram \citep{zechmeister2009} for the HARPS-N Radial Velocities (RVs), the $S_{HK}$ and the BIS values (Fig. \ref{fig:periodogram}). In each of the three datasets, we found a significant peak corresponding to a False Alarm Probability - FAP $<0.1$\% at a frequency $f \approx 0.07$~d$^{-1}$ and a period $P \approx 13.66$~d, most likely associated with the rotational period of the star $P_{\mathrm{rot}}$, as discussed further in the next sections.

\subsection{TESS}
\label{sec:tess}

\begin{figure}[!htbp]
    \centering
    \includegraphics[width=0.45\textwidth]{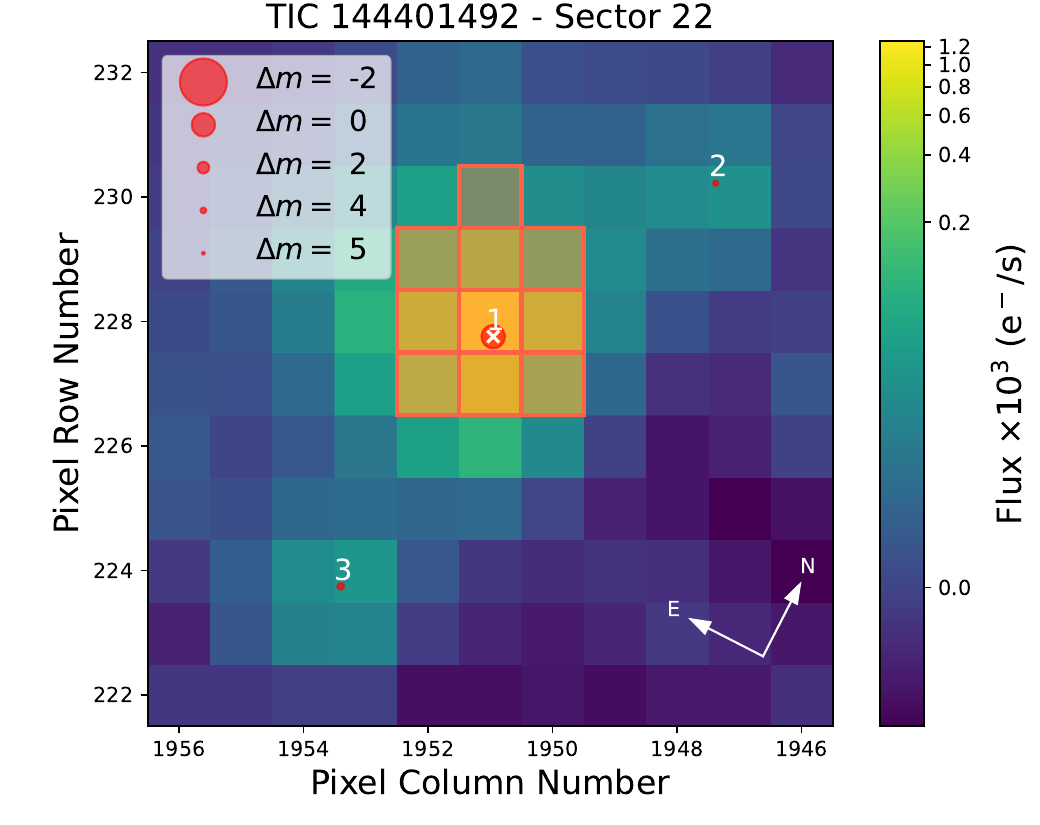}
    \caption{TESS aperture field around the target star TOI-1803. (Image generated with \texttt{tpfplotter}, \citet{aller2020}).\label{fig:tessaperture}}    
\end{figure}

The Transiting Exoplanet Survey Satellite (TESS, \citealt{ricker2015}) is a spacecraft designed to discover new exoplanets using the transit method. It features four identical refractive cameras that together provide a field of view spanning 24x96 degrees.
We used TESS Full Frame Images (FFIs) of Sectors 22 and 49 to extract the light curves of TOI-1803. Sector 22 was observed between February 18th, 2020, and March 18th, 2020 with a 30-minute cadence, while Sector 49 was observed between February 26, 2022, and March 26, 2022, with a 10-minute cadence. We extracted the raw light curves from FFIs by using the pyPATHOS pipeline (see \citealt{nardiello2019,nardiello2021}) and we corrected them by applying the cotrending basis vectors obtained by \citet{nardiello2020}. We computed the GLS of the data from both sectors, 22 and 49, revealing a major peak at $\sim 12.54$\,d and $\sim 13.91$\,d (see Fig.~\ref{fig:periodogram}, fourth and fifth panel). 
In Fig. \ref{fig:tessaperture} we show the field of view for the TESS space telescope.

\begin{figure*}[!htbp]
    \centering
    \includegraphics[width=0.49\textwidth]{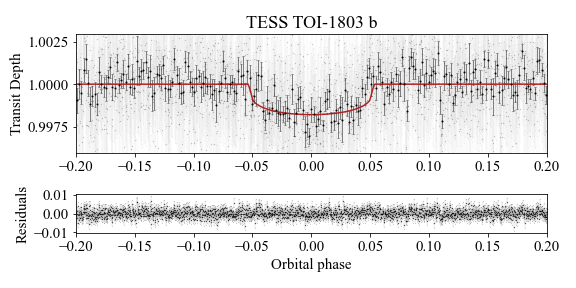}
    \includegraphics[width=0.49\textwidth]{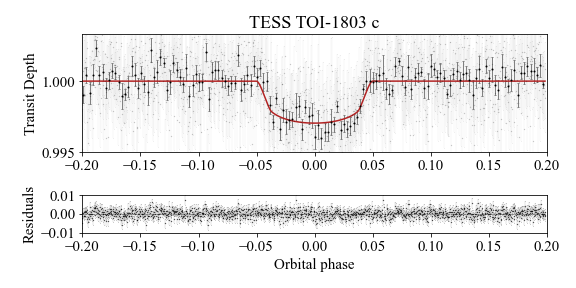}
    \caption{\textbf{Top}: TOI-1803 TESS phase-folded detrended observations. The red line is the best model for TESS observations of planets -b (left) and -c (right). \textbf{Bottom}: Residuals from both models (see \textbf{Sect.} \ref{sec:phot_rv}).}
    \label{fig:tess_phase}
\end{figure*}

We included the TESS observations in our analysis, where the stellar activity and instrumental signals have been modeled using the bi-weight function within the \texttt{wotan} package \citep{hippke2019} after masking the transits of the two planets. The phase-folded TESS light curves are shown in Fig.~\ref{fig:tess_phase}. 

\subsection{CHEOPS}
CHEOPS is an ESA S-class space mission launched in 2019. It consists of a 32cm primary mirror telescope designed to perform ultra-high-precision photometry of bright stars \citep{benz2021,fortier2024}. CHEOPS observed TOI-1803 during seven visits: four for planet b and three for planet c (see Tab \ref{tab:cheops_obs}). All the CHEOPS observations were processed using the CHEOPS Data Reduction Pipeline (DRP) \citep{hoyer2020} version 14.1.2. We used the DEFAULT aperture with a radius of 25 px. As a sanity check, we also modeled the PSF-modelled PIPE\footnote{\url{https://github.com/alphapsa/PIPE}}\footnote{{\url{https://ascl.net/code/v/3958}}} \citep{szabo2021,morris2021,brandeker2022} light curves. Since the difference between the DRP and the PIPE extractions did not lead to significant differences, we used the DRP light curves throughout this paper.

\begin{table*}[!htpb]
\caption[]{Log of the CHEOPS observations}
\label{tab:cheops_obs}
\centering
\begin{tabular}{ccccc}
    \toprule
    id \# & ObsID & File Key & Start Date (UTC) & Visit duration [hours]\\
    \midrule
    1 & 1693021 & CH\_PR100031\_TG047401\_V0300 & 2022-01-10 16:59 & 23.13 \\
    2 & 1701998 & CH\_PR100031\_TG048101\_V0300 & 2022-01-23 10:22 & 11.34 \\
    3 & 1702871 & CH\_PR100031\_TG048201\_V0300 & 2022-01-27 21:04 & 21.81 \\
    4 & 1738539 & CH\_PR100031\_TG050301\_V0300 & 2022-02-22 13:32 & 10.32 \\
    5 & 1773891 & CH\_PR100031\_TG049301\_V0300 & 2022-03-27 08:41 & 11.34 \\
    6 & 1774381 & CH\_PR100031\_TG050401\_V0300 & 2022-04-02 09:01 & 10.20 \\
    7 & 1787579 & CH\_PR100031\_TG049401\_V0300 & 2022-04-21 13:03 & 11.34 \\

    \bottomrule
\end{tabular}
\end{table*}


\begin{figure*}[!htbp]
    \centering
    \includegraphics[width=0.49\textwidth]{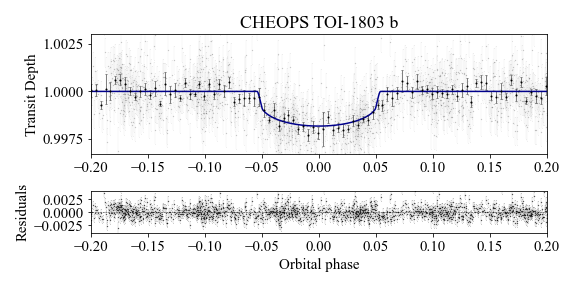}
    \includegraphics[width=0.49\textwidth]{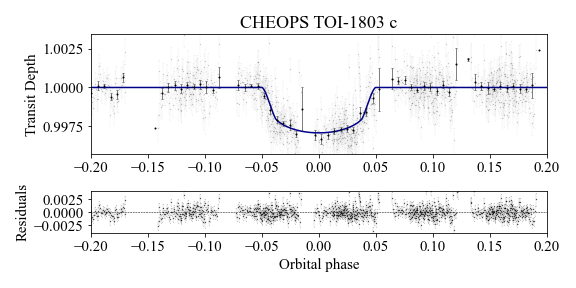}
    \caption{\textbf{Top}: TOI-1803 CHEOPS phase-folded detrended observations. The blue line is the best-fit model for CHEOPS observations of planets -b (left) and -c (right). \textbf{Bottom}: Residuals from both models (see Sec. \ref{sec:phot_rv}).}
    \label{fig:cheops_phase}
\end{figure*}

The observations from CHEOPS were detrended using \texttt{cheope}\footnote{\url{https://github.com/tiziano1590/cheops_analysis-package}}, an optimized \texttt{Python} tool \citep[which uses \texttt{pycheops} as backend described in][]{maxted2021}, to derive the planetary signal from the CHEOPS data frames. The best detrending model, based on the Bayes factor, is selected automatically by \texttt{cheope} using \texttt{lmfit} \citep{newville2014} and the Bayesian model is run with \texttt{emcee} \citep{foreman2013}.

\subsection{ASAS-SN}
\label{sec:asas-sn}

The All Sky Automated Survey for SuperNovae (ASAS-SN) is a program that searches for new supernovae and other transient astronomical phenomena. It consists of 20 robotic telescopes distributed worldwide, able to survey the entire sky approximately once every day.

To understand the peak at around $\sim 13.6$ days found in the TESS and HARPS-N data sets and better identify the stellar activity signal, we used almost five years of data from ASAS-SN \citep{shappee2014,kochanek2017}, spanning from May 2018 to July 2023. The ASAS-SN images have a resolution of 8 arcsec/pixel ($\sim$15$\arcsec$ FWHM PSF), and the observations of TOI-1803 were conducted in the Sloan $g-$band. We computed the GLS periodogram, which shows the main peak at 13.63 days, see Fig. \ref{fig:periodogram}.

The peaks observed in the ASAS-SN dataset confirm the evidence of stellar activity noticed also with HARPS-N and both TESS sectors. We interpreted these peaks as the rotation period of the host star. Combining the results of the HARPS-N, TESS, and ASAS-SN datasets we can infer a stellar rotation period of $P_{\mathrm{rot}} = 13.4 \pm 0.6$ days.

\subsection{Ground-based Light Curve Follow-up (TFOP)}
\label{sec:ground_based}

Several ground-based photometric observations have been gathered during the transit of TOI-1803\,c. These observations show partial transits or not enough out-of-transit signal, or they present strong TTVs. As such, they were employed in the TTV analysis but not in the combined photometric and spectroscopic time series fit (see Sect. \ref{sec:phot_rv}) as they did not lead to any significant improvement in the orbital and physical parameters of the planet, while adding the complication of the inclusion of TTV and additional limb darkening coefficients to already computationally expensive modelling. 

\begin{figure}[h]
    \centering
    \includegraphics[width=0.5\textwidth]{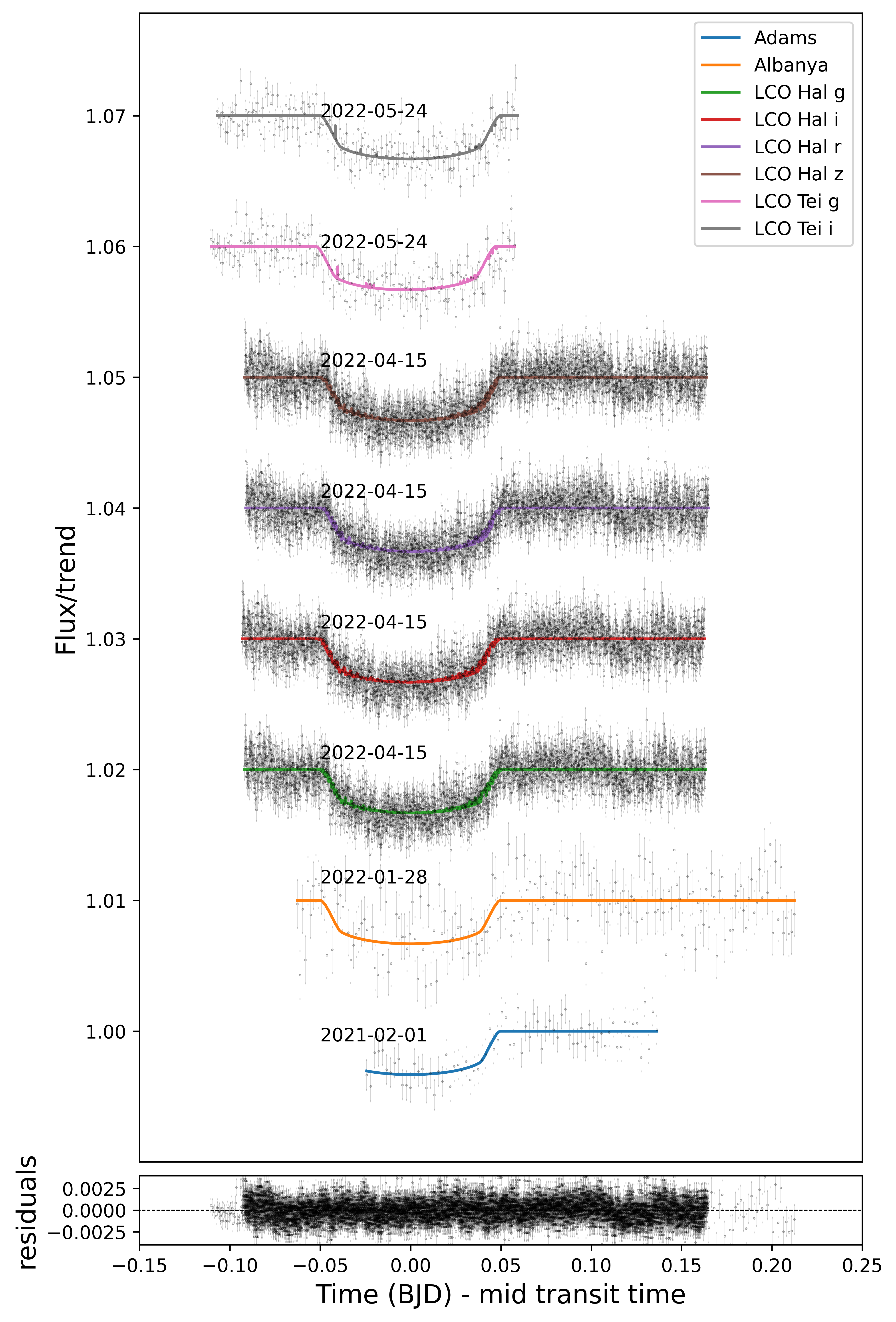}
    \caption{Light curves for each ground-based observation. The light curves have been stacked for better visibility and centered for each mid-transit time.}
    \label{fig:curves_ground}
\end{figure}

We used the resources of the \textit{TESS} Follow-up Observing Program \citep[TFOP;][]{collins:2019}\footnote{https://tess.mit.edu/followup} to collect four additional ground-based observations of TOI-1803\,c (Fig.~\ref{fig:curves_ground}). We used the {\tt TESS Transit Finder}, which is a customised version of the {\tt Tapir} software package \citep{Jensen:2013}, to schedule our transit observations. 

We observed a full transit of TOI-1803 c on UTC 2022 April 15 simultaneously in Sloan $g'$, $r'$, $i'$, and Pan-STARRS $z$-short from the Las Cumbres Observatory Global Telescope \citep[LCOGT;][]{Brown:2013} 2\,m Faulkes Telescope North at Haleakala Observatory on Maui, Hawai'i (Hal). The telescope is equipped with the MuSCAT3 multi-band imager \citep{Narita:2020}. We observed another full transit of TOI-1803 c in alternating Sloan $g'$ and $i'$ band filters on UTC 2022 May 24 from the LCOGT 1\,m network node at Teide Observatory on the island of Tenerife (Tei). The 1\,m telescopes are equipped with $4096\times4096$ SINISTRO cameras having an image scale of $0\farcs389$ per pixel, resulting in a $26\arcmin\times26\arcmin$ field of view. All images were calibrated by the standard LCOGT {\tt BANZAI} pipeline \citep{McCully:2018} and differential photometric data were extracted using {\tt AstroImageJ} \citep{Collins:2017}.

Another egress observation was made from Sherman, TX, USA from the Adams Observatory 0.61m telescope, which sits on top of Austin College's science building, the IDEA Center using Cousins I band on UT 2021 February 1. The telescope is equipped with an FLI ProLine detector that has an image scale of $0\farcs38$ pixel$^{-1}$, resulting in a $26\arcmin\times26\arcmin$ field of view. The images were calibrated and differential photometric data were extracted using {\tt AstroImageJ}.

Finally, another full transit was observed on UTC 2022 January 28 from the Observatori Astron\`{o}mic Albany\`{a} (Albanya) 0.41m Meade ACF catadioptric telescope, which is located in Albany\'{a}, Girona Spain. The telescope is equipped with a Moravian G4-9000 camera that has an image scale of $1\farcs44$ per $2\times2$ binned pixel resulting in an $36\arcmin\times36\arcmin$ field of view. The images were calibrated and differential photometric data were extracted using {\tt AstroImageJ}.

\begin{figure}[!htbp]
    \centering
    \includegraphics[width=0.5\textwidth]{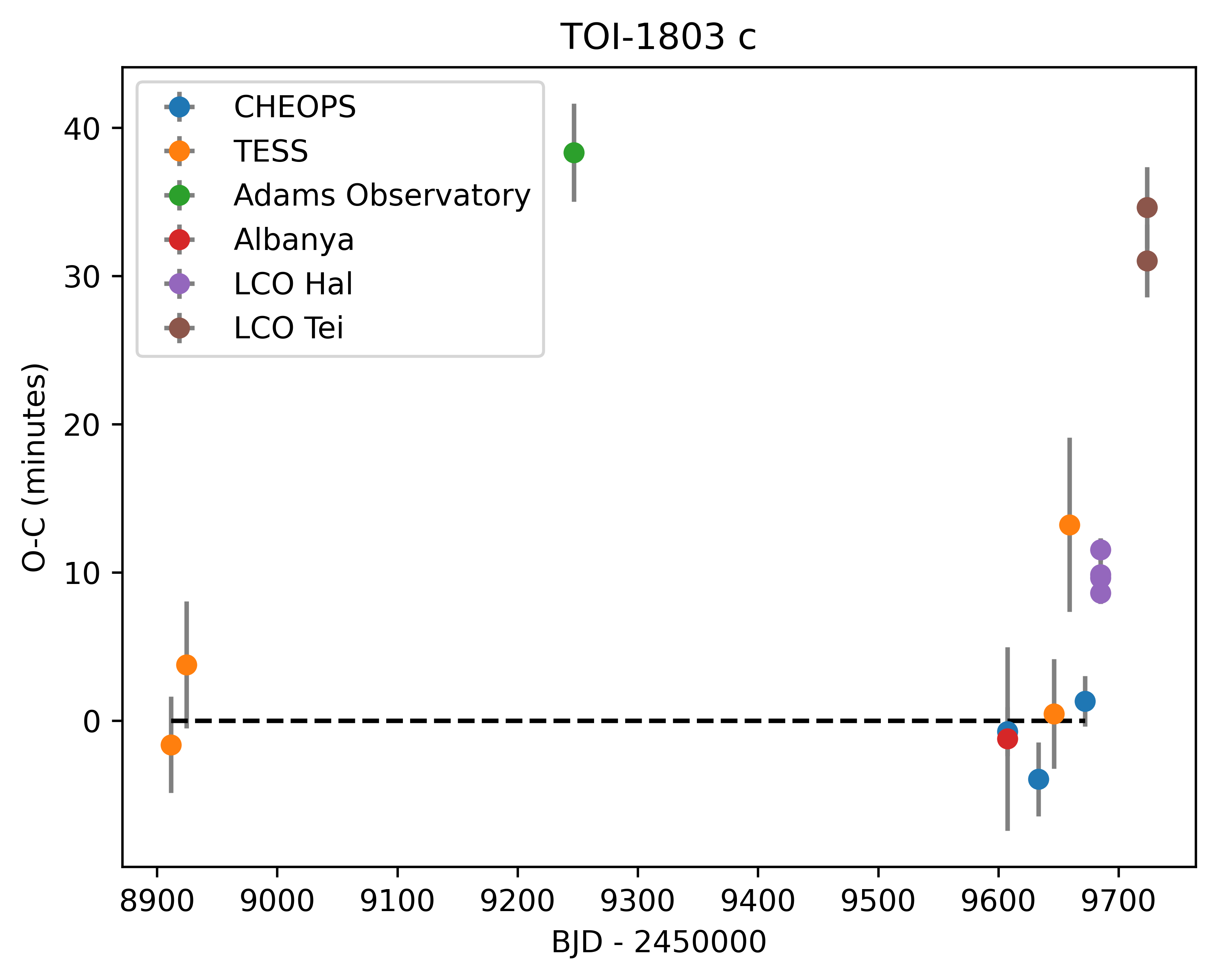}
    \caption{O-C diagram for TOI-1803 c including the ground-based observations.}
    \label{fig:O-C_ground}
\end{figure}

The transit times of the ground-based observations have been extracted by analysing each transit separately (transit and detrend modelling). The resulting O-C diagram is shown in Fig.~\ref{fig:O-C_ground}, which exhibits the hint of a TTV signal. 

\section{Stellar parameters}
\label{sec:star_pars}

We used the \texttt{Specmatch-emp}\footnote{\url{https://github.com/samuelyeewl/specmatch-emp}}, a spectral analysis package for the extraction of stellar parameters \citep{Yee2017}. After re-formatting the co-added spectrum to a compatible format \citep{Hirano2018}, this code compares it with a library of over 400 spectra of stars of all types with well-determined physical parameters. A minimization and interpolation calculation provides estimates of temperature \teff, logarithmic surface gravity \logg, metallicity \feh, as well as star mass \mstar, star radius \rstar, and age. For further details, we refer the reader to, e.g., \citet{Fridlund2020} and references therein. 

With astroARIADNE \citep{vines022}, we performed a Bayesian model averaging of four stellar atmospheric model grids from {\tt Phoenix~v2}, \citep{husser2013}, {\tt {Bt-Settl}}, {\tt {Bt-Cond}}, and  {\tt {Bt-NextGen}}  \citep{Allard+2012, Hau+1999}, for stars with  $T_\mathrm{eff}$ convolved with various filter response functions. We used magnitudes from {\it Gaia} DR3 ($G, G_{\rm BP}, G_{\rm RP}$), {\it WISE} (W1-W2), $J, H, K_S$  from {\it 2MASS}, and  Johnson $B$ and $V$  from APASS (see Table~\ref{table:starphotometry}). The parallax was also taken from {\it Gaia} DR3 applying the parallax offset of \cite{Lindegren2021}. 
The software interpolates model grids of \teff, \logg~and \feh~assuming distance, extinction  ($A_V$), and stellar radius as free parameters.
The maximum line-of-sight value from the dust maps of \citet{Schlegel:1998} was used as an upper limit of $A_V$. 

We then applied the IDL package Spectroscopy Made Easy (SME) which synthesizes a model of individual absorption lines in the observed spectrum based on several well-determined stellar atmospheric models \citep{Valenti1996, Piskunov2017}. It utilizes atomic and molecular parameters from the VALD database \citep{Piskunov1995}. In the case of TOI-1803, we applied the Atlas12 model grid \citep{Kurucz2014} for the synthesis. Following again schemes outlined in e.g. \citet{Fridlund2020} and references therein, we kept the turbulent velocities \vmac~and \vmic~fixed at the empirical values found in the literature \citep{Gray2008}. We then find \vsini to be $<$~1.0\,$\pm$\,0.5 km/s. Using SME to fit several hundred TiO lines with \teff as the only free parameter, we then find \teff = 4687\,$\pm$\,65 K, in very good agreement with the \teff\ of astroARIADNE (\teff\ 4692\,$\pm$\,110 K), which is consistent with the result of Specmatch (\teff\ 4788\,$\pm$\,110 K). 

Using a Markov-Chain Monte Carlo (MCMC) modified infrared flux method \citep{Blackwell1977,Schanche2020}, we determined the stellar radius of TOI-1803. We constructed spectral energy distributions (SEDs) using stellar atmospheric models from three catalogs \citep{Kurucz1993,Castelli2003,Allard2014} with priors defined from our spectral analysis. Finally, we calculated the bolometric flux of TOI-1803 by comparing our computed synthetic and the observed broadband photometry in the following bandpasses: {\it Gaia} $G$, $G_\mathrm{BP}$, and $G_\mathrm{RP}$, 2MASS $J$, $H$, and $K$, and \textit{WISE} $W1$ and $W2$ \citep{Skrutskie2006,Wright2010,GaiaCollaboration2022}. We converted the stellar bolometric flux into effective temperature and angular diameter that are the MCMC-step parameters. These are determined to be 4790$\pm$31\,K and 0.055$\pm$0.0004\,mas. Thus, when combined with the offset-corrected \textit{Gaia} parallax \citep{Lindegren2021}, we retrieved the stellar radius. To robustly account for stellar atmospheric model uncertainties, we conducted a Bayesian modeling averaging of the \textsc{atlas} \citep{Kurucz1993,Castelli2003} and \textsc{phoenix} \citep{Allard2014} catalogs that resulted in the weighted averaged posterior distribution of the radius to be $R_\star=0.715\pm0.005\,R_{\odot}$.

The basic input set ($T_{\mathrm{eff}}$, [Fe/H], $R_{\star}$) along with their respective errors is finally used to determine the stellar mass $M_{\star}$ using two different stellar evolutionary models. 
A first estimate $M_{\star,1}=0.748\pm0.032\,M_{\odot}$ was computed via the isochrone placement algorithm \citep{bonfanti15,bonfanti16}, which interpolates the input values within pre-computed grids of PARSEC\footnote{\textsl{PA}dova and T\textsl{R}ieste \textsl{S}tellar \textsl{E}volutionary \textsl{C}ode: \url{http://stev.oapd.inaf.it/cgi-bin/cmd}} v1.2S \citep{marigo17} isochrones and tracks. A second estimate $M_{\star,2}=0.77\pm0.11\,M_{\odot}$ was inferred using the CLES \citep[Code Li\'{e}geois d'\`{e}volution Stellaire,][]{scuflaire08} code, which generates the best-fit evolutionary track according to the stellar input parameters and following the Levenberg-Marquardt minimization scheme \citep{salmon21}.
After that, the two mass outcomes were combined after checking their mutual consistency using the $\chi^2$-based criterion presented in \citet{bonfanti21}, where the full statistical treatment is described in detail. As our final estimate, we obtained $M_{\star}=0.756_{-0.042}^{+0.050}\,M_{\odot}$. 

Given its quite slow evolution, the isochronal age of TOI-1803 is inconclusive. 
However, we can use the stellar rotation period and the effective temperature (see Tab.~\ref{tab:obspars}) to obtain a gyrochronological estimate of the age of the star. The relation by \citet{barnes10} returned an age $t_{\star}=0.9\pm0.1$ Gyr. Interpolating between the open cluster sequences in \texttt{gyro-interp}\footnote{\url{https://github.com/lgbouma/gyro-interp}} \citep{Bouma2023}, we find $t_{\star}$ = $1.40^{+0.12}_{-0.16}$ Gyr (1$\sigma$ uncertainties), or $1.40^{+0.23}_{-0.53}$ Gyr (2$\sigma$ uncertainties).  The asymmetry in the age posterior is due to the stalled spin-down of K dwarfs near this age.

The relevant stellar parameters are listed in Table~\ref{table:starphotometry}

\section{Photometric and radial velocity analysis}
\label{sec:phot_rv}

To calculate the orbital and physical parameters, we run \texttt{PyORBIT}\footnote{\url{https://github.com/LucaMalavolta/PyORBIT}} \citep{malavolta2016,malavolta2018} on each data set. For the photometric data, we used CHEOPS observations (listed in Tab. \ref{tab:cheops_obs}) and TESS Sector 22 and 49 with a transit model of planets -b and -c. We adopted a quadratic law, defining $u_1$, $u_2$ for the limb darkening (LD) coefficients but we re-parameterized them as $q_1$, $q_2$ as fitting parameters, following \citet{kipping2013}. We assumed a Gaussian prior distribution over the LD coefficients for all the instruments, using a bi-linear interpolation of the limb darkening profile defined in \citet{claret2017,claret2021}. The priors set on the coefficients $q_1$, and $q_2$ are reported in Table~\ref{tab:obspars}. Moreover, we used RV measurements interpreting them with up to three planetary components. We also included activity indexes to characterize the stellar activity in the RV dataset. Photometric and spectroscopic data were modelled simultaneously. Finally, we tested both circular and eccentric models for the planets.  Additionally, we added a jitter term (see Tab \ref{tab:off_jit}) for both the photometric and the spectroscopic datasets to absorb unaccounted sources of errors.

To include the effects of stellar activity on radial velocity measurements, we relied on Gaussian processes (GPs), which are a powerful tool for modelling and understanding complex systems (see \citet{aigrain2023} for a detailed review).
In particular, we used the so-called multidimensional GP, where the radial velocities and activity proxies are described as a combination of an underlying Gaussian process $G(t)$ and its first derivative $\dot{G(t)}$ \citep{rajpaul2015}. For the GP, we relied on the quasi-periodic covariance kernel :
\begin{equation}
\gamma (t_i, t_j) = \exp{ \left \{-\frac{\sin^2{[\pi(t_i - t_j)/ P_\mathrm{rot}]}}{2 \mathrm{w} ^2} - \frac{(t_i-t_j)^2}{2 P_\mathrm{dec}^2} \right \} }
\end{equation}
where $P_\mathrm{rot}$ is equivalent to the rotation period of the star, $\mathrm{w}$ is the inverse of the harmonic complexity, also denoted as the coherence scale, and it controls the complexity of the signal within a period; $P_\mathrm{dec}$ is usually associated with the decay time scale of the active regions \citep[e.g.][]{grunblatt2015,nardiello2022,mantovan2024}.

To model stellar activity using multidimensional GPs, we used RVs, the $S_{HK}$-index dataset, and the bisector inverse slope (BIS) span. These indicators are related to the presence and strength of different types of activity on the star's surface. These data are used to constrain the underlying GP model that generates the radial velocity variations associated with stellar activity.
The multidimensional GP modelling takes the final form:

\begin{align*}
 \Delta{\rm RV} &= V_{\rm c} G(t) + V_{\rm r} \dot{G(t)} \\
 S_{HK} &= L_{\rm c} G(t) + L_{\rm r} \dot{G(t)} \\
 {\rm BIS} &=  B_{\rm c} G(t) + B_{\rm r} \dot{G(t)} \\
\end{align*}

where the letters with the $c$ and $r$ underscores denote the amplitude coefficient of the underlying GP and its first derivative, respectively, for each of the associated datasets. 
Differently from other analyses in the literature, we include the first derivative of the GP in the modelling of the $S_{HK}$ index. The posterior of the associate coefficient is consistent with zero within $1\sigma$, confirming that the contribution of this term is negligible.

We verified that without the use of GPs, we would not be able to extract planetary signals with an expected semi-amplitude of a few meters per second, as detailed in the next paragraphs.

 \begin{table}[!htb]
   \caption[]{Stellar properties of TOI-1803}
     \label{table:starphotometry}
     \small
     \centering
       \begin{tabular}{lcc}
         \hline
         \noalign{\smallskip}
         Parameter   &  \object{TOI-1803} & Source  \\
         \noalign{\smallskip}
         \hline
         \noalign{\smallskip}
$\alpha$ (J2000)          &   11 52 11.17	& {\it Gaia} DR3 \citep{gavras2023}    \\
$\delta$ (J2000)          &   +35 10 18.43  & {\it Gaia} DR3 \citep{gavras2023}  \\
$\mu_{\alpha}$ (mas yr$^{-1}$)  &    -86.286$\pm$0.014  & {\it Gaia} DR3 \citep{gavras2023}  \\
$\mu_{\delta}$ (mas yr$^{-1}$)  &    3.210$\pm$0.013  & {\it Gaia} DR3 \citep{gavras2023}  \\
RV     (km s$^{-1}$)            &    2.10$\pm$0.60   & {\it Gaia} DR3 \citep{gavras2023}   \\
$\pi$  (mas)             &    8.4016$\pm$0.0168 & {\it Gaia} DR3 \citep{gavras2023}  \\
V (mag)                  &    11.87$\pm$0.12     & Tycho-2 \citep{hog2000}   \\ 
$B-V$ (mag)                &    1.35$\pm$0.30 & Tycho-2 \citep{hog2000}  \\
$G_{\rm BP}$ (mag)         &    12.2990$\pm$0.0031  & {\it Gaia} DR3 \citep{gavras2023}  \\
$G$ (mag)                  &    11.7680$\pm$0.0028  & {\it Gaia} DR3 \citep{gavras2023}  \\
$G_{\rm RP}$ (mag)         &    11.0868$\pm$0.0039  & {\it Gaia} DR3 \citep{gavras2023}  \\
$J_{\rm 2MASS}$ (mag)    &   10.283$\pm$0.02  & 2MASS \citep{cutri2003}  \\
$H_{\rm 2MASS}$ (mag)    &   9.771$\pm$0.016  & 2MASS \citep{cutri2003}  \\
$K_{\rm 2MASS}$ (mag)    &   9.658$\pm$0.019  & 2MASS \citep{cutri2003}  \\
$W1$ (mag)    &   9.609$\pm$0.023  & WISE \citep{Wright2010}  \\
$W2$ (mag)    &   9.683$\pm$0.020  & WISE \citep{Wright2010}  \\
$\log g$                 &  4.62$\pm$0.10     & {\it Gaia} DR3 \citep{gavras2023} \\ 
$T_{\rm eff}$ (K)        &  4687$\pm$65       & This paper (spec) (Sect. \ref{sec:star_pars}) \\   
${\rm [Fe/H]}$ (dex)     &  0.014$\pm$0.060     & This paper (Sect. \ref{sec:star_pars}) \\ 
Mass ($M_{\odot}$)       &    0.756$^{+0.050}_{-0.042}$   & This paper (Sect. \ref{sec:star_pars}) \\
Radius ($R_{\odot}$)     &    0.715$\pm$0.005   & This paper (Sect. \ref{sec:star_pars}) \\
Luminosity ($L_{\odot}$)     &    0.222$\pm$0.013   & This paper (Sect. \ref{sec:star_pars}) \\
\vsini (km/s)            & $<$~1.0\,$\pm$\,0.5  & This paper (Sect. \ref{sec:star_pars}) \\
P$_{\mathrm{rot}}$ (days) & 13.659$^{+0.029}_{-0.033}$ & This paper (Sect. \ref{sec:phot_rv}) \\
Age  (Gyr)               &    $1.40^{+0.12}_{-0.16}$ & This paper (Sect. \ref{sec:star_pars}) \\
         \noalign{\smallskip}
         \hline
      \end{tabular}

\end{table}

We performed the simultaneous fits using \pyde\footnote{\url{https://github.com/hpparvi/PyDE}.} with 64000 generations for the determination of the initial point and \texttt{emcee}\footnote{\url{https://github.com/mcfit/emcee}} applying an MCMC chain with 200000 steps and 132 walkers \citep{parviainen2016,foreman2013}. At first, we performed a joint fit assuming a third planetary signal with the following boundary conditions: orbital period $P$ between 1 and 1000 days, radial velocity semi-amplitude $K$ between 0.001 and 100 m/s, and eccentricity between 0 and 0.5. We left the stellar rotation period as a free parameter between 12.0 and 15.0 days, choosing the range after the periodogram analysis (see Tab. \ref{tab:obspars}). This final model was consistent with the absence of a third planetary signal, resulting in a $K<0.8$ m/s and an unconstrained orbital period. As a consequence, to interpret the RV dataset, we continued with only two planetary components (planet -b and -c).

We also performed a test to assess the eccentricity of the two planetary orbits. We tested two models, the first one with circular orbits and a second one assuming $e \ge 0$ as a free parameter with uniform $\mathcal{U}$(0, 0.5) prior. 
With this configuration we estimated an eccentricity of $e_b =0.094_{-0.065}^{+0.095}$ and $e_c = 0.19_{-0.13}^{+0.15}$, respectively, for planet -b and -c.
The selection of the best model has been estimated with the Bayesian Information Criterion \citep[BIC;][]{schwarz1978}. BIC is based on Bayesian inference and provides a means of comparing different models by calculating the relative probability of each model given the observed data.
The BIC value is calculated using the following formula:

\begin{equation}
\indent BIC = -2 \cdot \log{\mathcal{L}} + k \cdot \log{n}
\end{equation}

where $\mathcal{L}$ is the likelihood of the data given the model, $k$ is the number of parameters in the model, and $n$ is the sample size.

\begin{figure}[!htbp]
    \centering
    \includegraphics[width=0.45\textwidth]{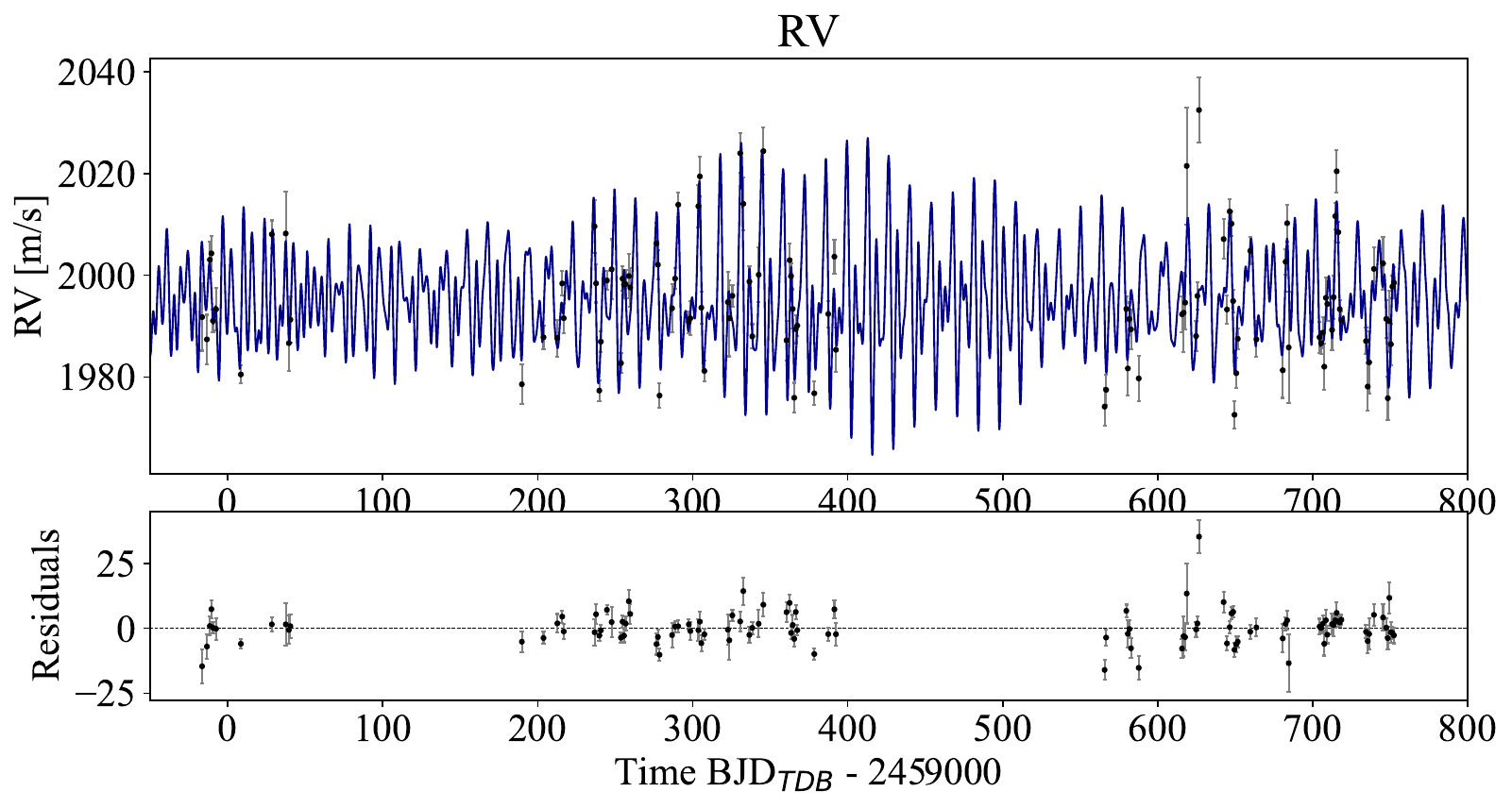}
    \includegraphics[width=0.45\textwidth]{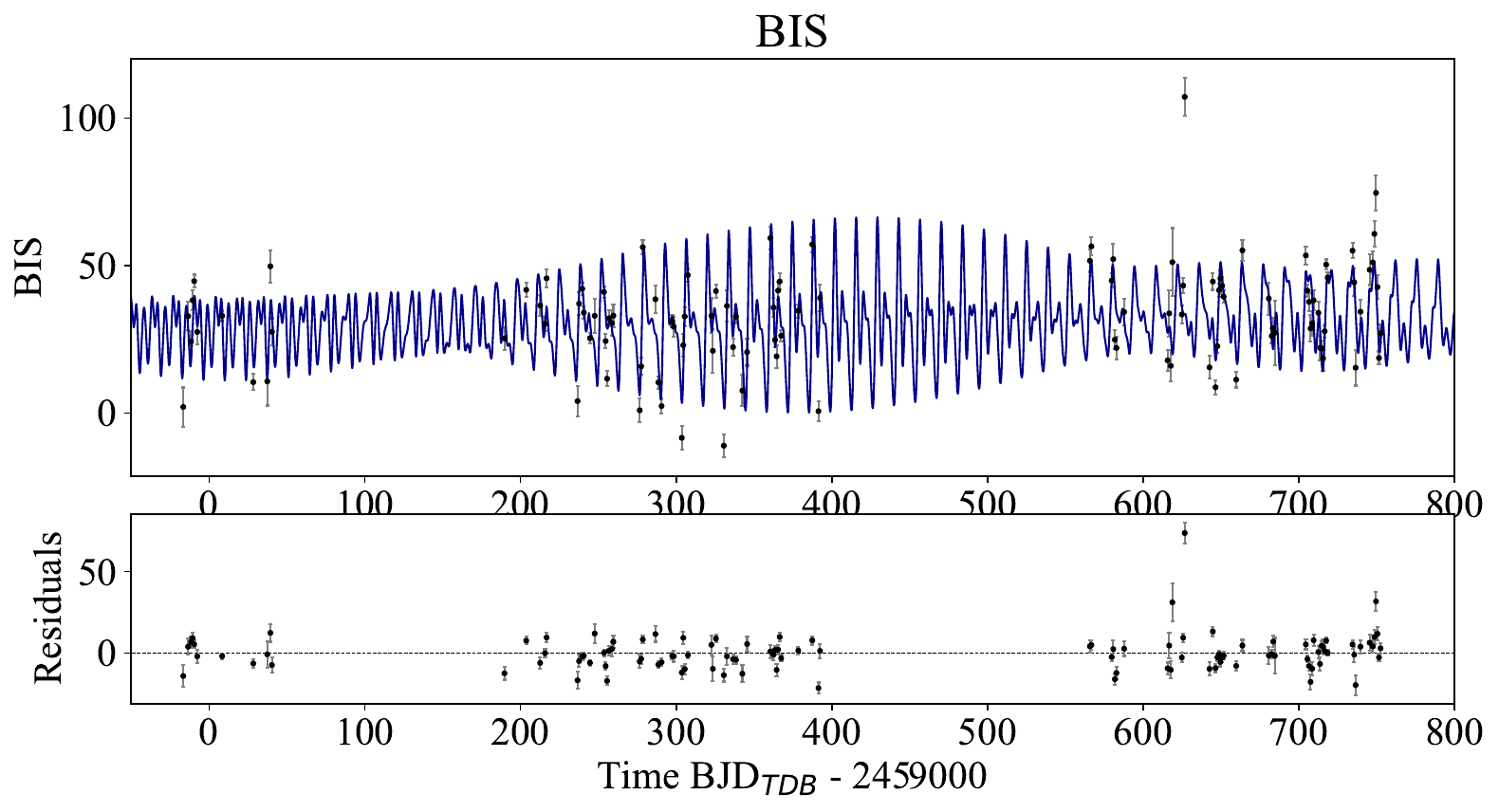}
    \includegraphics[width=0.45\textwidth]{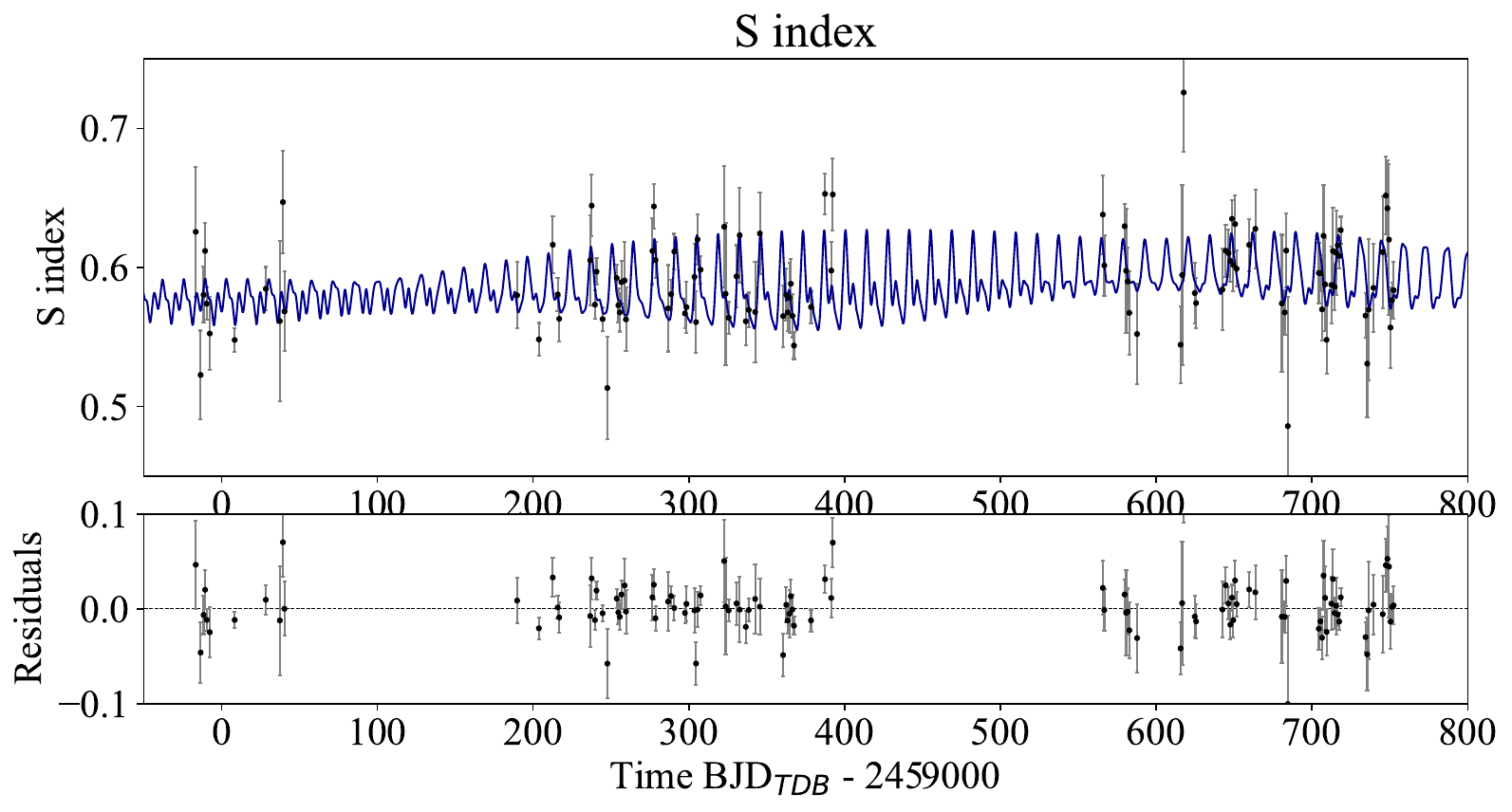}
    \caption{In blue is shown the best-fit model using multidimensional GP. The input vector for the algorithm has been constructed using radial velocity (top panel), bisector (middle panel), and S-index data (bottom panel). For each data set, the residuals from the models are shown}
    \label{fig:multiGP}
\end{figure}

The BIC value is then compared across different models, with the model having the lowest BIC value usually favored. This means that the best model is the one that provides the best balance between the data points and the number of fitting parameters \citep[e.g.][]{heller2019}. The BIC values for the model assuming a circular orbit and the eccentric model are, respectively, BIC$_{\mathrm{circ}}=-60864$ and BIC$_{\mathrm{ecc}}=-60126$, with a difference of BIC$_{\mathrm{ecc}}$ - BIC$_{\mathrm{circ}} = 738$.
As the difference in BIC values is greater than 15 (BIC rule of thumb as in Appendix E of \citealt{BICrule}.),
the preferred model is the one with circular orbits.
From two independent analyses, using a Nested sampling optimizer with \texttt{dynesty} \citep{speagle2020,skilling2004}, we also computed the logarithmic Bayes Factor between the model with $e = 0$ and $ e > 0$, resulting in $\log\mathcal{B} = logE_{e=0} - logE_{e>0} = 32.3$, with $E$ defined as the Bayesian evidence, in favor of the model with circular orbits. We computed the best-fit values and the associated error by taking the 15-th and 84-th percentiles for the associated posterior distributions. The inferred parameters are reported in Table \ref{tab:obspars}, the amplitudes for the multidimensional GP are reported in Table \ref{tab:gp_ampl} while the jitter and offset values for the photometric and spectroscopic datasets are reported in Table \ref{tab:off_jit}. The best-fit model using the GP is shown in Fig. \ref{fig:multiGP}. 

The Bayes factor suggests that the circular orbit should be considered favoured, however, a detailed dynamic simulation (see section \ref{sec:ttv}) suggests that the eccentric Keplerian orbit would not be totally excluded. The radial velocity (and then the masses) for the two planetary components are still consistent between both scenarios. Given the hint of a TTV signal, for completeness, we also show the orbital parameters obtained by the joint fit assuming eccentric orbits in Table \ref{tab:obspars_ecc}.

\begin{table*}[!htbp]
\caption{Priors and results for the modeling of planets b and c from the analysis of the photometric and radial velocities time series. Circular orbits.} 
\label{tab:obspars}
\centering          
\begin{tabular}{l c c c}     
\hline\hline     
 \multicolumn{4}{c}{Combined Radial Velocities - Multidimensional GP - Photometric fit} \rule{0pt}{2ex} \rule[-0.9ex]{0pt}{0pt} \\ 
\hline    
Parameter & Unit & Prior & Value \rule{0pt}{2.2ex} \rule[-0.9ex]{0pt}{0pt}\\ 
\hline
\multicolumn{4}{c}{Star} \rule{0pt}{2.2ex} \rule[-0.9ex]{0pt}{0pt}\\ 
\hline  
   Stellar density ($\rho_{\star}$) & $\rho_{\sun}$ & $\mathcal{N}$(2.00, 0.13) & 2.03$\pm$0.13 \rule{0pt}{2.2ex} \rule[-1.2ex]{0pt}{0pt}\\
   CHEOPS Limb Darkening ($q_{1, CHEOPS}$) &    & $\mathcal{U}$(0, 1) & $0.497_{-0.079}^{+0.085}$ \\
   CHEOPS Limb Darkening ($q_{2, CHEOPS}$) &    & $\mathcal{U}$(0, 1) & $0.461_{-0.034}^{+0.039}$ \\
   TESS Limb Darkening ($q_{1, TESS}$) &    & $\mathcal{U}$(0, 1) & $0.49_{-0.09}^{+0.10}$ \\
   TESS Limb Darkening ($q_{2, TESS}$) &    & $\mathcal{U}$(0, 1) & $0.38_{-0.033}^{+0.039}$ \\
\hline
\multicolumn{4}{c}{Planet b} \rule{0pt}{2.2ex} \rule[-0.9ex]{0pt}{0pt}\\ 
\hline    
Parameter & Unit & Prior & Value \rule{0pt}{2.2ex} \rule[-0.9ex]{0pt}{0pt}\\ 
\hline 
    Orbital period ($P_{\rm b}$) & days & $\mathcal{U}$(6.2, 6.4) & 6.293287$\pm$0.000028 \\
    Central time of transit ($T_{\rm 0,b}$) & BTJD$^a$ & $\mathcal{U}$(1904,1905) & 1904.6237$\pm$0.0027 \\
    Impact parameter ($b$) &  & $\mathcal{U}$(0, 1) & 0.34$^{+0.09}_{-0.14}$ \\
    $R_p/R_*$ & & $\mathcal{U}$(0.00001, 0.5) & 0.038$\pm$0.001\\   
    Planetary Radius ($R_p$) & $R_{\mathrm{\oplus}}$ & ... & $2.99 \pm 0.08$ \\
    $a/R_{\star}$ &  & ... & $18.2 \pm 0.4$\\
    Semi-major axis $a$ & AU  & ... & $0.060 \pm 0.001$ \\
    Radial Velocity semi-amplitude (K) & m/s & $\mathcal{U}$(0, 10) & 4.4$\pm$1.0 \\ 
    Inclination ($i$) & deg & ... & $88.9 \pm 0.5$ \\
    Transit duration ($T_{14}$) & days & ... & $0.108 \pm 0.003$  \\
    Planetary mass ($M_p$) & $M_{\mathrm{\oplus}}$ & ... & $10.3 \pm 2.5$ \\
    Planetary density ($\rho_p$) & $\rho_{\mathrm{\oplus}}$ & ... & $0.39 \pm 0.10$ \\
    Instellation ($F_i$) & W m$^{-2}$ & ... & (7.9$\pm$1.6)$\cdot 10^4$\\
    Equilibrium Temperature ($T_{\rm eq}$) & K$^b$ & ... & $747 \pm 11$ \\
\hline
\multicolumn{4}{c}{Planet c} \rule{0pt}{2.2ex} \rule[-0.9ex]{0pt}{0pt}\\ 
\hline    
Parameter & Unit & Prior & Value \rule{0pt}{2.2ex} \rule[-0.9ex]{0pt}{0pt}\\ 
\hline 
   Orbital period ($P_{\rm c}$) & days & $\mathcal{U}$(12.8, 13.0) & 12.885779$\pm$0.000036\\
   Central time of transit ($T_{\rm 0,c}$) & BTJD$^a$ & $\mathcal{U}$(1911,1912) & 1911.6747$\pm$0.0019 \\
   Impact parameter ($b$) &  & $\mathcal{U}$(0, 1) & 0.779$^{+0.014}_{-0.015}$\\
   $R_p/R_*$ & & $\mathcal{U}$(0.00001, 0.5) & 0.055$\pm$0.001\\
   Planetary Radius ($R_p$) & $R_{\mathrm{\oplus}}$ & ... & $4.29 \pm0.08$ \\ 
   $a/R_{\star}$ &  & ... &  $29.3 \pm 0.6$\\
   Semi-major axis $a$ & AU & ... & $0.097 \pm 0.002$\\
   Radial Velocity semi-amplitude (K) & m/s & $\mathcal{U}$(0, 10) & 2.1$^{+1.1}_{-1.0}$ \\ 
   Inclination ($i$) & deg & ... & $88.48 \pm 0.06$ \\
   Transit duration ($T_{14}$) & days & ... & $0.0998 \pm 0.0015$ \\
   Planetary mass ($M_p$) & $M_{\mathrm{\oplus}}$ & ... & $6.0 \pm 3.0$ \\
   Planetary density ($\rho_p$) & $\rho_{\mathrm{\oplus}}$ & ... & $0.076 \pm 0.038$ \\
   Instellation ($F_i$) & W m$^{-2}$ & ... & (3.0$\pm$0.6)$\cdot 10^4$\\
   Equilibrium Temperature ($T_{\rm eq}$) & K$^b$ & ... & $588 \pm 10$ \\
\hline
\multicolumn{4}{c}{Activity}\\ 
\hline  
 Stellar rotation period ($P_{\rm rot}$) & days & $\mathcal{U}$(12.0, 15.0) & 13.659$^{+0.029}_{-0.033}$ \\
 Decay time ($P_{\rm dec}$) & days & $\mathcal{U}$(10, 2000) & 113$^{+28}_{-24}$ \\
 Coherence scale (w) &   & $\mathcal{N}$(0.35, 0.035) & 0.388$^{+0.026}_{-0.025}$ \\
\bottomrule
\end{tabular}
\tablefoot{\tablefoottext{a}{TESS Barycentric Julian Date ($\mathrm{BJD_{TDB}}$ - 2457000).} \tablefoottext{b}{Computed as $T_{\rm eq} = T_\star \left(\frac{R_{\star}}{2a} \right)^{1/2}\left[f (1-A_b) \right]^{1/4}$, assuming a Bond albedo $A_b = 0.3$ and $f = 1$.}}
\end{table*}

The orbital periods of TOI-1803\,b, and TOI-1803\,c were determined to be $6.29329 \pm 0.00003$ and $12.88578 \pm 0.00004$ days, respectively, and the radii were determined to be $2.99 \pm 0.08\,R_{\oplus}$ and $4.29 \pm 0.08\,R_{\oplus}$, respectively. Moreover, we obtained a stellar activity radial velocity semi-amplitude of $14.8 \pm 2.0$m\,s$^{-1}$, whose modeling is crucial to extract the planetary components with semi-amplitude of $4.4 \pm 1.0$\,m\,s$^{-1}$ and $2.1 \pm 1.0$\,m\,s$^{-1}$ for, respectively, planet -b and -c. From the RV semi-amplitudes we obtain the planetary masses, which are $10.3 \pm 2.5\,M_{\oplus}$ and $6.0 \pm 3.0\,M_{\oplus}$ for planet -b and -c, respectively.

The combination of photometric and radial velocity data allowed for a characterization of the two mini Neptunes, including the determination of their orbital periods, sizes with a relative error of 2.7\% and 1.9\%, and masses with a relative error of 24\% and 50\% for planet -b and -c, which corresponds to a $4\sigma$ and $2\sigma$ mass detection, respectively.

\begin{figure*}[!htbp]
    \centering
    \includegraphics[width=0.45\textwidth]{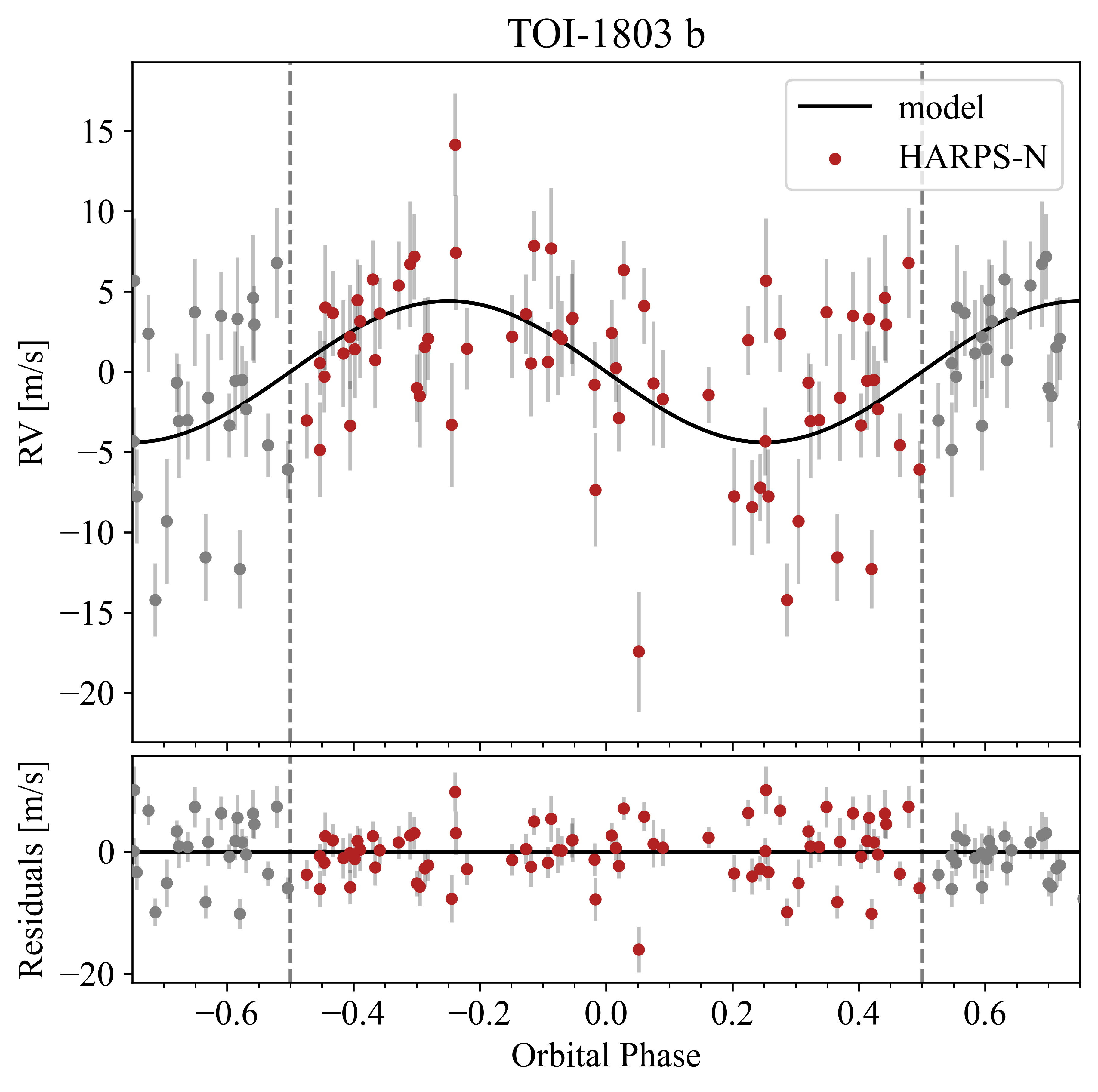}
    \includegraphics[width=0.45\textwidth]{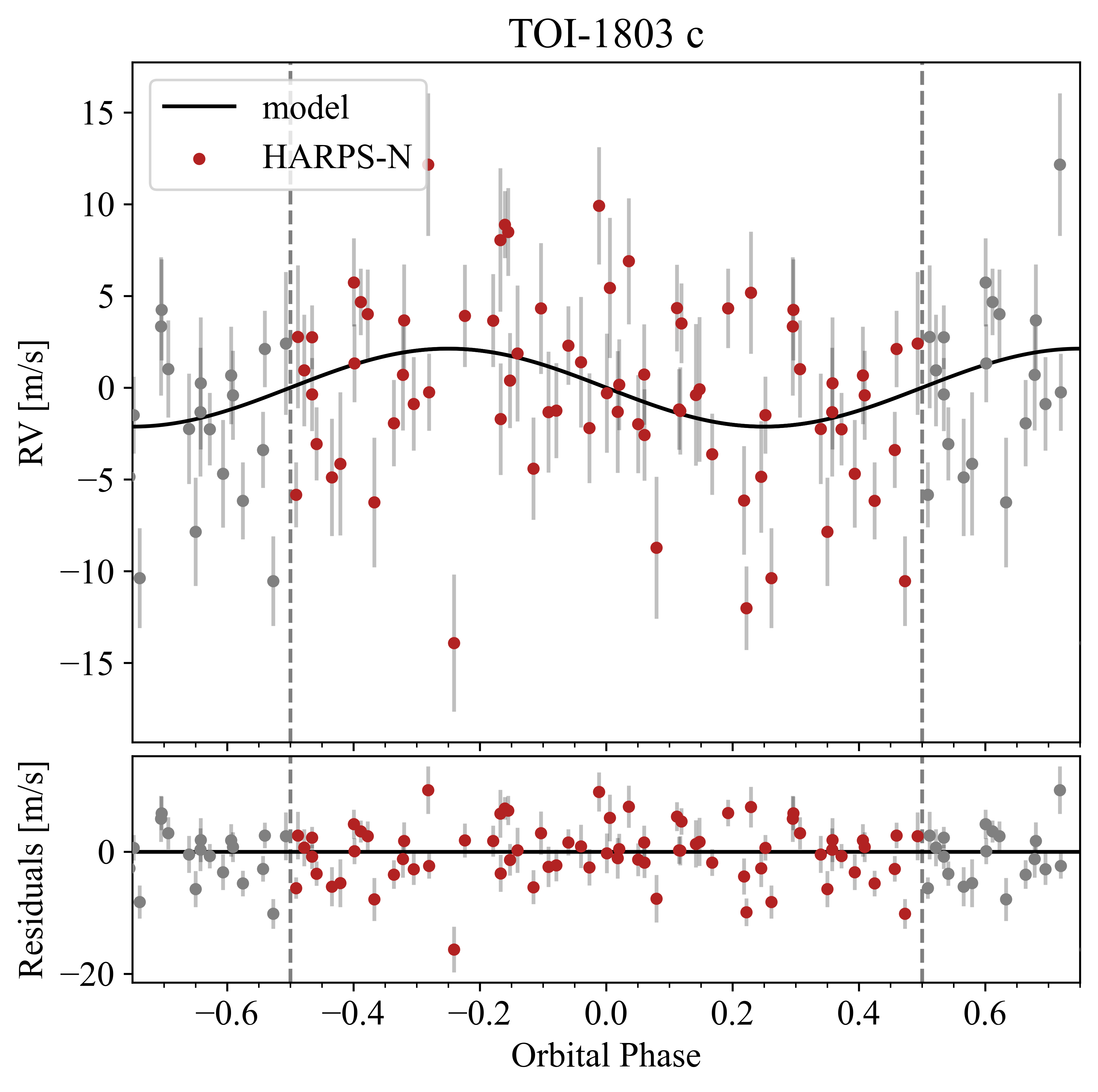}
    \caption{Phase-folded radial velocity components for the two exoplanets -b and -c. From the planetary RV component, it is possible to measure the masses of the planets -b and -c within, respectively, $4.1\sigma$ and $2.0\sigma$}
    \label{fig:semiamplitudes}
\end{figure*}

The modeling of the stellar activity using the multidimensional GP approach allowed us to extract the planetary RV components (Fig. \ref{fig:semiamplitudes}). Radii and masses were computed taking into account the stellar mass and radius with the associated uncertainties, as reported in Table \ref{table:starphotometry}. The derived bulk densities of the two planets are $0.39 \pm 0.10\,\rho_{\oplus}$ and $0.076 \pm 0.038\,\rho_{\oplus}$ for, respectively planet -b and -c. Given these results, TOI-1803\,c is among the least dense of the known Neptunian exoplanets.

\subsection{TTV analysis}
\label{sec:ttv}

The periods obtained from the previous section show a commensurability close to 2:1, 
hinting at a possible mean motion resonance (MMR) of the first order.
In this configuration, due to the strong gravitational interaction between the planets,
we could expect an enhanced and
anti-correlated TTV signal of the two bodies.
We measured the mid-transit times ($T_{0}$s) of both planets with \texttt{PyORBIT} (see Table~\ref{tab:transit_times}).
We evaluated the possible TTV signal through the observed - calculated (O-C) diagram,
computed as the difference between the observed (O) $T_{0}$s  and the predicted (C) ones
from the linear ephemeris (reference times and periods in Table~\ref{tab:obspars}).

We used both space-based and ground-based observations for this analysis. We re-fit all the transit lightcurves with PyORBIT, as it has been done in the previous section, but leaving all the single mid-transit times as free parameters.

\begin{table}[!htpb]
\caption[]{List of transit times ($T_0$s) determined with \texttt{PyORBIT}.}
\label{tab:transit_times}
\centering
\begin{tabular}{lccl}
    \hline
    \hline
    planet & $T_0$ & $\sigma_{T_0}$ & telescope \\
           & ($\mathrm{BJD_{TDB}}$) & (\textbf{days})       \\
    \hline
    b & 2458904.61853 & 0.00507 & TESS\\
      & 2458910.91509 & 0.00568 & TESS\\
      & 2458917.21893 & 0.00801 & TESS\\
      & 2458923.50230 & 0.00405 & TESS\\
      & 2459590.60255 & 0.00274 & CHEOPS\\
      & 2459603.18267 & 0.00339 & CHEOPS\\
      & 2459647.23118 & 0.00647 & TESS\\
      & 2459659.80664 & 0.00818 & TESS\\
      & 2459666.10569 & 0.00225 & CHEOPS\\
      & 2459691.27953 & 0.00242 & CHEOPS\\
    \hline
    c & 2458911.67372 & 0.00226 & TESS\\
      & 2458924.56324 & 0.00297 & TESS\\
      & 2459246.86370 & 0.00210 & Adams\\
      & 2459607.51370 & 0.00760 & Albanya\\
      & 2459607.50634 & 0.00115 & CHEOPS\\
      & 2459633.27566 & 0.00173 & CHEOPS\\
      & 2459646.16450 & 0.00257 & TESS\\
      & 2459659.05914 & 0.00408 & TESS\\
      & 2459671.93665 & 0.00119 & CHEOPS\\
      & 2459684.82708 & 0.00051 & LCO Hal ($z_s$)\\
      & 2459684.82754 & 0.00041 & LCO Hal ($r'$)\\
      & 2459684.82870 & 0.00100 & LCO Hal ($g'$)\\
      & 2459684.82983 & 0.00062 & LCO Hal ($i'$)\\
      & 2459723.50000 & 0.0019  & LCO Tei ($i'$)\\
      & 2459723.50080 & 0.0035  & LCO Tei ($g'$)\\
      
    \hline
\end{tabular}
\end{table}

We decided to run a dynamic analysis to further investigate the possible
orbital configuration that could produce the observed O-Cs.
We used the dynamical code \trades\footnote{\url{https://github.com/lucaborsato/trades}}
\citep{borsato2014, borsato2019, borsato2021a, borsato2021b, nascimbeni2023A&A...673A..42N, Borsato2024A&A...689A..52B},
which allows us to integrate the planetary orbits and 
simultaneously fit the $T_{0}$s and the RVs.
At the time of writing, \trades{} does not implement a stellar activity model, 
but simply it adds in quadrature a jitter term to the RV uncertainties. To include the RV dataset in this analysis, we subtracted the stellar activity component from the RV dataset using the parameters obtained with \texttt{PyORBIT} in Sect. \ref{sec:phot_rv}.\par

We used as planetary fitting parameters
the mass ratio ($M_\mathrm{p}/M_\star$),
the periods ($P$),
the mean longitudes ($\lambda$)\footnote{
$\lambda = \mathcal{M} + \omega + \Omega$,
where $\mathcal{M}$ is the mean anomaly,
$\omega$ is the argument of periastron (or pericenter),
and $\Omega$ is the longitude of the ascending node.
},
and fixed the longitude of ascending node $\Omega = 180^\circ$ \citep[following][]{winn2010exop.book...55W,borsato2014},
for both planets.

We tested a circular circular configuration, 
favoured by the $\Delta$BIC of the \texttt{PyORBIT} analysis,
and an eccentric one fitting eccentricities ($e$) and
the argument or pericentres ($\omega$) 
in the form $\sqrt{e}\cos\omega$ and $\sqrt{e}\sin\omega$.

All the parameters have been defined in astrocentric coordinates and
they are osculating parameters at the reference time $2458904\ \mathrm{BJD_{TDB}}$.
Even if we removed the stellar activity model from the RV data set, we
also fitted an RV jitter term ($\sigma_\mathrm{j}$) in $\log_{2}$ and an RV offset (RV$_\gamma$).
All the parameters have uniform uninformative priors (see Table~\ref{tab:trades_parameters})
based on the physical wide parameters,
in particular, the masses have been bounded between 0.1 and $100\ M_\oplus$.\par
For both configurations, we first run \trades{} with \pyde{} for 54000 steps and 54 configurations.
Then we took the best-fit configuration from \pyde{} and 
we used it  to generate 54 initial walkers\footnote{
The walkers are initialised as a tight Gaussian centred on the best-fit solution.}
for \texttt{emcee} and run it for 500\,000 steps.
We used a thinning factor of 100 and a burning phase of 250\,000 steps,
well after the chains reached the convergence following the 
Gelman-Rubin \citep{gelmanrubin1992}, Geweke \citep{geweke1991} criteria,
auto-correlation function \citep{AffineInvariantGoodmanWeare2010},
and visual inspection.
From the posterior distribution, we computed the best-fit configuration
as the maximum-a-posteriori probability (MAP), 
that is the parameter set at the maximum of the log-probability.

\begin{figure*}[!htbp]
    \centering
    \includegraphics[width=0.99\columnwidth]{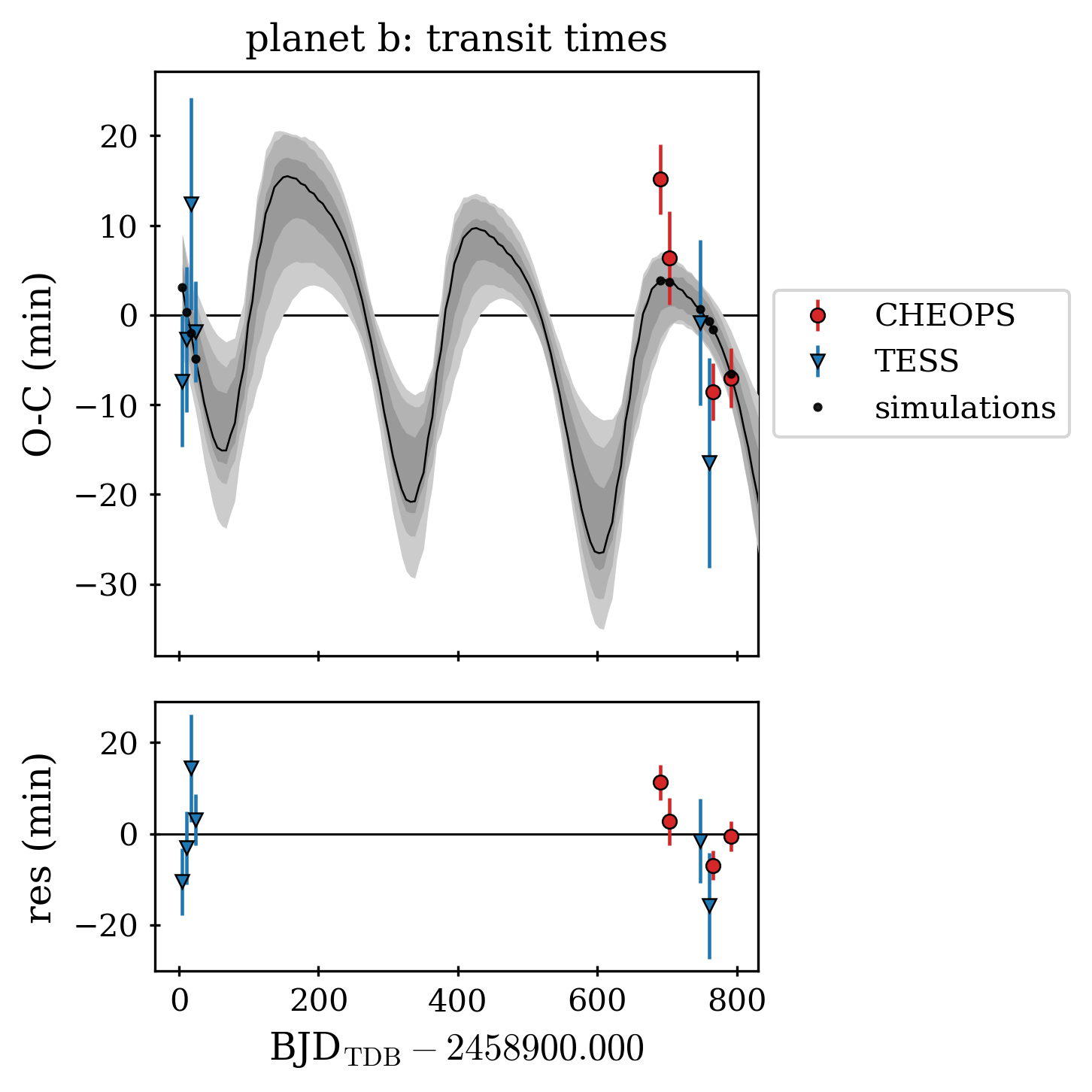}
    \includegraphics[width=0.99\columnwidth]{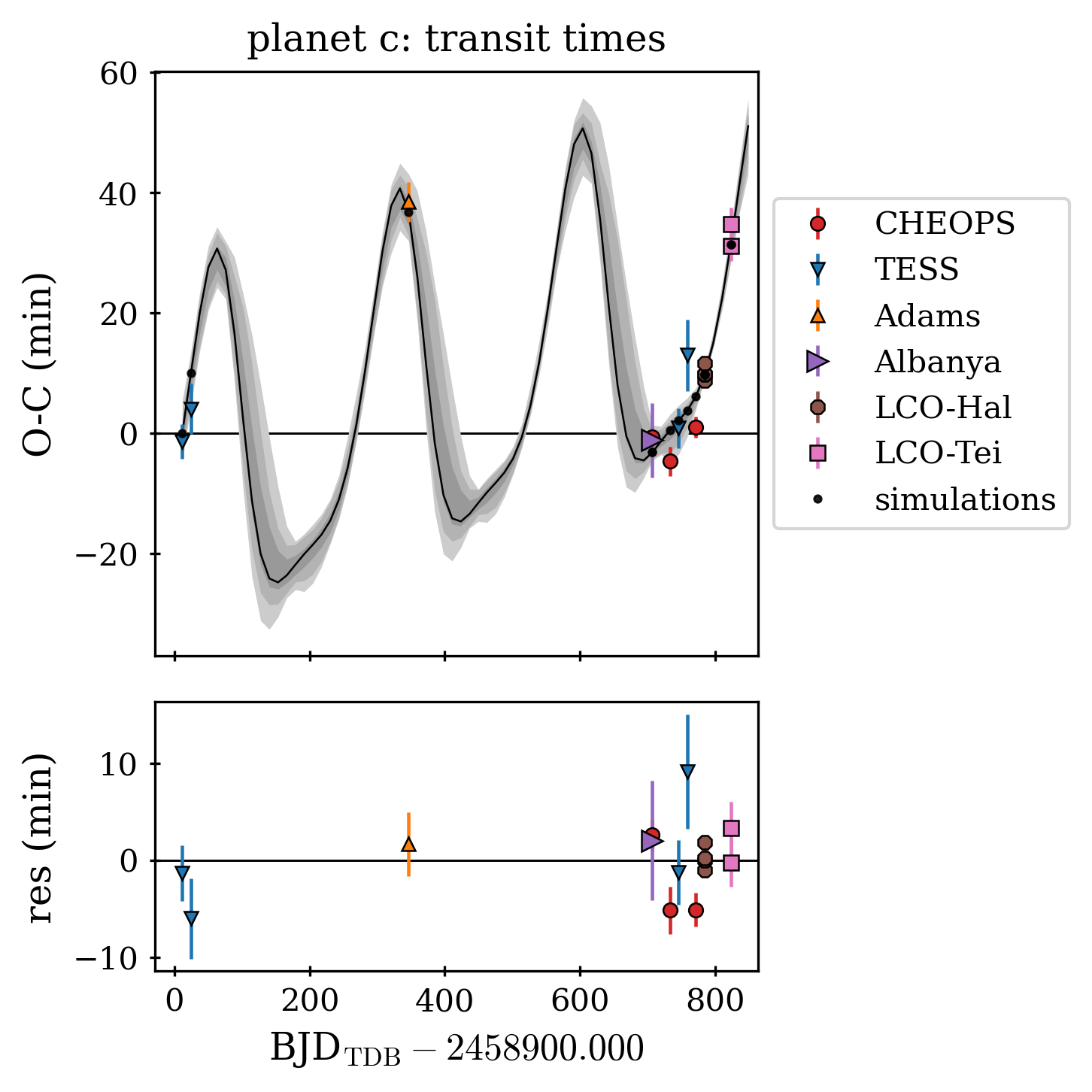}
    \caption{O-C diagram for the planet -b \textbf{(\textit{left}) and -c (\textit{right})} comparing 
    the observations 
    \textbf{(different marker and colour)}
    with \textsc{TRADES} simulations (black circles).
    The black line is the over-sampled best-fit model and
    the grey 
    \textbf{shaded areas are the one, two, and three $\sigma$ computed from}
    100 samples drawn from the posterior distribution
    .
    }
    \label{fig:trades_oc_bc}
\end{figure*}

\begin{figure}[!htbp]
    \centering
    \includegraphics[width=0.99\columnwidth]{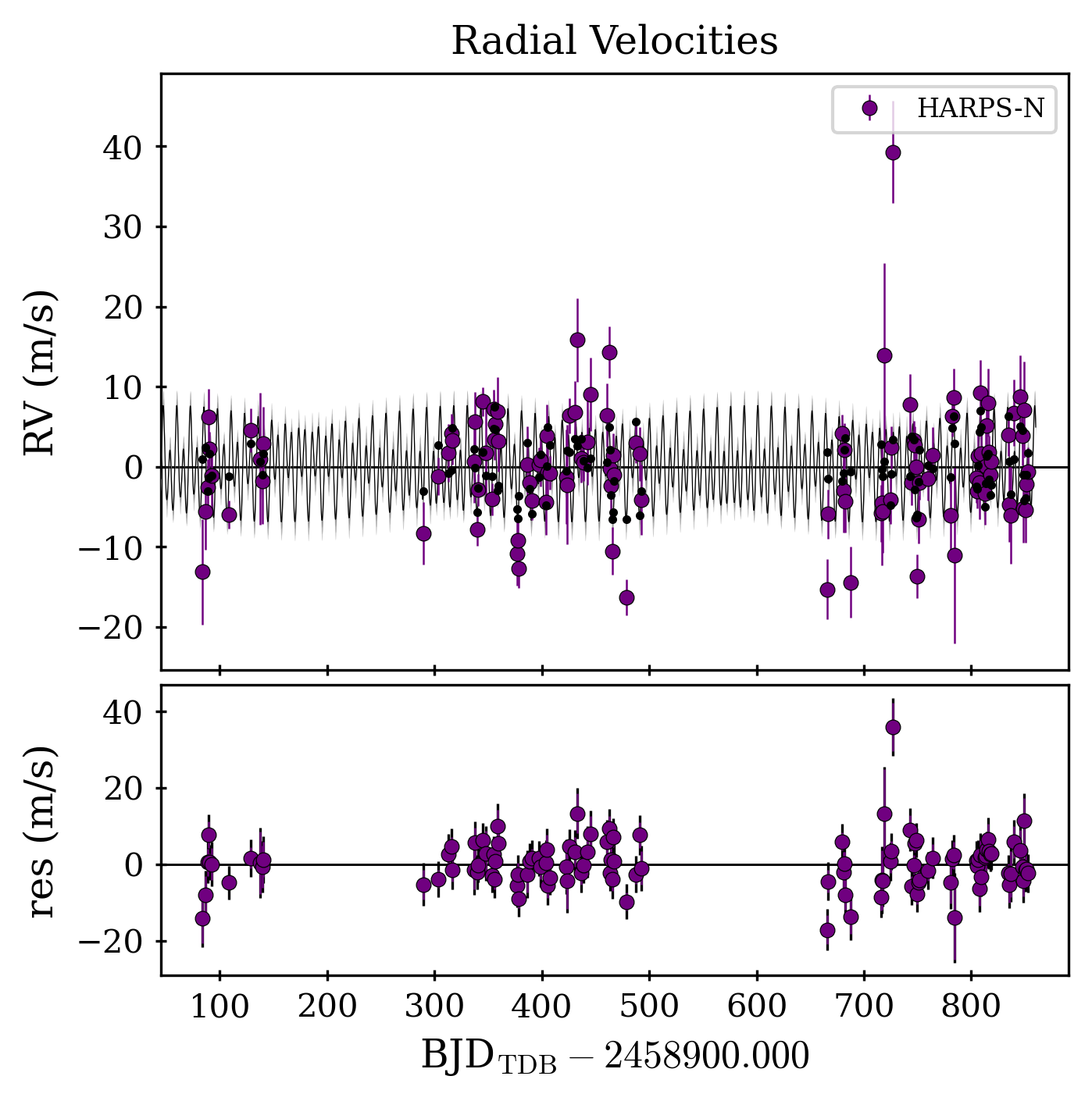}
    \caption{RV plot (\textit{upper panel}) with observations as purple circles, 
    \texttt{TRADES} simulations as black
    circles.
    The black line is the over-sampled best-fit model and
    the grey
    \textbf{regions computed as for Fig.~\ref{fig:trades_oc_bc}.}
    Residuals plot (\textit{lower panel}) where the black error bars are the observed RV errors with the jitter term added in quadrature.
    }
    \label{fig:trades_rv}
\end{figure}

The uncertainties have been computed as the High-Density Interval (HDI,
or high posterior density\footnote{
\trades{} implements the code \texttt{hpd} within \texttt{PyAstronomy},
available at \url{https://github.com/sczesla/PyAstronomy}.
}) at the $68.27\%$ of the posterior distribution, which is the equivalent of the confidence intervals
in the case of Gaussian distribution.
We computed the physical posteriors of the masses multiplying 
the posterior of $M_\mathrm{p}/M_\star$ 
by a Gaussian distribution of the stellar mass (Tab.~\ref{tab:obspars})
and computed the HDI, while the MAP of the masses has been computed multiplying the posterior for the fixed value of the stellar mass.
As previously, we computed the $\Delta \mathrm{BIC}$ between 
the circular and eccentric case,
and we found that it strongly favours the eccentric case 
($\Delta \mathrm{BIC} = \mathrm{BIC_{e > 0}} - \mathrm{BIC_{e=0}} < - 200$).
See the summary of the parameters in Table~\ref{tab:trades_parameters},
the O-C models of both planets in Fig.~\ref{fig:trades_oc_bc} and 
the RV plot in Fig.~\ref{fig:trades_rv}.

The computed dynamical masses $M_\mathrm{dyn}$ of 
$10.32_{-0.48}^{+1.66}\ M_\oplus$ and $8_{-2}^{+1}\ M_\oplus$
of planets b and c, respectively, are consistent  at $z$-score\footnote{We defined the Z-score as
    $z = \frac{|M_\mathrm{kep} - M_\mathrm{dyn}|}{
        \sqrt{ \max|\sigma_{M_\mathrm{kep}}^{\pm}|^{2} + \max|\sigma_{M_\mathrm{dyn}}^{\pm}|^{2}}}$,
    where the subscripts kep and dyn stand for keplerian and dynamical, respectively,
    and the $\sigma^{\pm}$ is the asymmetric error.
    } $z=1\sigma$

with values \textbf{$M_\mathrm{kep}$} computed 
in Section~\ref{sec:phot_rv} from the combined analysis of the photometry and of RV. 
Since we subtracted the stellar activity signal in this analysis, the error bars may be underestimated. 
This occurs because the possible degeneracy between planetary and stellar signals is not taken into account, 
especially in this case where the planetary period of planet -c is very close to the stellar rotation period.

\par

\begin{table}
    \caption{\textbf{Best-fit parameters (MAP and HDI) 
    for the eccentric configuration of the dynamical analysis with \trades.}} 
    \label{tab:trades_parameters}      
    \centering                          
    \begin{tabular}{l c c}        
    \hline\hline                 
     & MAP (HDI) & Prior \\     
    \hline                        
    
    $\frac{M_\plb}{M_\star}\, \left(\frac{M_\odot}{M_\star}\right) \times 10^{-6}$ & $42.32_{-0.81}^{+5.96}$ & \unif{0.39}{437.00} \\
    $P_\plb$~(days)                  & $6.291709_{-0.000055}^{+0.000431}$ & \unif{5.5}{8.0} \\
    $\sqrt{e_\plb}\cos\omega_\plb$   & $0.2040_{-0.0128}^{+0.0024}$ & \unif{-\sqrt{0.5}}{\sqrt{0.5}} \\
    $\sqrt{e_\plb}\sin\omega_\plb$   & $0.106_{-0.025}^{+0.026}$ & \unif{-\sqrt{0.5}}{\sqrt{0.5}} \\
    $\lambda_\plb\, (^\circ) $       & $0.053_{-0.051}^{+0.418}$ & \unif{0}{360} \\

    $\frac{M_\plc}{M_\star}\, \left(\frac{M_\odot}{M_\star}\right) \times 10^{-6}$ & $35_{-7}^{+4}$ & \unif{0.39}{437.00} \\
    $P_\plc$~(days)                  & $12.89385_{-0.00033}^{+0.00056}$ & \unif{11}{14} \\
    $\sqrt{e_\plc}\cos\omega_\plc$   & $0.1827_{-0.0200}^{+0.0092}$ & \unif{-\sqrt{0.5}}{\sqrt{0.5}} \\
    $\sqrt{e_\plc}\sin\omega_\plc$   & $-0.283_{-0.014}^{+0.052}$   & \unif{-\sqrt{0.5}}{\sqrt{0.5}} \\
    $\lambda_\plc\, (^\circ) $       & $296.40_{-0.42}^{+1.80}$ & \unif{0}{360} \\
    $\log_{2}\sigma_\mathrm{jitter}$ & $2.013_{-0.173}^{+0.089}$ & \unif{-49.83}{6.64} \\
    $\mathrm{RV}_\gamma\, (\mathrm{m\,s^{-1}})$ & $1995.46_{-0.30}^{+0.42}$ & \unif{-3021}{7035} \\

    $M_\plb\, (M_\oplus)$            & $10.32_{-0.48}^{+1.66}$ & (derived) \\
    $e_\plb$                         & $0.0527_{-0.0062}^{+0.0015}$ & (derived) \\
    $\omega_\plb\, (^\circ)$         & $27_{-6}^{+7}$ & (derived) \\
    $\mathcal{M}_\plb\, (^\circ)$    & $153_{-7}^{+6}$ & (derived) \\
    $M_\plc\, (M_\oplus)$            & $8_{-2}^{+1}$ & (derived) \\
    $e_\plc$                         & $0.1135_{-0.0318}^{+0.0064}$ & (derived) \\
    $\omega_\plc\, (^\circ)$         & $-57_{-3}^{+4}$ & (derived) \\
    $\mathcal{M}_\plb\, (^\circ)$    & $174_{-4}^{+3}$ & (derived) \\
    $\sigma_\mathrm{jitter}\, (\mathrm{m\,s^{-1}})$ & $4.04_{-0.47}^{+0.24}$ & (derived) \\
    \hline                                   
    \end{tabular}
    \tablefoot{
    All the parameters have been defined at the reference time $2458904\ \mathrm{BJD_{TDB}}$.
    $\mathcal{M}$ is the mean anomaly. 
    $\Omega$ has been set to $180^\circ$ for both planets.
    }
\end{table}

\subsection{Stability analysis}

\begin{figure*}
    \centering
	\includegraphics[width=\textwidth]{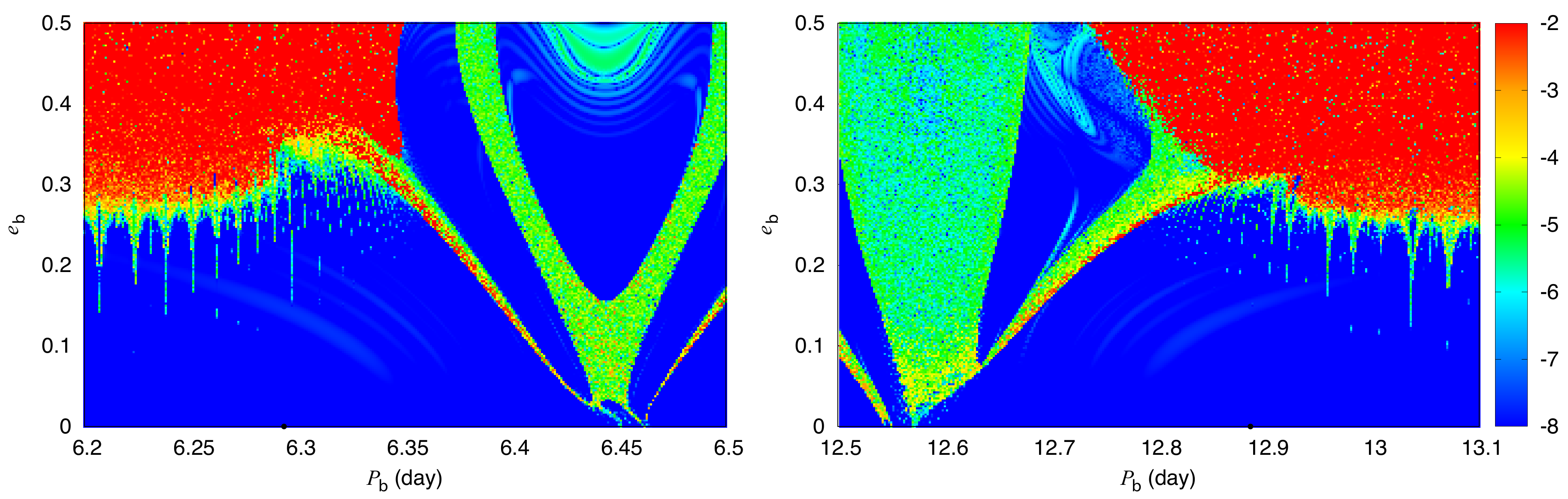}
    \caption{Stability analysis of the TOI-1803 planetary system. For fixed initial conditions (Table~\ref{tab:obspars}), the parameter space of the system is explored by varying the orbital period and the eccentricity of planet-b (left panel) and planet-c (right panel). The step size is $0.0025$ in the eccentricities, $0.001$~day in the orbital period of planet-b, and $0.002$~day in the orbital period of planet-c. For each initial condition, the system is integrated over 5000~yr, and a stability indicator is calculated, which involves a frequency analysis of the mean longitude of the inner planet. The chaotic diffusion is measured by the variation in the frequency (see text). Red points correspond to highly unstable orbits, while blue points correspond to orbits that are likely to be stable on Gyr timescales. The black dots show the values of the best-fit solution (Table~\ref{tab:obspars}).}
    \label{figSA}
\end{figure*}

The TOI-1803 system is composed by two close-in mini Neptunes ($M_b \approx 10.3\,M_\oplus$, $M_c \approx 6.0\,M_\oplus$) near a 2:1 mean motion resonance ($P_c / P_b \approx 2.048$, Table~\ref{tab:obspars}).
To get a clear view of the system dynamics, we performed a stability analysis in a similar way as for other planetary systems \citep[eg.][]{Correia_etal_2005, Correia_etal_2010}.
The system is integrated on a regular 2D mesh of initial conditions in the vicinity of the best fit (Table~\ref{tab:obspars}).
We used the symplectic integrator SABAC4 \citep{Laskar_Robutel_2001}, with a step size of $5 \times 10^{-4} $~yr and general relativity corrections.
Each initial condition is integrated for 5000~yr, and a stability indicator, $\Delta = |1-n'/n|$, is computed. 
Here, $n$ and $n'$ are the main frequency of the mean longitude of the planet over 2500~yr and 5000~yr, respectively, calculated via frequency analysis \citep{Laskar_1990, Laskar_1993PD}. 
In Fig.~\ref{figSA}, the results are reported in color: orange and red represent strongly chaotic unstable trajectories; yellow indicates the transition between stable and unstable regimes; green correspond to moderately chaotic trajectories, but stable on Gyr timescales; cyan and blue give extremely stable quasi-periodic orbits.

We explore the stability of the system by varying the orbital period and the eccentricity of the inner planet (Fig.~\ref{figSA}, left) and of the outer planet (Fig.~\ref{figSA}, right), respectively.
We observe that the best-fit solution from Table~\ref{tab:obspars} and \ref{tab:trades_parameters} are
completely stable (black dots in Fig.~\ref{figSA}), even if we increase the eccentricities up to 0.1. 
As a by-product of the analysis, \trades{} outputs the Hill stability \citep{sundman1913} of the parameter set
through the so-called AMD-Hill criterion \citep[Eq. 26,][]{Petit2018A&A...617A..93P},
based on the angular momentum deficit \citep[AMD,][]{Laskar1997A&A...317L..75L,Laskar2000PhRvL..84.3240L,LaskarPetit2017A&A...605A..72L}
We found that the entire posterior distribution is AMD-Hill stable for both the circular and eccentric configurations.
In addition, we verify that the system is outside the 2:1 mean motion resonance, which corresponds to the large stable structure in the middle of the figures.
We also note that the system is ``wide'' of the resonance as predicted by tidal evolution models.
Indeed, the inner planet is close enough to the star to undergo strong tidal interactions that drive the period ratio to a value above the exact resonance \citep[eg.][]{Lissauer_etal_2011K, Delisle_Laskar_2014}.
We conclude that the TOI-1803 planetary system presented in Table~\ref{tab:obspars} is realistic and supple to the uncertainties in determining the eccentricities.

The tides raised by the star inside planet b can produce significant heating of its interior, leading to a tidal circularization of its orbit. Unfortunately, its rheology is unknown because we do not know any details about its internal structure. Nevertheless, assuming it is similar to those of Uranus or Neptune, we adopt a modified tidal quality factor of $Q^{\prime} =10^{5}$ \citep[][Sect.~5.4]{ogilvie2014} and an eccentricity of 0.1 that yield an internally dissipated power of $1.4 \times 10^{16}$~W that may affect the internal dynamics of the planet. Such a power has been computed employing the constant-time-lag tidal model of \citet{leconte2010}, where we computed the product by the Love number $k_{2}$ of the planet and its tidal time lag $\Delta t$ using the formula $k_{2} \Delta t = (2/3) Q^{\prime}/n$, where $n = 2\pi/P_{\rm orb}$ is the orbital mean motion of the planet with $P_{\rm orb}$ being its orbital period.  The e-folding time scales for the decay of the eccentricities of the orbits of planets b and c can be estimated employing the same model and turn out to be 15.4 Gyr and 33.5 Gyr, respectively, which is much longer than the age of the system, thus indicating that tides did not have time to circularize initially eccentric orbits if we assume $Q^{\prime} = 10^{5}$ for those planets.

\section{Planetary formation and evolution}\label{sec:planetary_formation}
\begin{table}
    \centering
    \caption{Input parameters used to run the modified GroMiT code to produce the synthetic planetary populations shown in Fig. \ref{fig:synthetic_population}. Half the simulations adopt pebble sizes of 1$\,$cm and the other half of 1$\,$mm.}
    \label{tab:sim_params}
    \begin{tabular}{lcc}
    \hline \hline
    \multicolumn{2}{c}{Simulation Parameters}\\
    \hline
    \multicolumn{1}{l}{N$^\circ$ of simulations} & \multicolumn{1}{c}{6$\times10^4$}\\
    \multicolumn{1}{l}{Seed formation time} & \multicolumn{1}{c}{0.1--1$\, \times \, 10^6$\,yr} \\
    \multicolumn{1}{l}{Disk lifetime} & \multicolumn{1}{c}{$5 \times 10^6$\,yr} \\
    \hline
    \multicolumn{2}{c}{Star, Planet \& Disk properties} \\
    \hline
      \multicolumn{1}{l}{Stellar Mass} & \multicolumn{1}{c}{0.756$\,$M${_\odot}$}\\
      \multicolumn{1}{l}{Seed Mass} & \multicolumn{1}{c}{0.01$\,$M${_\oplus}$} \\
      \multicolumn{1}{l}{Initial envelope mass} & \multicolumn{1}{c}{0.0$\,$M${_\oplus}$} \\
      \multicolumn{1}{l}{Initial semimajor axis} & \multicolumn{1}{c}{0.1--30.0$\,$au} \\
      \multicolumn{1}{l}{Disk characteristic radius} & \multicolumn{1}{c}{30.0$\,$au} \\
      \multicolumn{1}{l}{Temperature @ 1$\,$au} & \multicolumn{1}{c}{150$\,$K} \\
      \multicolumn{1}{l}{Surface density @ 30$\,$au} & \multicolumn{1}{c}{490\,kg m$^{-2}$} \\
      \multicolumn{1}{l}{Disk accretion coefficient, $\alpha$} & \multicolumn{1}{c}{0.01}\\
      \multicolumn{1}{l}{Turbulent viscosity, $\alpha$$_\nu$}& \multicolumn{1}{c}{0.0001}\\
      \multicolumn{1}{l}{Pebble size}& \multicolumn{1}{c}{1$\,$cm \& 1$\,$mm}\\
    \hline    
    \end{tabular}
\end{table}

We took advantage of the characterization of TOI-1803's planets to look into the formation history of the planetary system and the possible nature of its native protoplanetary disk. Our investigation combines population synthesis simulations in the framework of the pebble accretion scenario with the Monte Carlo exploration of the interior structure of both planets. The population synthesis simulations are performed by means of a modified Monte Carlo version of the GroMiT code \citep{polychroni2023}, which models the formation track of planets through the treatment for the growth and migration of solid planets/planetary cores from \cite{johansen2019} and that of gaseous planets from \cite{tanaka2020}. The solids-to-gas ratio of the protoplanetary disk, which sets the local abundance of pebbles, is computed based on the treatment from \cite{turrini2023}. The Monte Carlo simulations of the interior structure of the two planets are based on the equations of \cite{lopez2014}.

\begin{figure*}
\centering
\includegraphics[width=\textwidth]{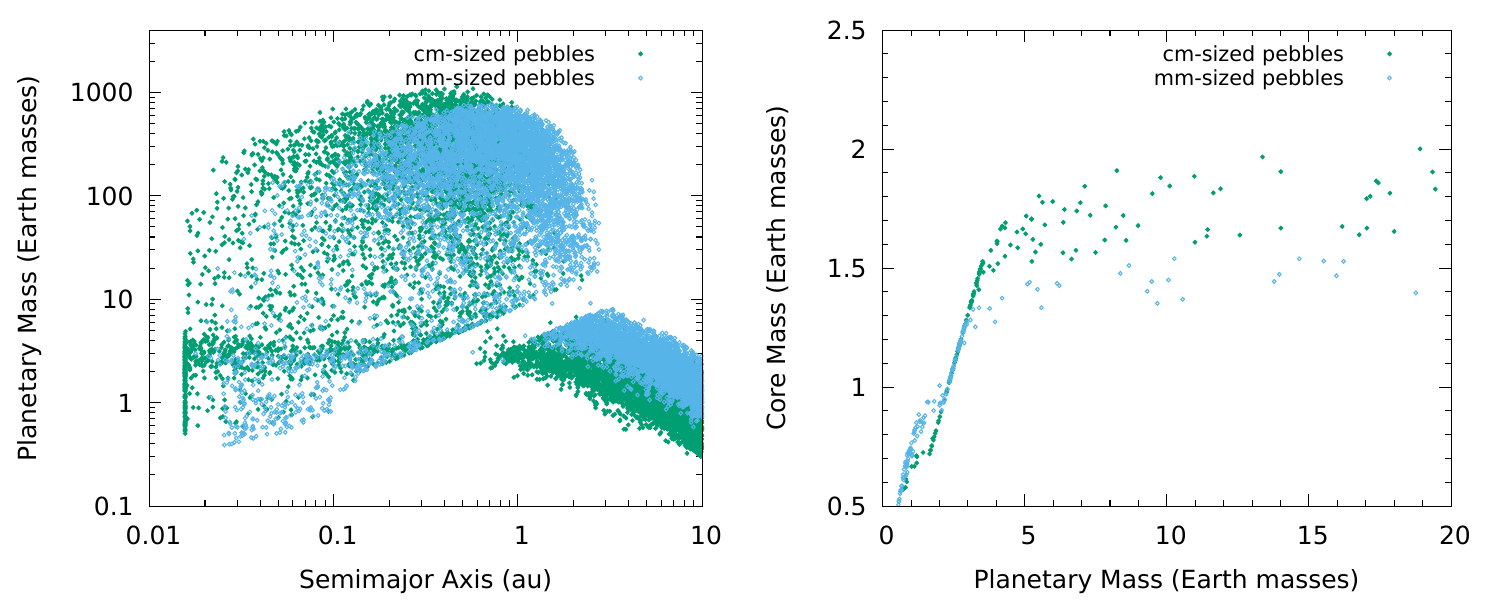}
\caption{Synthetic populations of planets produced by the Monte Carlo version of GroMiT using mm-sized (blue symbols) and cm-sized (green) pebbles. \textit{Left:} final planetary masses and semimajor axes of the planetary populations. \textit{Right:} core masses (pebble isolation masses) of the planets with final semimajor axes between 0.05 and 0.1 au and planetary masses between 0 and 20 M$_\oplus$ as a function of the final planetary mass.}\label{fig:synthetic_population}
\end{figure*}

In the population synthesis simulations, we assume a protoplanetary disk with a characteristic radius of 30 au and gas mass of 0.03 M$_\odot$. This disk mass is set by computing the mass of a minimum mass solar nebula with the same extension and multiplying it by the ratio between the masses of TOI-1803 and the Sun (see \citealt{turrini2021,turrini2023} for details). We consider both cm-sized and mm-sized pebbles with pebble density of 1.5 g/cm$^{3}$, i.e. an equal-part mixture of silicates and ices with 50\% porosity. The temperature profile of the disk, which determines the positions of the ice snowlines and the midplane solids-to-gas ratio profile, is set to T=$150~\rm{K}\left(\rm{R}/{1~\rm{au}}\right)^{-0.5}$ to account for the colder stellar temperature. We perform a total of 60000 Monte Carlo runs injecting in the disk 30000 planetary seeds of 0.01 M$_\oplus$ for each pebble size uniformly distributed between 0.1 and 30 au. The seeds are uniformly implanted in the disk within 0 and 1 Myr and we track their growth and migration across the lifetime of the disk, which is assumed to be 5 Myr. The resulting synthetic planetary populations are shown in Fig. \ref{fig:synthetic_population} (left plot).

Due to the uncertainty on the masses of TOI-1803 b and c, we select from our synthetic planetary population all planets with a final semimajor axis comprised between 0.05 and 0.1 au and mass comprised between 0 and 20 M$_\oplus$. As shown by the right plot of Fig. \ref{fig:synthetic_population}, all the selected planets are characterized by small cores no more massive than 1.5-2 M$_\oplus$. It must be pointed out, however, that these values are based solely on the pebble isolation masses across the circumstellar disk. GroMiT's tracks do not yet account for the possible accretion of planetesimals by the growing cores nor for the enhanced disk gas metallicity due to the sublimation of ices at the crossing of the snowlines by the inward drifting pebbles. To test whether TOI-1803's planets could have formed by pebble accretion, we explored the core-envelope ratios compatible with the observed planetary radii.

\begin{figure}
\centering
\includegraphics[width=\columnwidth]{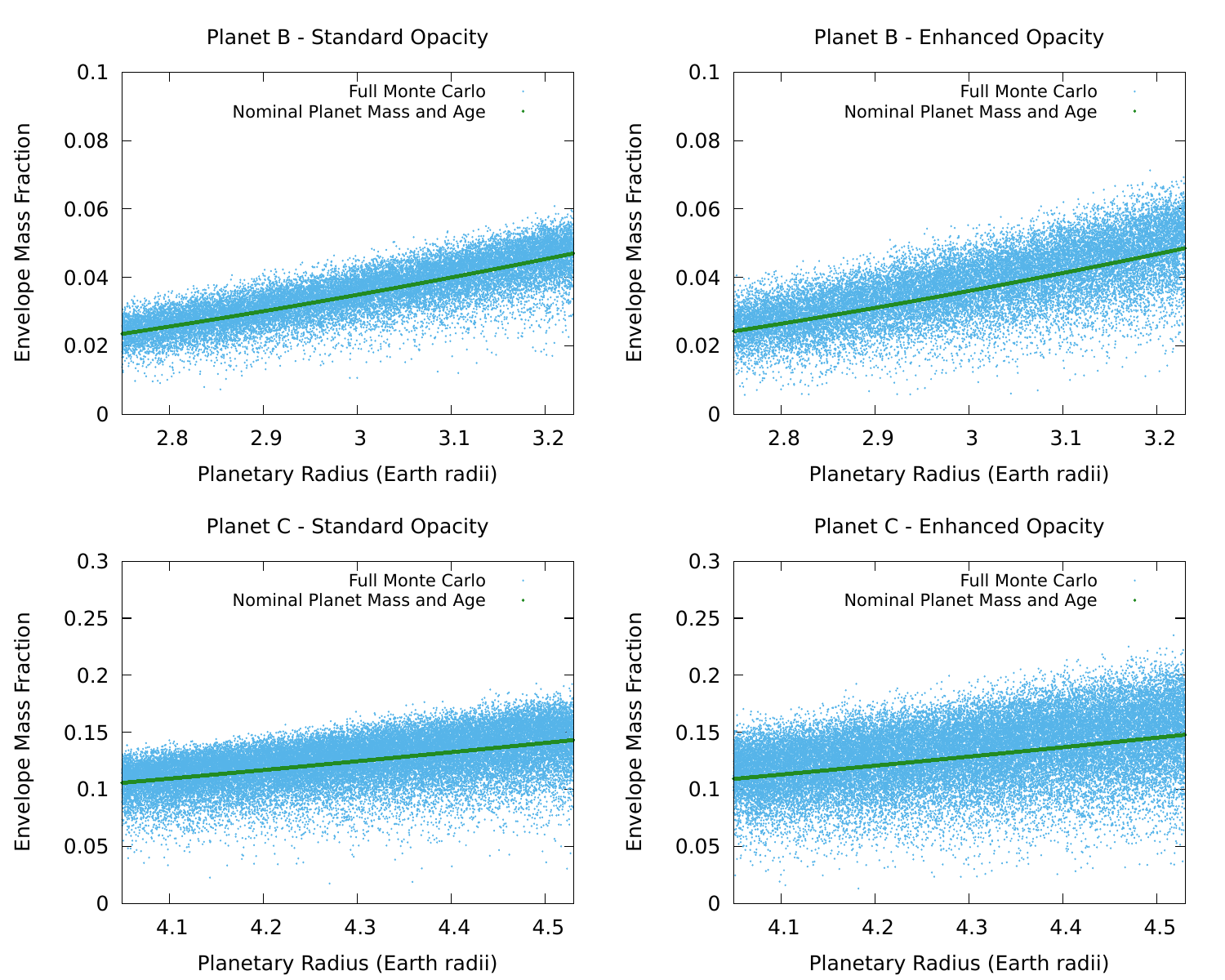}
\caption{Envelope mass fractions of planet b and c as a function of the real planetary radius for both the assumptions of standard and enhanced opacity of the planetary envelopes. The green symbols show the envelope mass fraction--planetary radius curves emerging from the Monte Carlo simulation when the nominal planetary masses and stellar age are adopted. The blue symbols show the same curves when we account for the uncertainties on both planetary masses and stellar ages.}\label{fig:envelope_vs_radius}
\end{figure}

\begin{figure}
\centering
\includegraphics[width=\columnwidth]{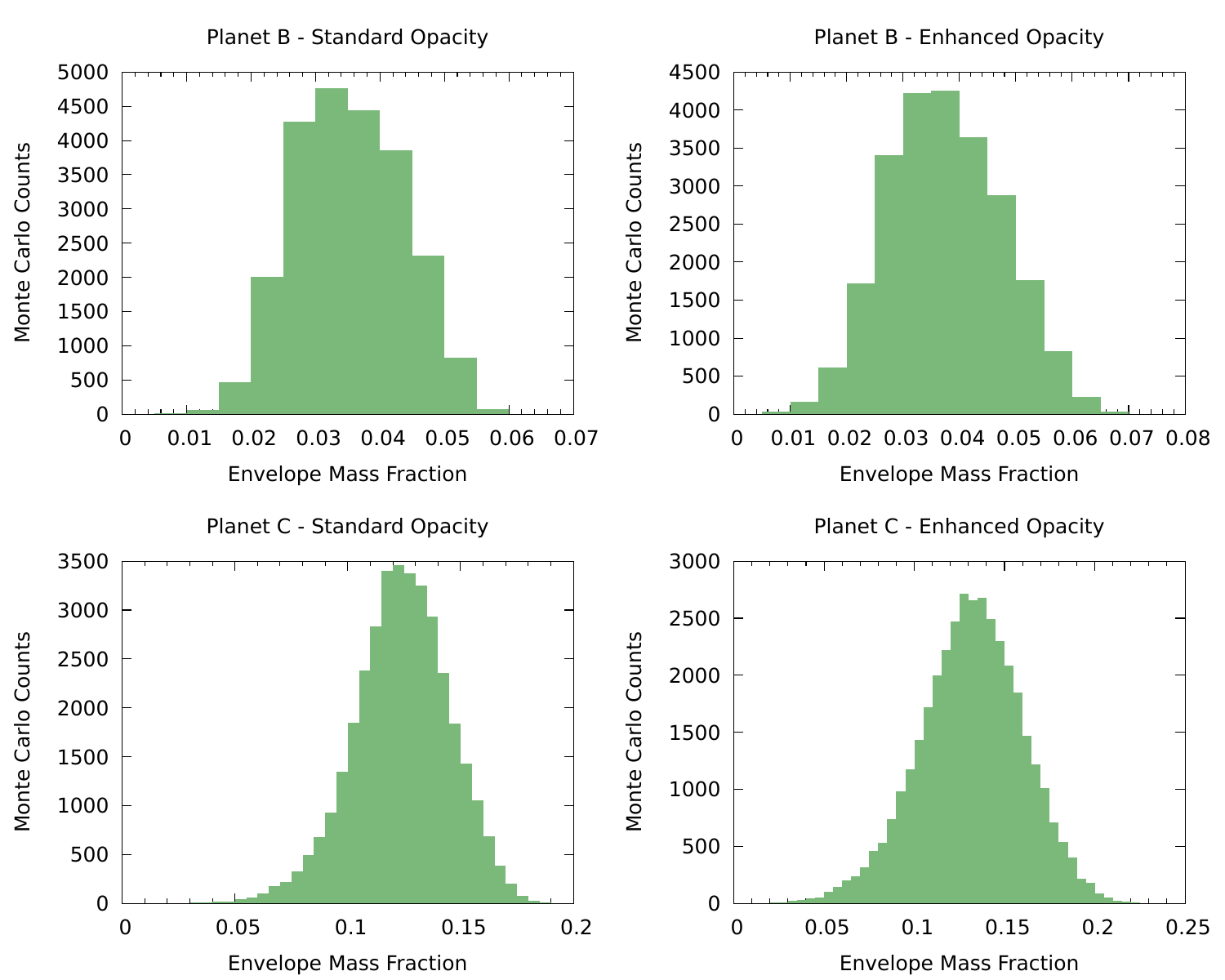}
\caption{Histograms of the envelope mass fractions of planets b and c resulting from the full Monte Carlo sampling from Fig. \ref{fig:envelope_vs_radius} for both the assumptions of standard and enhanced opacity of the planetary envelopes. The modal envelope mass fraction of planet b falls between 3-4\%, while that of planet c between 12-14\%.}\label{fig:envelope_histogram}
\end{figure}

We run two sets of $10^6$ Monte Carlo simulations of the interior structure for each planet based on the equations from \cite{lopez2014} for both the cases of standard and enhanced envelope opacity. The first set adopts the nominal planetary masses and stellar age and randomly extracts the core-to-envelope ratio. The second set runs a full Monte Carlo investigation accounting also for the uncertainty on the planetary masses and stellar age. In both sets of simulations, we select only those planets whose radii fall within $3~\sigma$ from the observed radii. Fig. \ref{fig:envelope_vs_radius} compares the curves resulting from both sets and both opacity assumptions in the planetary radius--envelope mass fraction space, while Fig. \ref{fig:envelope_histogram} shows the distribution of the envelope mass fractions that can fit the observed planetary radii.

As can be seen, fitting the observed radii requires envelope mass fractions of a few percent for planet b (modal peak: 3-4\%) and of about ten percent for planet c (modal peak: 12-14\%), i.e. significantly larger cores than those resulting purely from the pebble isolation masses computed by GroMiT. If the planets formed in a pebble-dominated circumstellar disk, their envelope should be composed of high-metallicity gas due to the enrichment in heavy elements of the disk gas by the ices sublimating from the inward drifting pebbles. The observations of the Juno mission for Jupiter open the possibility that such high-metallicity envelopes can mimic the effects of larger cores \citep{wahl2017,stevenson2020}. Due to the solar metallicity of TOI-1803, however, such a scenario appears to point to both planets having been more massive in the past. 

The accretion of high-metallicity disk gas can result in envelope metallicities \citep{schneider2021} broadly consistent with the average mass-metallicity trend observationally estimated (albeit with large uncertainties) for gas-rich planets \citep{thorngren2016}. Following these results, we use the mass-metallicity trend from \cite{thorngren2016} to get a first-order estimate of the envelope metallicity in the mass range of TOI-1803's planets. 
The resulting 40x solar metallicity requires planets b and c to have been 50\% and 30\% more massive, respectively, for their envelopes to provide enough heavy elements to mimic the effects of larger cores on their planetary radii. 

It must be noted that, due to the uncertainty of the observational data, planetary metallicity can be a factor of a few higher or lower than that estimated with the mass-metallicity trend from \cite{thorngren2016}. As such, without additional constraints it is not possible to rule out higher envelope metallicities that would not require the two planets to have been bigger in the past, although this would require more extreme planet formation scenarios than those investigated by \cite{schneider2021}. Independently on this, due to the inner orbits of the two planets, the two most relevant snowlines for the high-metallicity envelope are those of water and refractory organic carbon \citep{turrini2021,pacetti2022}, suggesting that in the pebble-rich disk scenario, the atmospheres of the two planets should be dominated by oxygen and carbon \citep{schneider2021}.

Alternatively, both planets could have formed in a circumstellar disk containing comparable amounts of pebbles and planetesimals, resulting in both planets experiencing planetesimal impacts during their migration and the growth of their cores not being limited by the pebble isolation mass. Due to the close to resonant architecture, it is plausible that the system formed by convergent migration and that planet c, due to its outer orbit, crossed disk regions already depleted of planetesimals by planet b, thus explaining its smaller core. 
In such a scenario, the resulting envelopes would be significantly enriched in more refractory elements due to planetesimal impacts \citep{turrini2021,pacetti2022} and outgassing from the cores.

The two formation scenarios are thus associated to different atmospheric compositions for the two planets. The pure pebble accretion scenario predicts atmospheres highly enriched in oxygen and carbon and poor of more refractory elements, with a super-stellar C/O ratio \citep{booth2019,schneider2021}. The hybrid pebble-planetesimal formation scenario predicts, instead, atmospheric compositions characterized by more elements and with a stellar or sub-stellar C/O ratio \citep{turrini2021,pacetti2022,fonte2023}. As discussed in Section \ref{sec:JWST}, JWST observations can discriminate among these end-member scenarios. Further constraints could be derived by characterizing the atmospheric C/S ratio, as this ratio is expected to be significantly larger in the pure pebble scenario than in the hybrid pebble-planetesimal one \citep{turrini2021,pacetti2022,crossfield2023}.

\section{Internal structure modelling}
\label{sec:internal_structure}

\begin{figure*}
    \centering
    \includegraphics[width=\linewidth]{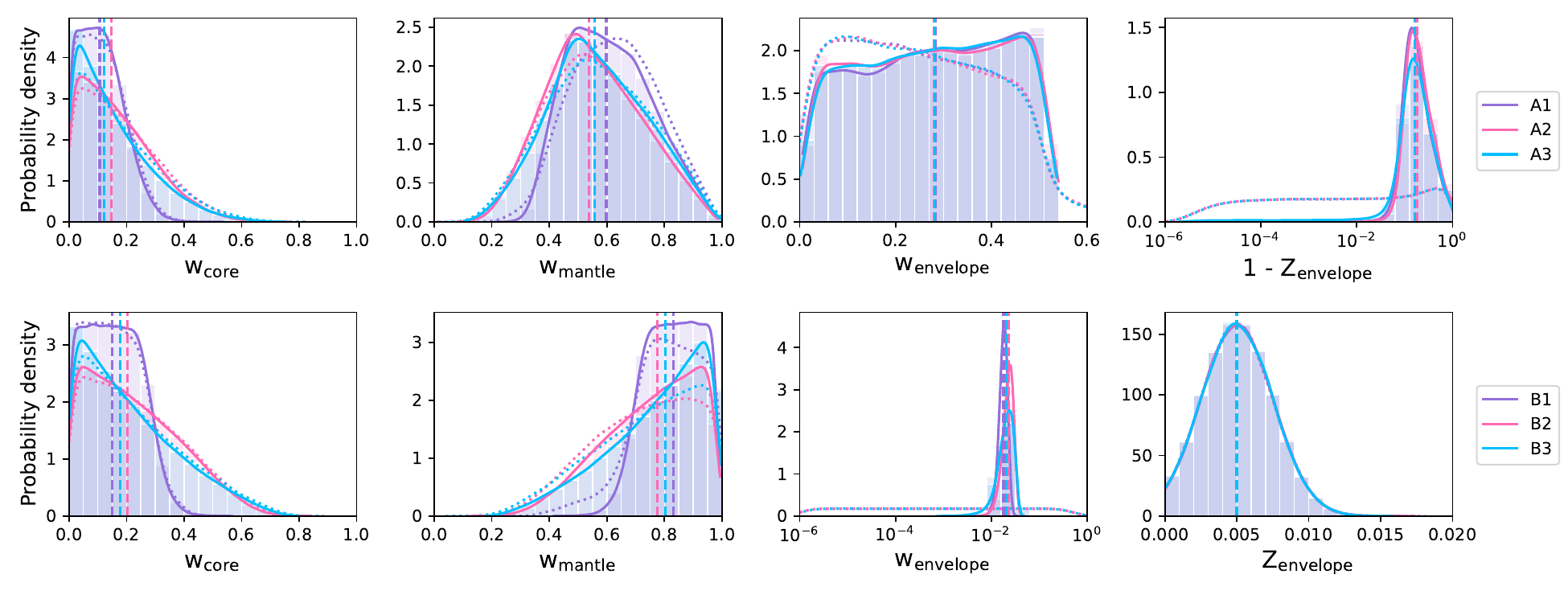}
    \caption{Posterior distributions of the interior structure of TOI-1803 b. We show the mass fractions of the inner core, mantle, and envelope layer as well as the mass fraction of water in the envelope. The different colors show three different priors for the planetary Si/Mg/Fe ratios, stellar (purple), iron-enriched (pink) and sampled uniformly from a simplex (blue). The top row shows the results when assuming a formation scenario outside the ice line, the bottom row is compatible with a formation inside the ice line. The dashed line shows the median of each distribution, while the priors are shown as dotted lines.}
    \label{fig:int_struct_b}
\end{figure*}

\begin{figure*}
    \centering
    \includegraphics[width=\linewidth]{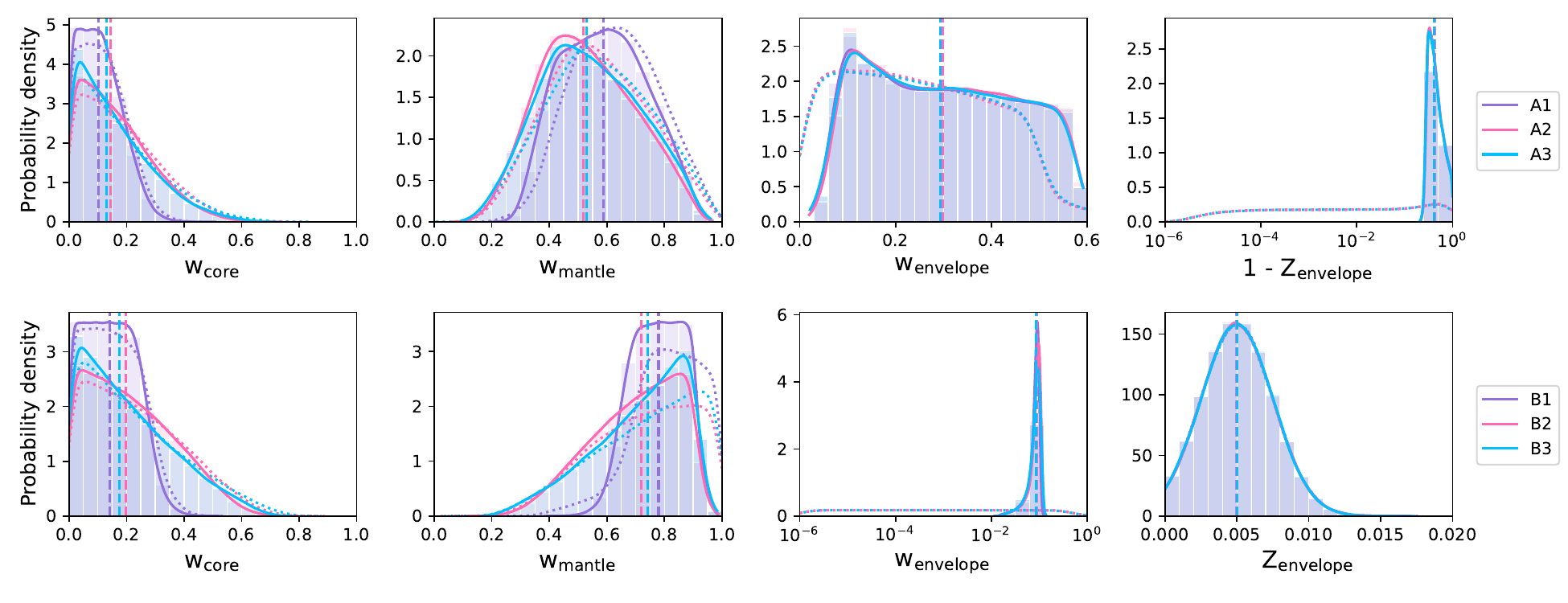}
    \caption{Same as Figure \ref{fig:int_struct_b} but for TOI-1803 c.}
    \label{fig:int_struct_c}
\end{figure*}

In Section \ref{sec:planetary_formation}, we found that fitting the observed radii requires a core mass fractions of a few percent for planet b and around ten percent for planet c. In this section, we apply a more sophisticated modeling framework \texttt{plaNETic}\footnote{\url{https://github.com/joannegger/plaNETic}} \citep{Egger+2024} to the two planets and compare and contrast the results. \texttt{plaNETic} is based on a neural network trained on the internal structure model from BICEPS \citep{Haldemann+2024}. This neural network is then used as the forward model in a full-grid accept-reject sampling algorithm. Each planet is modeled as a three-layered structure of an inner iron core with up to 19\% silicon, a mantle composed of oxidized Si, Mg, Fe, and a volatile layer containing a uniform mixture of H/He and water.

In the case of a multi-planetary system such as TOI-1803, all planets are modeled simultaneously. We ran in total six models with different priors, which influence the resulting posteriors. On the one hand, we use two different prior options for the water content on the planet, one compatible with a formation outside the ice line where water is readily available to be accreted (option A), and one compatible with a formation scenario inside the ice line, assuming water can only be accreted through the accreted gas (option B). On the other hand, we use three different prior options for the planetary Si/Mg/Fe ratios, one where we assume they match the stellar ones \citep[e.g.][]{Thiabaud+2015}, one assuming an iron-enriched scenario \citep[e.g.][]{adibekyan2021} and one where the molar fractions of Si, Mg and Fe are sampled uniformly from a simplex, with an upper limit of 0.75 for Fe. For more details on \texttt{plaNETic} and these chosen priors, we refer the reader to \cite{Egger+2024}.

The resulting posterior distributions of the most important parameters are shown in Figures \ref{fig:int_struct_b} and \ref{fig:int_struct_c}. Further, Tables \ref{tab:internal_structure_results_b} and \ref{tab:internal_structure_results_c} in the appendix summarise the obtained posterior distributions. As expected for sub-Neptunes, we do not properly constrain the core and mantle mass fractions and a majority of the posteriors are very close to the chosen priors. An exception is the mantle mass fraction of planet c, where we find a slight tendency towards lower mantle mass fractions as compared to the priors in question. We further find that a very wide range of envelope mass fractions is possible for the water-enriched case A. These envelopes generally show high metallicity values, with median values of around 80\% for planet b and around 60\% for planet c. For a formation scenario inside the ice line, we find that the planets are expected water-poor envelopes with tightly constrained mass fractions with medians of around 2\% for planet b and around 9\% for planet c.

\section{Characterisation with JWST/NIRSpec}\label{sec:JWST}

\begin{figure}
    \centering
    \includegraphics[width=0.5\textwidth]{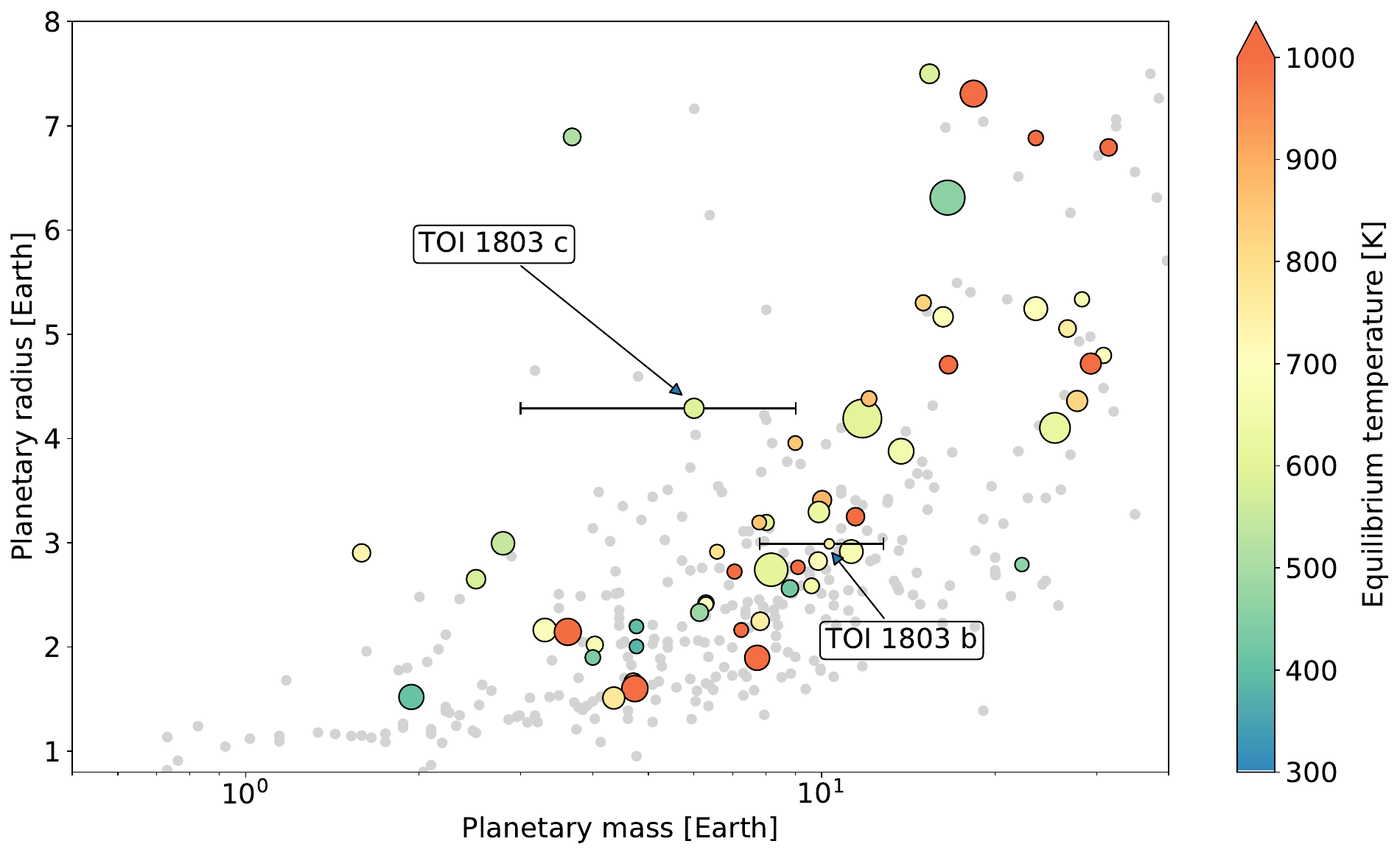}
    \caption{Mass-Radius diagram with all the planetary candidates with $M < 30 M_{\oplus}$ and $R < 8 R_{\oplus}$ in the TEPCat catalogue \citep{southworth2011}. The color bar represents the equilibrium temperature of the planet when the object has TSM > 80, while the others are colored in grey.}
    \label{fig:tsm-plot}
\end{figure}

The atmospheric characterization of the TOI-1803 system could lead to some significant hints about its planetary formation. As shown in Fig. \ref{fig:tsm-plot} planet c is particularly interesting, since it has one of the lowest densities among the other sub-Neptunian exoplanets. The Transmission Spectroscopy Metric (TSM, \citealt{kempton2018}) of the two planets are TSM$_b = 43$ and TSM$_c = 170$, which suggests that the outer planet is particularly suitable for transit spectroscopy.

TOI-1803\,c is indeed an ideal candidate for precise transmission spectroscopy and atmospheric characterization, with a focus on the carbon-to-oxygen (C/O) ratio. This ratio is essential for understanding planetary formation mechanisms and grasping planetary atmosphere composition, shedding light on volatile content and atmospheric chemistry. Within protoplanetary disks, where planets form, the C/O ratio influences the condensation of volatile compounds. A higher C/O ratio favors the formation of carbon-rich species such as carbon monoxide (CO) and methane (CH4), which can impact the composition of planetary atmospheres \citep{tabone2023}.
The availability of carbon and oxygen determines different chemical reactions as well as the stability of molecules of planetary atmospheres. Understanding the C/O ratio helps in predicting the composition and behavior of atmospheric constituents like carbon dioxide (CO$_2$), carbon monoxide (CO), methane (CH$_4$), and water (H$_2$O) \citep{keyte2023}.

To test the feasibility of TOI-1803\,c for atmospheric characterization using JWST, we assumed two different metallicities $Z=Z_{\odot}$ and $Z=10Z_{\odot}$, with $Z_{\odot}$ the solar metallicity. These two configurations represent the primary and secondary atmosphere, respectively. We assumed equilibrium chemistry as a function of temperature and pressure using FastChem \citep{stock2018} and three different C/O ratios, corresponding to a sub-solar ratio of 0.25, solar ratio of 0.5, and a super-solar ratio of 1.0. We used FastChem within TauREx3 \citep{alrefaie2021,waldmann2015a,waldmann2015b} using the \texttt{taurex-fastchem}\footnote{\url{https://pypi.org/project/taurex-fastchem}} plugin. TauREx is a retrieval code that uses a Bayesian approach to infer atmospheric properties from observed data, utilizing a forward model to generate synthetic spectra by solving the radiative transfer equation throughout the atmosphere. We used all the possible gas contributions within FastChem and cross-sections from the ExoMol catalog \citet{tennyson2013,tennyson2020} \footnote{\url{https://www.exomol.com/data/molecules/}}.

\begin{figure*}
    \centering
    \includegraphics[width=0.45\hsize]{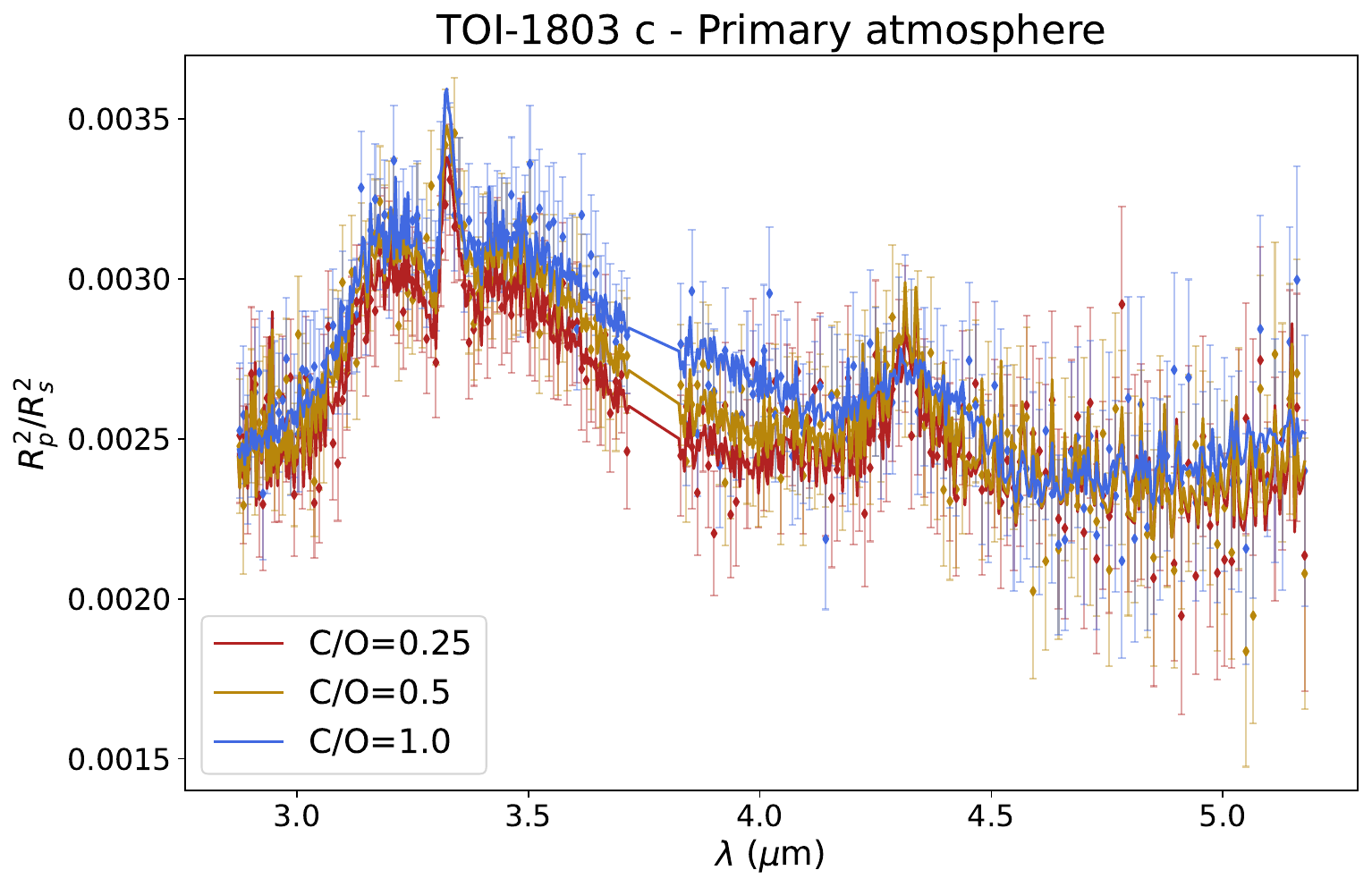}
    \includegraphics[width=0.45\hsize]{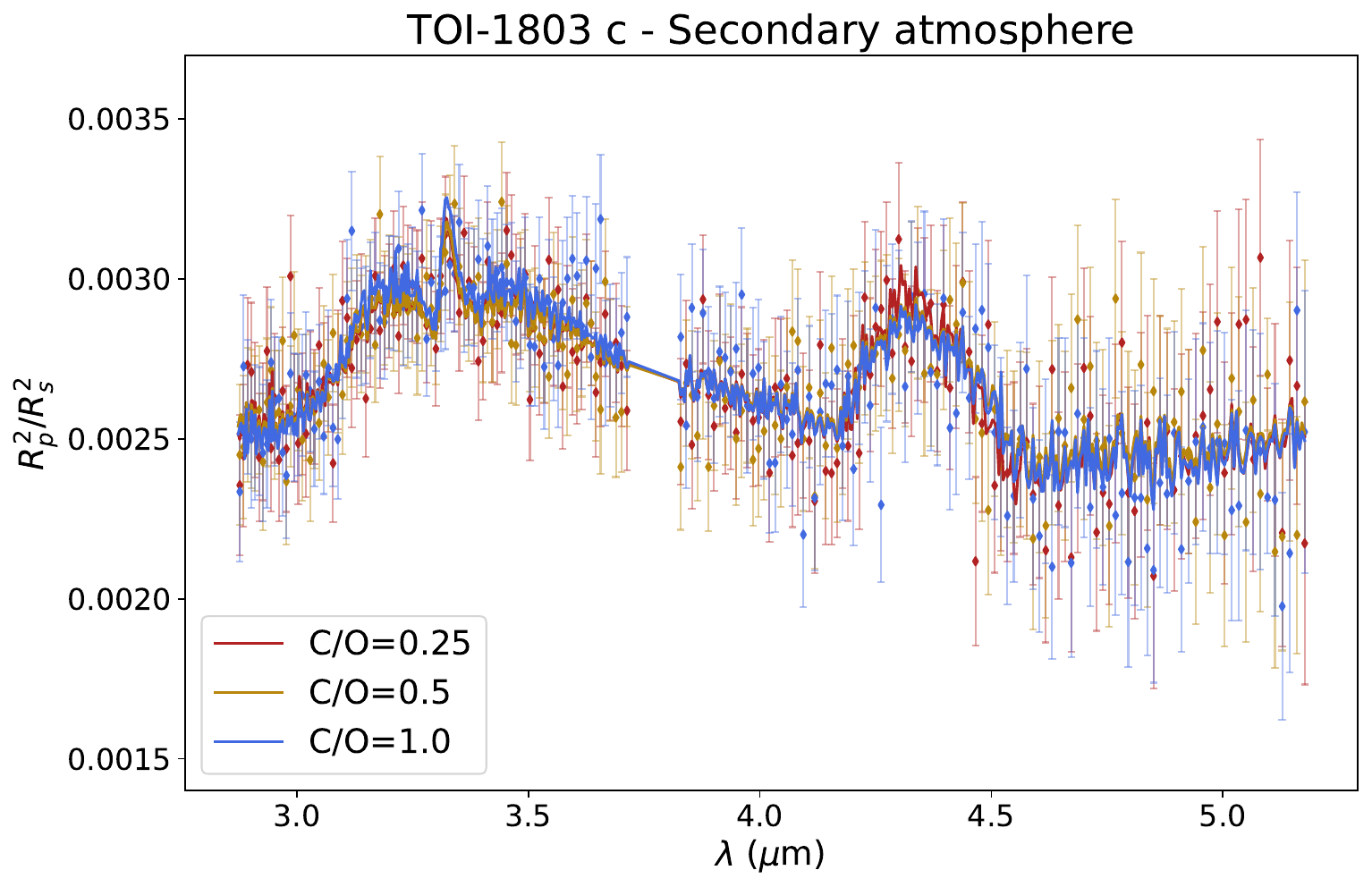}
    \caption{NIRSpec observation simulation using the g395h disperser with F290LP filter (scatter points) and best-fit models from TauREx (lines). The three colors indicate three different C/O configurations: C/O = 0.25 in red, C/O = 0.5 in yellow, and C/O = 1.0 in blue. On the \textbf{left} panel are displayed the spectra for assuming a primary atmosphere scenario, while on the \textbf{right} are ones assuming a secondary atmosphere. See details of simulations (metallicity, observation time, etc.) in main text.}
    \label{fig:retrieval-spectra}
\end{figure*}

After generating the transmission spectra using TauREX+FastChem, we simulated a JWST observation using Pandexo \citep{batalha2017}, a software tool specifically developed for the JWST mission. 
The software allows users to model and simulate various atmospheric scenarios, incorporating factors such as atmospheric composition, temperature profiles, and molecular opacities. We simulated a NIRSpec observation in \texttt{bots} mode, using the \texttt{s1600a1} aperture with \texttt{g395h} disperser, \texttt{sub2048} subarray, \texttt{nrsrapid} read mode, and \texttt{F290LP} filter. We simulated one single transit and an observation \textbf{$1.75 T_{14} = 4.191$ hours} long to ensure a robust baseline coverage. We fixed this instrumental configuration for all three scenarios. In Fig. \ref{fig:retrieval-spectra}, we show the resulting spectra for the different C/O ratios and their best-fit models.

We performed three atmospheric retrievals on the JWST/NIRSpec simulations using a Nested Sampling algorithm with the \texttt{multinest} \citep{feroz2009} library with 1000 live points. We fitted three parameters: the radius of the planet $R_p$, the equilibrium temperature of the atmosphere $T_{\rm eq}$, the metallicity $\log{Z}$, and the C/O ratio.

\begin{figure}[!htbp]
    \centering
    \includegraphics[width=\hsize]{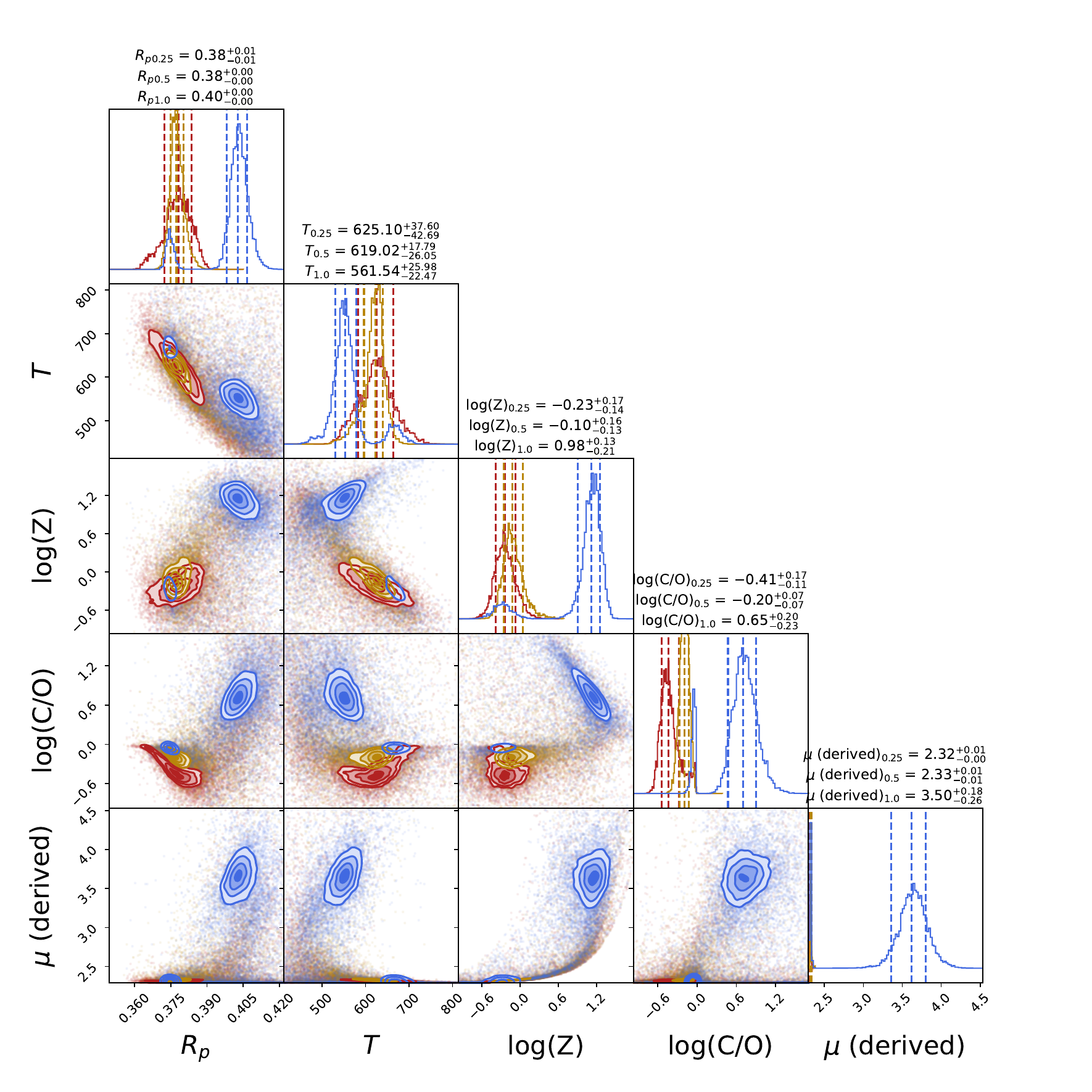}
    \includegraphics[width=\hsize]{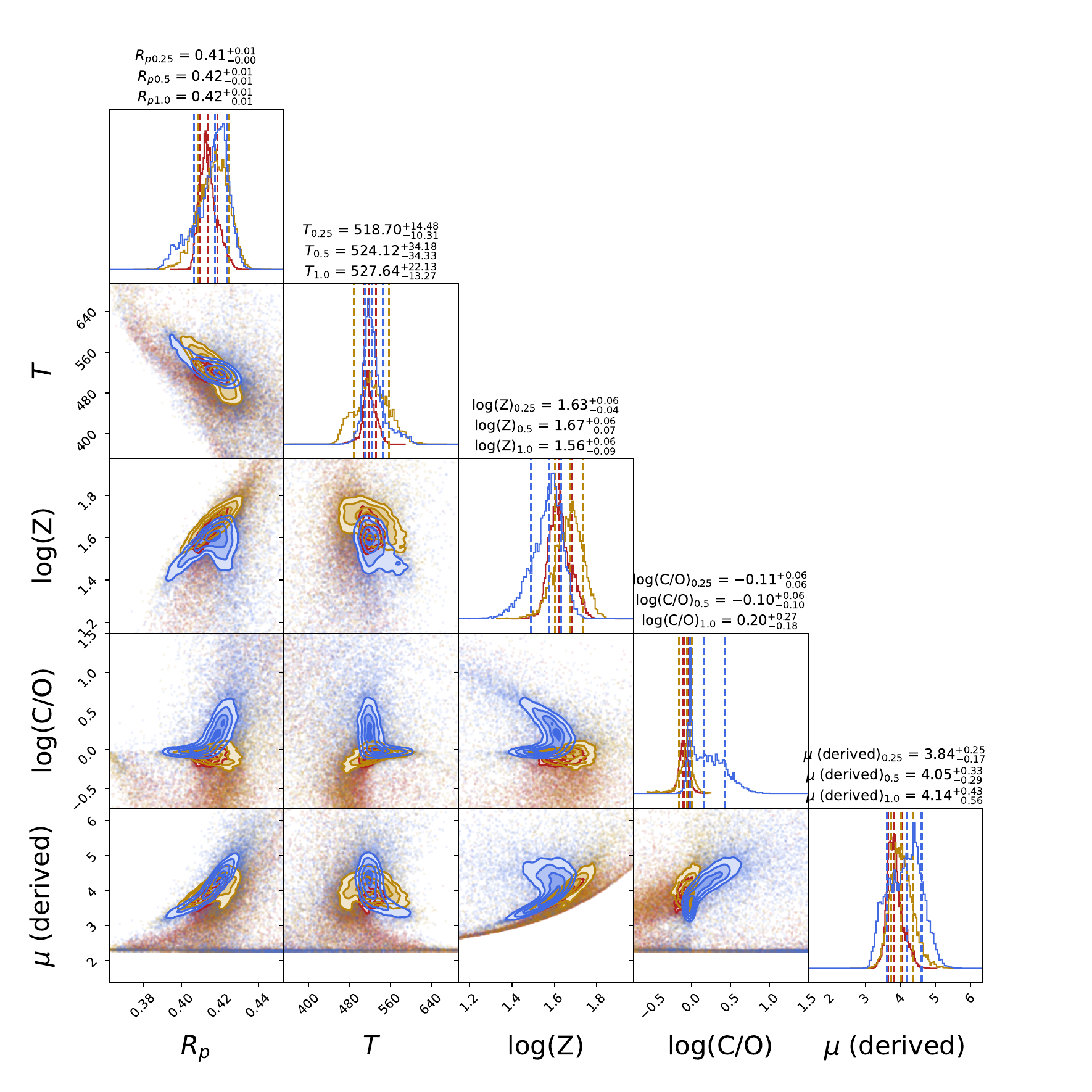}
    \caption{Posterior distributions for the three different C/O assumptions. C/O = 0.25 in red, C/O = 0.5 in yellow, and C/O = 1.0 in blue. On the \textbf{top} are displayed the posterior distributions for the primary atmosphere scenario, while on the \textbf{bottom} are the ones for the secondary atmosphere scenario.}
    \label{fig:retrieval-posteriors}
\end{figure}

Using NIRSpec with the \texttt{g395h} disperser, with the wavelength range 3.82\,$\mu$m - 5.18\,$\mu$m, we can distinguish between a light, primary atmosphere with $Z = Z_{\odot}$ and a heavy, secondary atmosphere with $Z = 10 Z_{\odot}$. Moreover, we can estimate a C/O ratio for both atmospheric assumptions within $\sim 2 \sigma$ error bars (for more details, see Fig. \ref{fig:retrieval-posteriors}). 

The results of atmospheric retrievals confirm and quantify the feasibility of atmospheric characterization using NIRSpec. Furthermore, these results demonstrate that TOI-1803\,c is an excellent candidate for comprehensive atmospheric analysis, to measure the C/O ratio and, therefore, to constrain planet formation theories for this system.

\begin{table}[h]
\caption{Retrieval results for the two atmospheric assumptions.}
\centering
\begin{tabular}{p{1.53cm}| p{1.8cm}| p{1.8cm} |p{1.92cm}}
\hline \hline
\multicolumn{4}{c}{Primary atmosphere} \\
\hline \hline
Parameter & C/O = 0.25 & C/O = 0.5 & C/O = 1.0 \rule{0pt}{2.3ex} \rule[-1ex]{0pt}{0pt} \\
\hline 
& & &\\
$R_p$ $(R_J)$ &  $0.38 \pm 0.01$ & $0.38 \pm 0.01$ & $0.40\pm0.01$ \rule{0pt}{2.3ex} \rule[-1.25ex]{0pt}{0pt} \\

$T_{\rm eq}$ (K) & $625 \pm 40$ & $619 \pm 26$ & $562 \pm 26$ \rule{0pt}{2.3ex} \rule[-2.2ex]{0pt}{0pt} \\

$\log Z$ & $-0.23 \pm 0.17$ & $-0.10 \pm 0.16$ & $0.98 \pm 0.21$ \rule{0pt}{2.3ex} \rule[-2.2ex]{0pt}{0pt} \\

C/O & $0.39 \pm 0.15$ & $0.63 \pm 0.10$ & $4.5 \pm 2.4$ \rule{0pt}{2.3ex} \rule[-2.7ex]{0pt}{0pt} \\

$\mu$ (derived) & $2.32 \pm 0.01$ &  $2.33 \pm 0.01$ & $3.50_{-0.26}^{+0.18}$ \rule{0pt}{2.3ex} \rule[-1.5ex]{0pt}{0pt} \\
\hline \hline
\multicolumn{4}{c}{Secondary atmosphere} \\
\hline \hline
Parameter & C/O = 0.25 & C/O = 0.5 & C/O = 1.0 \rule{0pt}{2.3ex} \rule[-1ex]{0pt}{0pt} \\
\hline 
& & &\\
$R_p$ $(R_J)$ &  $0.41 \pm 0.01$ & $0.42 \pm 0.01$ & $0.42\pm0.01$ \rule{0pt}{2.3ex} \rule[-1.25ex]{0pt}{0pt} \\

$T_{\rm eq}$ (K) & $519 \pm 15$ & $524 \pm 34$ & $528 \pm 22$ \rule{0pt}{2.3ex} \rule[-2.2ex]{0pt}{0pt} \\

$\log Z$ & $1.63 \pm 0.06$ & $1.67 \pm 0.07$ & $1.56 \pm 0.09$ \rule{0pt}{2.3ex} \rule[-2.2ex]{0pt}{0pt} \\

C/O & $0.78 \pm 0.11$ & $0.79 \pm 0.18$ & $1.58 \pm 0.99$ \rule{0pt}{2.3ex} \rule[-2.7ex]{0pt}{0pt} \\

$\mu$ (derived) & $3.84_{-0.17}^{+0.25}$ &  $4.05_{-0.29}^{+0.33}$ & $4.14_{-0.56}^{+0.43}$ \rule{0pt}{2.3ex} \rule[-1.5ex]{0pt}{0pt} \\
\hline
\end{tabular}
\label{tab:retrieval-results}
\end{table}

\section{Discussion and Conclusion}
\label{sec:discussion_conclusion}

In this work, we showed a comprehensive study of the TOI-1803 system, describing the properties of two mini Neptunes with a period ratio slightly larger than the ratio corresponding to a 2:1 resonance commensurability. Moreover, it is worth noting that the low density of TOI-1803 c, one of the least dense mini Neptunes known, makes it an interesting target for atmospheric studies (Fig. \ref{fig:mr_relationship}). As such, the extended atmosphere of this planet could be useful for transmission spectroscopy allowing it to discriminate a primary atmosphere from a secondary atmosphere and, finally give information on its C/O ratio.

By combining photometric observations from TESS and CHEOPS with radial velocity measurements from HARPS-N, we managed to infer the physical and orbital parameters of both planets. The results revealed that the two mini Neptunes around TOI-1803 have orbital periods of 6.29 and 12.89 days, and radii of $2.99 \pm 0.08$ and $4.29 \pm 0.08$ $R_{\oplus}$ respectively. Given the previous joint-analysis, the preferred model with circular orbits estimates the masses of the planets $M_b = 10.3 \pm 2.5$ $M_{\oplus}$ and $M_c = 6.0 \pm 3.0$ $M_{\oplus}$.  

Commonly, systems in a 2:1 resonance are thought to be stable and long-lived, as the planets are in a specific orbital configuration that allows them to maintain consistent and stable orbital parameters. If there is gravitational interaction between planets within a multi-planet system, TTVs become apparent in the observations. Importantly, this effect is most pronounced if the planets are very close to each other or reside in MMR. This phenomenon emphasizes complex dynamical interactions in planetary systems and provides essential information about the masses, orbits, and evolution of planet systems \citep{agol2005,holman2005,steffen2007}. 
No clear TTVs are detected in either TESS or high-precision CHEOPS light curves, but it was clearly observed in the O-C diagram of the planet -c after adding the ground-based light curves (see Sect. \ref{sec:ground_based}). Additional transit observations, planned at specific TTV phase, will help examine this system more in-depth.

The TOI-1803 system investigation has provided important insights into the nature and dynamics of mini Neptunes. We demonstrated JWST observations can characterize the atmospheric type of the external planet, and would give robust constraints to formation and evolution theories. 
Specifically, TOI-1803 c is an inflated mini Neptune with unique opportunities to study its atmospheric properties. Further details would be unveiled by future examinations with more sophisticated instruments (using both space telescopes such as JWST or ground-based facilities). Such findings highlight the complexity and diversity of exoplanetary systems and emphasize the need for continued exploration of systems like TOI-1803.

\begin{acknowledgements}
CHEOPS is an ESA mission in partnership with Switzerland with important contributions to the payload and the ground segment from Austria, Belgium, France, Germany, Hungary, Italy, Portugal, Spain, Sweden, and the United Kingdom. The CHEOPS Consortium would like to gratefully acknowledge the support received by all the agencies, offices, universities, and industries involved. Their flexibility and willingness to explore new approaches were essential to the success of this mission. CHEOPS data analysed in this article will be made available in the CHEOPS mission archive (\url{https://cheops.unige.ch/archive_browser/}). 
LBo, GBr, VNa, IPa, GPi, RRa, GSc, VSi, and TZi acknowledge support from CHEOPS ASI-INAF agreement n. 2019-29-HH.0. 
TZi acknowledges NVIDIA Academic Hardware Grant Program for the use of the Titan V GPU card and the Italian MUR Departments of Excellence grant 2023-2027 “Quantum Frontiers”. 
C.B. acknowledges support from the Swiss Space Office through the ESA PRODEX program. 
This work has been carried out within the framework of the NCCR PlanetS supported by the Swiss National Science Foundation under grants 51NF40\_182901 and 51NF40\_205606. 
ML acknowledges support of the Swiss National Science Foundation under grant number PCEFP2\_194576. 
A. S. acknowledges support from the Swiss Space Office through the ESA PRODEX program. 
YAl acknowledges support from the Swiss National Science Foundation (SNSF) under grant 200020\_192038. 
The Belgian participation to CHEOPS has been supported by the Belgian Federal Science Policy Office (BELSPO) in the framework of the PRODEX Program, and by the University of Li\'{e}ge through an ARC grant for Concerted Research Actions financed by the Wallonia-Brussels Federation. 
L.D. thanks the Belgian Federal Science Policy Office (BELSPO) for the provision of financial support in the framework of the PRODEX Programme of the European Space Agency (ESA) under contract number 4000142531. 
ABr was supported by the SNSA. 
ACMC acknowledges support from the FCT, Portugal, through the CFisUC projects UIDB/04564/2020 and UIDP/04564/2020, with DOI identifiers 10.54499/UIDB/04564/2020 and 10.54499/UIDP/04564/2020, respectively. 
This work has been supported by the PRIN-INAF 2019 “Planetary systems at young ages (PLATEA)”. 
We acknowledge financial support from the Agencia Estatal de Investigaci\`{o}n of the Ministerio de Ciencia e Innovaci\`{o}n MCIN/AEI/10.13039/501100011033 and the ERDF “A way of making Europe” through projects PID2019-107061GB-C61, PID2019-107061GB-C66, PID2021-125627OB-C31, and PID2021-125627OB-C32, from the Centre of Excellence “Severo Ochoa” award to the Instituto de Astrof\`{i}sica de Canarias (CEX2019-000920-S), from the Centre of Excellence “Mar\`{i}a de Maeztu” award to the Institut de Ci\'{e}ncies de l’Espai (CEX2020-001058-M), and from the Generalitat de Catalunya/CERCA programme. 
We acknowledge financial support from the Agencia Estatal de Investigaci\`{o}n of the Ministerio de Ciencia e Innovaci\`{o}n MCIN/AEI/10.13039/501100011033 and the ERDF “A way of making Europe” through projects PID2019-107061GB-C61, PID2019-107061GB-C66, PID2021-125627OB-C31, and PID2021-125627OB-C32, from the Centre of Excellence “Severo Ochoa'' award to the Instituto de Astrof\`{i}sica de Canarias (CEX2019-000920-S), from the Centre of Excellence “Mar\`{i}a de Maeztu” award to the Institut de Ci\'{e}ncies de l’Espai (CEX2020-001058-M), and from the Generalitat de Catalunya/CERCA programme. 
S.C.C.B. acknowledges support from FCT through FCT contracts nr. IF/01312/2014/CP1215/CT0004. 
ACC acknowledges support from STFC consolidated grant number ST/V000861/1, and UKSA grant number ST/X002217/1. 
P.E.C. is funded by the Austrian Science Fund (FWF) Erwin Schroedinger Fellowship, program J4595-N. 
This project was supported by the CNES. 
This work was supported by FCT - Funda\c{c}\~{a}o para a Ci\^{e}ncia e a Tecnologia through national funds and by FEDER through COMPETE2020 through the research grants UIDB/04434/2020, UIDP/04434/2020, 2022.06962.PTDC. 
O.D.S.D. is supported in the form of work contract (DL 57/2016/CP1364/CT0004) funded by national funds through FCT. 
B.-O. D. acknowledges support from the Swiss State Secretariat for Education, Research and Innovation (SERI) under contract number MB22.00046. 
This project has received funding from the Swiss National Science Foundation for project 200021\_200726. It has also been carried out within the framework of the National Centre of Competence in Research PlanetS supported by the Swiss National Science Foundation under grant 51NF40\_205606. The authors acknowledge the financial support of the SNSF. 
MF and CMP gratefully acknowledge the support of the Swedish National Space Agency (DNR 65/19, 174/18). 
D. D. acknowledges support from the TESS Guest Investigator Program grant 80NSSC23K0769.
DG gratefully acknowledges financial support from the CRT foundation under Grant No. 2018.2323 “Gaseousor rocky? Unveiling the nature of small worlds”. 
DP acknowledges the support from the Istituto Nazionale di Oceanografia e Geofisica Sperimentale (OGS) and CINECA through the program “HPC-TRES (High Performance Computing Training and Research for Earth Sciences)” award number 2022-05 as well as the support of the ASI-INAF agreement n2021-5-HH.1-2022. DT acknowledges the support from the European Research Council via the Horizon 2020 Framework Programme ERC Synergy “ECOGAL” Project GA-855130.
DRC acknowledges partial support from NASA Grant 18-2XRP18\_2-0007. This research has made use of the Exoplanet Follow-up Observation Program (ExoFOP; DOI: 10.26134/ExoFOP5) website, which is operated by the California Institute of Technology, under contract with the National Aeronautics and Space Administration under the Exoplanet Exploration Program. Some of the data presented herein were obtained at Keck Observatory, which is a private 501(c)3 non-profit organization operated as a scientific partnership among the California Institute of Technology, the University of California, and the National Aeronautics and Space Administration. The Observatory was made possible by the generous financial support of the W. M. Keck Foundation. The authors wish to recognize and acknowledge the very significant cultural role and reverence that the summit of Maunakea has always had within the Native Hawaiian community. We are most fortunate to have the opportunity to conduct observations from this mountain.
M.G. is an F.R.S.-FNRS Senior Research Associate. 
MNG is the ESA CHEOPS Project Scientist and Mission Representative, and as such also responsible for the Guest Observers (GO) Programme. MNG does not relay proprietary information between the GO and Guaranteed Time Observation (GTO) Programmes, and does not decide on the definition and target selection of the GTO Programme. 
MP acknowledge support from the European Union – NextGenerationEU (PRIN MUR 2022 20229R43BH) and the "Programma di Ricerca Fondamentale INAF 2023.
CHe acknowledges support from the European Union H2020-MSCA-ITN-2019 under Grant Agreement no. 860470 (CHAMELEON). 
KGI is the ESA CHEOPS Project Scientist and is responsible for the ESA CHEOPS Guest Observers Programme. She does not participate in, or contribute to, the definition of the Guaranteed Time Programme of the CHEOPS mission through which observations described in this paper have been taken, nor to any aspect of target selection for the programme. 
K.W.F.L. was supported by Deutsche Forschungsgemeinschaft grants RA714/14-1 within the DFG Schwerpunkt SPP 1992, Exploring the Diversity of Extrasolar Planets. 
This work was granted access to the HPC resources of MesoPSL financed by the Region Ile de France and the project Equip@Meso (reference ANR-10-EQPX-29-01) of the programme Investissements d'Avenir supervised by the Agence Nationale pour la Recherche. 
PM acknowledges support from STFC research grant number ST/R000638/1. 
This work was also partially supported by a grant from the Simons Foundation (PI Queloz, grant number 327127). 
NCSa acknowledges funding by the European Union (ERC, FIERCE, 101052347). Views and opinions expressed are however those of the author(s) only and do not necessarily reflect those of the European Union or the European Research Council. Neither the European Union nor the granting authority can be held responsible for them. 
S.G.S. acknowledge support from FCT through FCT contract nr. CEECIND/00826/2018 and POPH/FSE (EC). 
The Portuguese team thanks the Portuguese Space Agency for the provision of financial support in the framework of the PRODEX Programme of the European Space Agency (ESA) under contract number 4000142255. 
GyMSz acknowledges the support of the Hungarian National Research, Development and Innovation Office (NKFIH) grant K-125015, a a PRODEX Experiment Agreement No. 4000137122, the Lend\`{u}let LP2018-7/2021 grant of the Hungarian Academy of Science and the support of the city of Szombathely. 
V.V.G. is an F.R.S-FNRS Research Associate. 
JV acknowledges support from the Swiss National Science Foundation (SNSF) under grant PZ00P2\_208945. 
NAW acknowledges UKSA grant ST/R004838/1. 
TWi acknowledges support from the UKSA and the University of Warwick. 
KAC acknowledges support from the TESS mission via subaward s3449 from MIT.
Funding for the TESS mission is provided by NASA’s Science Mission Directorate. We acknowledge the use of public TESS data from pipelines at the TESS Science Office and at the TESS Science Processing Operations Center. Resources supporting this work were provided by the NASA High-End Computing (HEC) Program through the NASA Advanced Supercomputing (NAS) Division at Ames Research Center for the production of the SPOC data products. TESS data presented in this paper were obtained from the Mikulski Archive for Space Telescopes (MAST) at the Space Telescope Science Institute. This research has made use of the Exoplanet Follow-up Observation Program (ExoFOP; DOI: 10.26134/ExoFOP5) website, which is operated by the California Institute of Technology, under contract with the National Aeronautics and Space Administration under the Exoplanet Exploration Program.

Co-funded by the European Union (ERC, FIERCE, 101052347). Views and opinions expressed are however those of the author(s) only and do not necessarily reflect those of the European Union or the European Research Council. Neither the European Union nor the granting authority can be held responsible for them. This work was supported by FCT - Fundação para a Ciência e a Tecnologia through national funds and by FEDER through COMPETE2020 - Programa Operacional Competitividade e Internacionalização by these grants: UIDB/04434/2020; UIDP/04434/2020.

This work makes use of observations from the LCOGT network. Part of the LCOGT telescope time was granted by NOIRLab through the Mid-Scale Innovations Program (MSIP). MSIP is funded by NSF. This is paper is based on observations made with the MuSCAT3 instrument, developed by the Astrobiology Center and under financial supports by JSPS KAKENHI (JP18H05439) and JST PRESTO (JPMJPR1775), at Faulkes Telescope North on Maui, HI, operated by the Las Cumbres Observatory.

This work is partly supported by JSPS KAKENHI Grant Number JP24H00017 and JSPS Bilateral Program Number JPJSBP120249910.

This work made use of \texttt{tpfplotter} by J. Lillo-Box (publicly available in \url{www.github.com/jlillo/tpfplotter}), which also made use of the python packages \texttt{astropy}, \texttt{lightkurve}, \texttt{matplotlib} and \texttt{numpy}.

\end{acknowledgements}

\bibliographystyle{aa}
\bibliography{main}

\appendix

\section{Additional parameters}
The amplitude of the GP term (denoted by the underscore $c$) and its first derivative (denoted by underscore $b$) for the RV (coefficients $V_c$ and $V_r$), BIS (coefficients $B_c$ and $B_r$) and $S_HK$-index (coefficients $L_c$ and $L_r$) are reported in table \ref{tab:gp_ampl}. 
As pointed out by the referee, the posterior of $B_r$ is slightly pushing towards the lower boundary. We repeated the analysis after widening the boundaries of all the hyperparameters and the new results were well within $1\sigma$ from the values reported in Table \ref{tab:obspars}, with the exception of $B_r$ now being $-21.3_{-3.6}^{+3.3}$ ms$^{-1}$ ($0.8 \sigma$). For the sake of reproducibility of all the subsequent analyses in the paper relying on the values of \ref{tab:obspars}, we left the values of the original analysis.

\begin{table}[!htpb]
\caption[]{Multidimensional GP parameters \citep{rajpaul2015}.}
\label{tab:gp_ampl}
\centering
\resizebox{0.45\textwidth}{!}{\begin{tabular}{cccc}
    \toprule
    \toprule
    Parameter & Unit & Prior & Value\\
    \midrule
   RV $V_c$ & m s$^{-1}$ & $\mathcal{U}$(-20, +20) & $-0.5 \pm 1.1$ \rule{0pt}{2.2ex} \rule[-0.8ex]{0pt}{0pt}\\ 
   RV $V_r$ & m s$^{-1}$ & $\mathcal{U}$(0, +20) & 15.0$^{+2.1}_{-1.9}$ \rule{0pt}{2.2ex} \rule[-0.8ex]{0pt}{0pt}\\ 
   BIS $B_c$ & m s$^{-1}$ & $\mathcal{U}$(-20, +20) & $-4.2 \pm 1.7$ \rule{0pt}{2.2ex} \rule[-0.8ex]{0pt}{0pt}\\ 
   BIS $B_r$ & m s$^{-1}$ & $\mathcal{U}$(-20, +20) & -18.2$^{+1.9}_{-1.3}$ \rule{0pt}{2.2ex} \rule[-0.8ex]{0pt}{0pt}\\ 
   $L_c$ ($S_{HK}$) &  & $\mathcal{U}$(-20, +20) &  $-0.0168_{-0.0027}^{0.0024}$ \rule{0pt}{2.2ex} \rule[-0.8ex]{0pt}{0pt}\\ 
   $L_r$ ($S_{HK}$) &  & $\mathcal{U}$(-20, +20) & $-0.0050_{-0.0047}^{0.0049}$ \rule{0pt}{2.2ex} \rule[-0.9ex]{0pt}{0pt}\\ 
    \bottomrule
\end{tabular}}
\end{table}

\begin{table}[!htpb]
\caption[]{Offset and jitter for all the data sets.}
\label{tab:off_jit}
\centering
\resizebox{0.45\textwidth}{!}{\begin{tabular}{ccc}
    \toprule
    \toprule
    parameter & Prior & Value\\
    \midrule
    \multicolumn{3}{c}{CHEOPS} \\
    \cmidrule{1-3}
    jitter & $\mathcal{U}$(0.000007,  0.208507) & $0.00084\pm0.00002$ \\
    \cmidrule{1-3}
    \multicolumn{3}{c}{TESS S22} \\
    \cmidrule{1-3}
    jitter & $\mathcal{U}$(0.000024,  0.245700) & $0.00011_{-0.00006}^{+0.00010}$ \\
    \cmidrule{1-3}
    \multicolumn{3}{c}{TESS S49} \\
    \cmidrule{1-3}
    jitter & $\mathcal{U}$(0.000019,  0.196200) & $0.00012_{-0.00007}^{+0.00011}$ \\
    \cmidrule{1-3}
    \multicolumn{3}{c}{RV} \\
    \cmidrule{1-3}
    offset & $\mathcal{U}$(-8027, 12032) & $1995_{-0.81}^{+0.78}$ \\  
    jitter & $\mathcal{U}$(0.017,  1153) & $5.17_{-0.80}^{+0.82}$ \\
    \cmidrule{1-3}
    \multicolumn{3}{c}{BIS} \\
    \cmidrule{1-3}
    offset & $\mathcal{U}$(-10011, 10107) & $30 \pm 1.8$ \\  
    jitter & $\mathcal{U}$(0.017,  1153) & $10.2 \pm 1.0$ \\
    \cmidrule{1-3}
    \multicolumn{3}{c}{S$_{\mathrm{HK}}$-index} \\
    \cmidrule{1-3}
    offset & $\mathcal{U}$(-10000, 10000) & $0.586 \pm 0.006$ \\  
    jitter & $\mathcal{U}$(0.0001, 11.3390) & $0.0074_{-0.0036}^{+0.0033}$ \\
    \bottomrule
\end{tabular}}
\end{table}

\section{Combined fit using e>0}
In the following, we show the results from PyORBIT joint-fit equivalent to the results shown in Tab \ref{tab:obspars} but leaving the orbital eccentricity of both planets as a free parameter.
\begin{table*}[!htbp]
\caption{Priors and results for the modeling of planets b and c from the analysis of the photometric and radial velocities time series. Eccentric orbits.} 
\label{tab:obspars_ecc}
\centering          
\resizebox{0.63\textwidth}{!}{\begin{tabular}{l c c c}     
\hline\hline     
 \multicolumn{4}{c}{Combined Radial Velocities - Multidimensional GP - Photometric fit} \rule{0pt}{2ex} \rule[-0.9ex]{0pt}{0pt} \\ 
\hline    
Parameter & Unit & Prior & Value \rule{0pt}{2.2ex} \rule[-0.9ex]{0pt}{0pt}\\ 
\hline
\multicolumn{4}{c}{Star} \rule{0pt}{2.2ex} \rule[-0.9ex]{0pt}{0pt}\\ 
\hline  
   Stellar density ($\rho_{\star}$) & $\rho_{\sun}$ & $\mathcal{N}$(2.00, 0.13) & 2.01$\pm$0.13 \rule{0pt}{2.2ex} \rule[-1.2ex]{0pt}{0pt}\\
   CHEOPS Limb Darkening ($q_{1, CHEOPS}$) &    & $\mathcal{U}$(0, 1) & $0.487_{-0.078}^{+0.085}$ \\
   CHEOPS Limb Darkening ($q_{2, CHEOPS}$) &    & $\mathcal{U}$(0, 1) & $0.463_{-0.034}^{+0.040}$ \\
   TESS Limb Darkening ($q_{1, TESS}$) &    & $\mathcal{U}$(0, 1) & $0.48_{-0.09}^{+0.10}$ \\
   TESS Limb Darkening ($q_{2, TESS}$) &    & $\mathcal{U}$(0, 1) & $0.38_{-0.034}^{+0.041}$ \\
\hline
\multicolumn{4}{c}{Planet b} \rule{0pt}{2.2ex} \rule[-0.9ex]{0pt}{0pt}\\ 
\hline    
Parameter & Unit & Prior & Value \rule{0pt}{2.2ex} \rule[-0.9ex]{0pt}{0pt}\\ 
\hline 
    Orbital period ($P_{\rm b}$) & days & $\mathcal{U}$(6.2, 6.4) & 6.293264$\pm$0.000034 \\
    Central time of transit ($T_{\rm 0,b}$) & BTJD$^a$ & $\mathcal{U}$(1904,1905) & 1904.6262$\pm$0.0035 \\
    Impact parameter ($b$) &  & $\mathcal{U}$(0, 1) & 0.31$^{+0.22}_{-0.21}$ \\
    $R_p/R_*$ & & $\mathcal{U}$(0.00001, 0.5) & 0.0384$\pm$0.0016\\   
    Planetary Radius ($R_p$) & $R_{\mathrm{\oplus}}$ & ... & $3.00\pm0.12$ \\
    $a/R_{\star}$ &  & ... & $18.1 \pm 0.4$\\
    Semi-major axis $a$ & AU  & ... & $0.060 \pm 0.001$ \\
    Radial Velocity semi-amplitude (K) & m/s & $\mathcal{U}$(0, 10) & 4.5$\pm$1.1 \\ 
    Inclination ($i$) & deg & ... & $88.99 \pm 0.66$ \\
    $\sqrt{e}\sin\omega$ &  & $\mathcal{U}$(-1, +1) & $0.01_{-0.27}^{+0.18}$ \\
    $\sqrt{e}\cos\omega$ &  & $\mathcal{U}$(-1, +1) & $0.23_{-0.24}^{+0.16}$ \\
    Eccentricity ($e$) &  & ... & $0.105_{-0.071}^{+0.11}$ \\
    Argument of periastron ($\omega$) &  & ... & $5_{-61}^{+75}$ \\
    Transit duration ($T_{14}$) & days & ... & $0.109 \pm 0.010$  \\
    Planetary mass ($M_p$) & $M_{\mathrm{\oplus}}$ & ... & $10.6 \pm 2.6$ \\
    Planetary density ($\rho_p$) & $\rho_{\mathrm{\oplus}}$ & ... & $0.39 \pm 0.11$ \\
    Instellation ($F_i$) & W m$^{-2}$ & ... & (8.4$\pm$0.6)$\cdot 10^4$\\
    Equilibrium Temperature ($T_{\rm eq}$) & K$^b$ & ... & $714 \pm 12$ \\
\hline
\multicolumn{4}{c}{Planet c} \rule{0pt}{2.2ex} \rule[-0.9ex]{0pt}{0pt}\\ 
\hline    
Parameter & Unit & Prior & Value \rule{0pt}{2.2ex} \rule[-0.9ex]{0pt}{0pt}\\ 
\hline 
   Orbital period ($P_{\rm c}$) & days & $\mathcal{U}$(12.8, 13.0) & 12.885776$\pm$0.000036\\
   Central time of transit ($T_{\rm 0,c}$) & BTJD$^a$ & $\mathcal{U}$(1911,1912) & 1911.6748$\pm$0.0019 \\
   Impact parameter ($b$) &  & $\mathcal{U}$(0, 1) & 0.740$^{+0.089}_{-0.27}$\\
   $R_p/R_*$ & & $\mathcal{U}$(0.00001, 0.5) & $0.0537_{-0.0039}^{+0.0031}$\\
   Planetary Radius ($R_p$) & $R_{\mathrm{\oplus}}$ & ... & $4.19_{-0.30}^{+0.25}$ \\ 
   $a/R_{\star}$ &  & ... &  $29.19 \pm 0.6$\\
   Semi-major axis $a$ & AU & ... & $0.097 \pm 0.002$\\
   Radial Velocity semi-amplitude (K) & m/s & $\mathcal{U}$(0, 10) & 2.0$^{+1.0}_{-1.1}$ \\ 
   Inclination ($i$) & deg & ... & $88.48_{-0.11}^{+0.16}$ \\
   $\sqrt{e}\sin\omega$ &  & $\mathcal{U}$(-1, +1) & $0.16_{-0.44}^{+0.36}$ \\
   $\sqrt{e}\cos\omega$ &  & $\mathcal{U}$(-1, +1) & $0.15_{-0.38}^{+0.29}$ \\
   Eccentricity ($e$) &  & ... & $0.24_{-0.16}^{+0.17}$ \\
   Argument of periastron ($\omega$) &  & ... & $40_{-101}^{+74}$ \\
   Transit duration ($T_{14}$) & days & ... & $0.105_{-0.013}^{0.027}$ \\
   Planetary mass ($M_p$) & $M_{\mathrm{\oplus}}$ & ... & $5.7 \pm 3.0$ \\
   Planetary density ($\rho_p$) & $\rho_{\mathrm{\oplus}}$ & ... & $0.08 \pm 0.04$ \\
   Instellation ($F_i$) & W m$^{-2}$ & ... & (3.22$\pm$0.23)$\cdot 10^4$\\
   Equilibrium Temperature ($T_{\rm eq}$) & K$^b$ & ... & $561 \pm 10$ \\
\hline
\multicolumn{4}{c}{Activity}\\ 
\hline  
 Stellar rotation period ($P_{\rm rot}$) & days & $\mathcal{U}$(12.0, 15.0) & 13.657$^{+0.028}_{-0.030}$ \\
 Decay time ($P_{\rm dec}$) & days & $\mathcal{U}$(10, 2000) & 122$^{+29}_{-26}$ \\
 Coherence scale (w) &   & $\mathcal{N}$(0.35, 0.035) & 0.400$^{+0.026}_{-0.025}$ \\
\bottomrule
\end{tabular}}
\tablefoot{\tablefoottext{a}{TESS Barycentric Julian Date ($\mathrm{BJD_{TDB}}$ - 2457000).} \tablefoottext{b}{Computed as $T_{\rm eq} = T_\star \left(\frac{R_{\star}}{2a} \right)^{1/2}\left[f (1-A_b) \right]^{1/4}$, assuming a Bond albedo $A_b = 0.3$ and $f = 1$.}}
\end{table*}

\section{Mass-Radius relationship}

\begin{figure}[!htbp]
    \centering
    \includegraphics[width=0.45\textwidth]{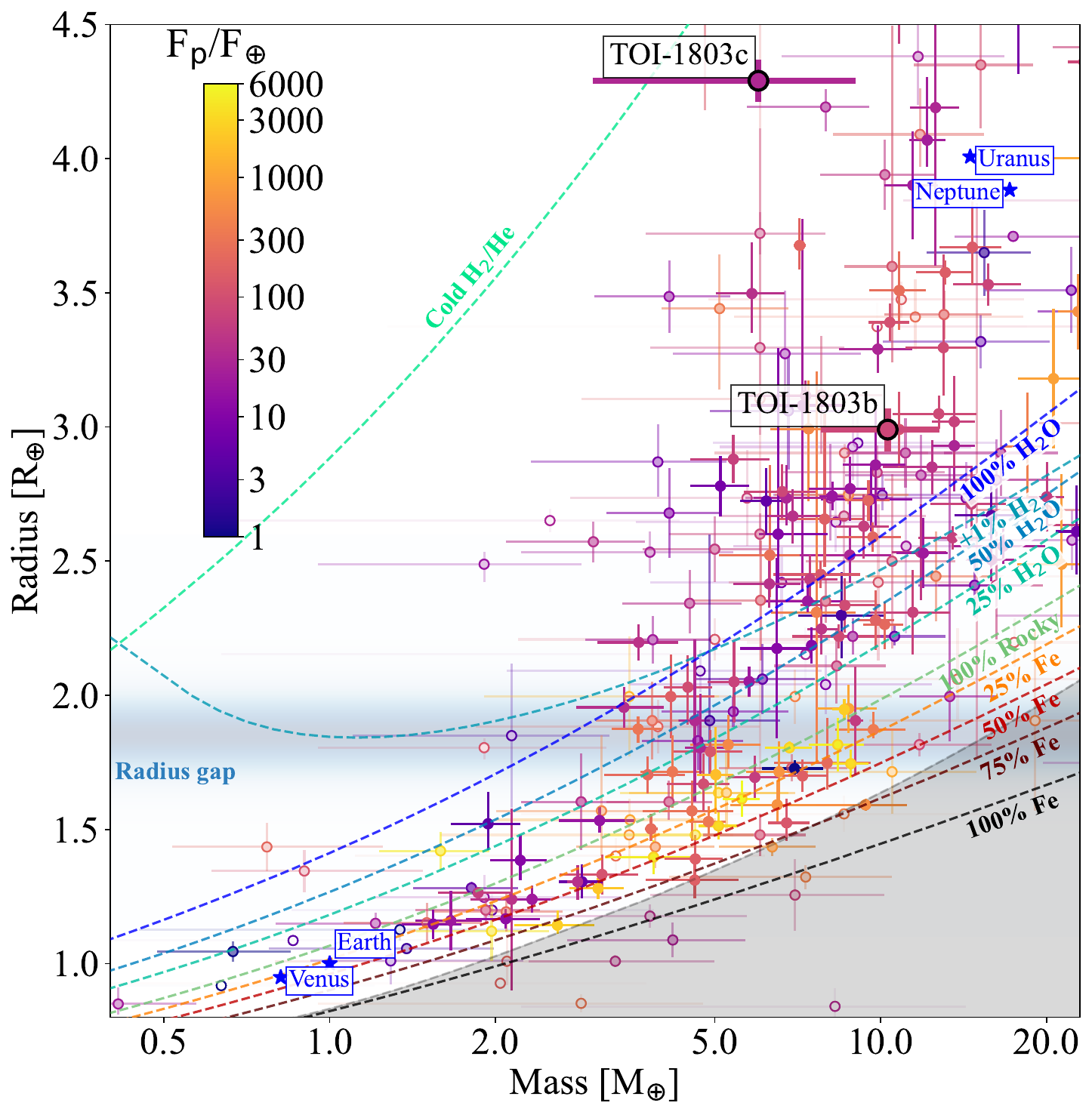}
    \caption{Mass-Radius diagram for all the known exoplanets with a radius $R < 4.5 R_{\oplus}$. We marked with labeled boxes the planets around TOI-1803. The composition models have been computed from \citet{zeng2016}.}
    \label{fig:mr_relationship}
\end{figure}

\begin{table*}
\renewcommand{\arraystretch}{1.5}
\caption{Results of the internal structure modelling for TOI-1803 b. w denotes mass fractions of the denoted layer with respect to the total planetary mass, while x means the molar fraction of the given element in the layer in question. Z$_\textrm{envelope}$ is the mass fraction of water in the envelope.}
\centering
\begin{tabular}{r|ccc|ccc}
\hline \hline
Water prior &              \multicolumn{3}{c|}{Formation outside iceline (water-rich)} & \multicolumn{3}{c}{Formation inside iceline (water-poor)} \\
Si/Mg/Fe prior &           Stellar (A1) &       Iron-enriched (A2) &      Free (A3) &
                           Stellar (B1) &       Iron-enriched (B2) &      Free (B3) \\
\hline
w$_\textrm{core}$ [\%] &        $10.7_{-7.2}^{+8.4}$ &    $14.8_{-10.4}^{+15.2}$ &    $12.3_{-9.0}^{+16.2}$ &
                           $15.0_{-10.2}^{+10.5}$ &    $20.3_{-14.3}^{+19.8}$ &    $17.7_{-12.9}^{+21.5}$ \\
w$_\textrm{mantle}$ [\%] &      $59.8_{-13.9}^{+16.5}$ &    $53.9_{-15.3}^{+19.1}$ &    $55.7_{-15.9}^{+19.3}$ &
                           $83.2_{-10.5}^{+10.2}$ &    $77.5_{-20.1}^{+14.5}$ &    $80.2_{-21.9}^{+13.2}$ \\
w$_\textrm{envelope}$ [\%] &    $28.3_{-18.4}^{+16.6}$ &    $28.1_{-18.0}^{+16.9}$ &    $28.2_{-17.9}^{+16.5}$ &
                           $1.8_{-0.4}^{+0.4}$ &    $2.3_{-0.6}^{+0.6}$ &    $2.1_{-0.7}^{+0.8}$ \\
\hline
Z$_\textrm{envelope}$ [\%] &        $83.3_{-18.4}^{+7.1}$ &    $81.5_{-20.3}^{+7.8}$ &    $83.6_{-18.8}^{+7.9}$ &
                           $0.5_{-0.2}^{+0.2}$ &    $0.5_{-0.2}^{+0.2}$ &    $0.5_{-0.2}^{+0.2}$ \\
\hline
x$_\textrm{Fe,core}$ [\%] &     $90.2_{-6.3}^{+6.6}$ &    $90.2_{-6.3}^{+6.6}$ &    $90.2_{-6.3}^{+6.6}$ &
                           $90.4_{-6.4}^{+6.5}$ &    $90.4_{-6.4}^{+6.5}$ &    $90.4_{-6.4}^{+6.5}$ \\
x$_\textrm{S,core}$ [\%] &      $9.8_{-6.6}^{+6.3}$ &    $9.8_{-6.6}^{+6.3}$ &    $9.8_{-6.6}^{+6.3}$ &
                           $9.6_{-6.5}^{+6.4}$ &    $9.6_{-6.5}^{+6.4}$ &    $9.6_{-6.5}^{+6.4}$ \\
\hline
x$_\textrm{Si,mantle}$ [\%] &   $42.1_{-9.3}^{+9.8}$ &    $36.6_{-11.5}^{+12.0}$ &    $30.6_{-23.0}^{+31.1}$ &
                           $42.0_{-9.3}^{+9.8}$ &    $36.7_{-11.3}^{+12.1}$ &    $33.9_{-23.7}^{+30.2}$ \\
x$_\textrm{Mg,mantle}$ [\%] &   $42.5_{-9.7}^{+9.8}$ &    $36.7_{-12.0}^{+12.3}$ &    $39.6_{-26.9}^{+35.8}$ &
                           $42.5_{-9.8}^{+9.8}$ &    $36.8_{-11.9}^{+12.3}$ &    $36.8_{-25.1}^{+31.2}$ \\
x$_\textrm{Fe,mantle}$ [\%] &   $15.3_{-9.9}^{+8.8}$ &    $24.4_{-16.6}^{+20.1}$ &    $19.0_{-14.9}^{+25.3}$ &
                           $15.4_{-10.0}^{+8.8}$ &    $24.1_{-16.4}^{+19.9}$ &    $20.3_{-14.9}^{+23.9}$ \\
\hline
\end{tabular}
\label{tab:internal_structure_results_b}
\end{table*}
\renewcommand{\arraystretch}{1.0}

\begin{table*}
\renewcommand{\arraystretch}{1.5}
\caption{Same as Table \ref{tab:internal_structure_results_b} but for TOI-1803 c.}
\centering
\begin{tabular}{r|ccc|ccc}
\hline \hline
Water prior &              \multicolumn{3}{c|}{Formation outside iceline (water-rich)} & \multicolumn{3}{c}{Formation inside iceline (water-poor)} \\
Si/Mg/Fe prior &           Stellar (A1) &       Iron-enriched (A2) &      Free (A3) &
                           Stellar (B1) &       Iron-enriched (B2) &      Free (B3) \\
\hline
w$_\textrm{core}$ [\%] &        $10.2_{-7.0}^{+8.4}$ &    $14.4_{-10.1}^{+14.9}$ &    $13.1_{-9.5}^{+15.9}$ &
                           $14.1_{-9.6}^{+9.8}$ &    $19.7_{-13.8}^{+18.7}$ &    $17.6_{-12.8}^{+20.4}$ \\
w$_\textrm{mantle}$ [\%] &      $58.7_{-15.7}^{+16.0}$ &    $51.7_{-15.8}^{+19.3}$ &    $52.9_{-16.8}^{+19.9}$ &
                           $77.9_{-10.0}^{+9.6}$ &    $71.9_{-19.0}^{+14.0}$ &    $74.2_{-20.8}^{+13.1}$ \\
w$_\textrm{envelope}$ [\%] &    $29.3_{-16.9}^{+18.8}$ &    $29.8_{-16.8}^{+18.6}$ &    $29.4_{-16.8}^{+18.8}$ &
                           $8.4_{-2.4}^{+1.2}$ &    $8.9_{-2.5}^{+1.5}$ &    $8.6_{-2.5}^{+1.7}$ \\
\hline
Z$_\textrm{envelope}$ [\%] &        $58.5_{-24.1}^{+10.6}$ &    $57.9_{-24.6}^{+10.7}$ &    $58.4_{-24.3}^{+10.7}$ &
                           $0.5_{-0.2}^{+0.2}$ &    $0.5_{-0.2}^{+0.2}$ &    $0.5_{-0.2}^{+0.2}$ \\
\hline
x$_\textrm{Fe,core}$ [\%] &     $90.3_{-6.4}^{+6.6}$ &    $90.3_{-6.4}^{+6.5}$ &    $90.3_{-6.4}^{+6.6}$ &
                           $90.3_{-6.4}^{+6.6}$ &    $90.3_{-6.4}^{+6.5}$ &    $90.4_{-6.4}^{+6.5}$ \\
x$_\textrm{S,core}$ [\%] &      $9.7_{-6.6}^{+6.4}$ &    $9.7_{-6.5}^{+6.4}$ &    $9.7_{-6.6}^{+6.4}$ &
                           $9.7_{-6.6}^{+6.4}$ &    $9.7_{-6.5}^{+6.4}$ &    $9.6_{-6.5}^{+6.4}$ \\
\hline
x$_\textrm{Si,mantle}$ [\%] &   $42.1_{-9.4}^{+9.8}$ &    $36.4_{-11.3}^{+12.1}$ &    $32.5_{-22.8}^{+30.4}$ &
                           $42.0_{-9.3}^{+9.8}$ &    $36.5_{-11.4}^{+12.1}$ &    $33.1_{-23.1}^{+29.7}$ \\
x$_\textrm{Mg,mantle}$ [\%] &   $42.5_{-9.7}^{+9.8}$ &    $36.4_{-11.8}^{+12.3}$ &    $36.5_{-25.0}^{+30.3}$ &
                           $42.5_{-9.7}^{+9.8}$ &    $36.5_{-11.8}^{+12.4}$ &    $36.4_{-24.6}^{+30.2}$ \\
x$_\textrm{Fe,mantle}$ [\%] &   $15.4_{-10.1}^{+8.8}$ &    $25.0_{-16.9}^{+19.8}$ &    $22.1_{-15.9}^{+24.1}$ &
                           $15.4_{-10.0}^{+8.8}$ &    $24.9_{-16.9}^{+19.7}$ &    $21.9_{-15.9}^{+24.0}$ \\
\hline
\end{tabular}
\label{tab:internal_structure_results_c}
\end{table*}
\renewcommand{\arraystretch}{1.0}

\end{document}